%% file: main.tex
\PassOptionsToPackage{dvipsnames}{xcolor}
\documentclass{article}  
\usepackage[T1]{fontenc}
\usepackage{graphicx} 
\usepackage{geometry}
\usepackage[disable]{todonotes}
\usepackage{amsmath}
\usepackage{amssymb}
\usepackage{amsmath,amssymb,amsfonts}
\usepackage[hidelinks]{hyperref}
\usepackage{geometry}
\usepackage{stmaryrd}
\usepackage{cleveref}
\usepackage{authblk}
\usepackage{etoolbox}

\usepackage{listings}
\usepackage{tikz}
\usepackage{bbm}
\usepackage{natbib}
\usepackage{bussproofs}
\usepackage{ifthen}
\usepackage{enumitem}
\usepackage{tikz-cd}
\usepackage{tabularx}
\usepackage{subcaption}
\usepackage{overpic}
\usepackage{amsthm}
\usetikzlibrary{positioning,arrows.meta}
\usepackage[dvipsnames]{xcolor}
\definecolor{Notecolor}{rgb}{0.0,0.0,0.0}

\usetikzlibrary{positioning,calc}

\newcommand{\CatDQBSL}{\mathbf{DQBS}_\Trace}
\newcommand{\CatDQBS}{\mathbf{DQBS}}

\newcommand{\CatAffCone}{\mathbf{AffCone}}
\newcommand{\CatStoch}{\mathbf{Stoch}}

\newcommand{\op}{\mathsf{op}}

\newcommand{\PSDpoly}{\PSD_{\mathrm{poly}}}
\newcommand{\PSDplot}{\PSD_{\mathrm{plot}}}
\newcommand{\PSDW}{\PSD_{\mathrm{weight}}}

\newcommand{\tensorunit}{\mathbf{1}}

\newcommand{\CatSmooth}{\mathbf{C}^\infty}

\newcommand{\CatSyn}{\mathbf{Syn}}

\newcommand{\CatFinTrace}{\mathbf{FinTrace}}
\newcommand{\CatFinTraceWeighted}{\mathbf{WFinTrace}}
\newcommand{\CatC}{\mathcal{C}}

\newcommand{\Hom}{\mathbf{Hom}}
\newcommand{\CatD}{\mathcal{D}}

\newcommand{\Realisation}{\mathcal{R}}
\newcommand{\Trace}{\mathcal{T}}
\newcommand{\Tracescore}{\mathcal{T}_\mathsf{score}}

\newcommand{\Affine}{\mathcal{A}}
\newcommand{\AffineW}{\mathbb{A}}

\newcommand{\normaliseF}{\mathcal{N}}

\newcommand{\Jet}{\mathcal{J}}
\newcommand{\FisherDirac}{\mathfrak{G}}
\newcommand{\Fisher}[1]{g^{#1}}
\newcommand{\GWFR}{\mathfrak{G}_{\text{WFR}}}
\newcommand{\Gout}{\mathfrak{G}_{\text{out}}}

\newcommand{\V}{\operatorname{V}}

\newcommand{\Ind}{\mathbb 1}
\newcommand{\Simplex}[1]{\Delta_{#1}}
\newcommand{\Simplexo}[1]{\Delta^\circ_{#1}}

\newcommand{\transp}{^{\mathsf T}}
\newcommand{\E}{\mathbb{E}}

\newcommand{\supp}{\textsf{supp}}
\newcommand{\Ctest}{C^{\infty}_{\text{test}}}

\newcommand{\PPL}{\mathcal L}
\newcommand{\PPLscore}{\mathcal L_{\text{score}}}
\newcommand{\RR}{\mathbb{R}}
\newcommand{\RRplus}{\mathbb{R}_{> 0}}

\newcommand{\RRc}{\mathbb{R}_c}
\newcommand{\RRs}{\mathbb{R}_s}
\newcommand{\R}{\mathcal{R}}
\newcommand{\Rc}{\mathcal{R}_c}
\newcommand{\Rs}{\mathcal{R}_s}
\newcommand{\Rsquare}{\mathcal{R}_\square}
\usepackage[bb=boondox]{mathalfa}
\newcommand{\inleft}{\mathsf{left}}
\newcommand{\inright}{\mathsf{right}}

\newcommand{\semisnneg}[1]{\begin{cases}
\true & \text{if } #1 \geq 0\\
\false & \text{otherwise}
\end{cases}
}

\newcommand{\semcoerce}{\mathbf{coerce}}

\newcommand{\Prim}{\mathsf{Prim}}

\newcommand{\im}{\operatorname{im}}

\newcommand{\OmegaUL}{\Omega_\PPL}
\newcommand{\OmegaULR}{\Omega_{\PPL_\RR}}

\newcommand{\Mstd}{\mathcal M_{\mathrm{std}}}
\newcommand{\Ddisc}{\mathcal D_{\mathrm{disc}}}

\newcommand{\true}{\mathsf{true}}
\newcommand{\false}{\mathsf{false}}

\newcommand{\unitset}{\{\star\}}
\newcommand{\unitsetelem}{\star}

\newcommand{\isnneg}[1]{[#1 \geq 0]}

\newcommand{\emptyctx}{\varnothing}
\newcommand{\emptytype}{\mathbb{0}}
\newcommand{\unittype}{\mathbb{1}}

\newcommand{\coerce}[1]{\mathsf{coerce}(#1)}
\newcommand{\unitelem}{\star}
\newcommand{\NN}{\mathbb{N}}
\newcommand{\Bool}{\mathsf{Bool}}
\newcommand{\sem}[1]{\llbracket #1 \rrbracket}

\newcommand{\kto}{\rightsquigarrow}

\newcommand{\Syn}{\mathbf{Syn}}

\newcommand{\Prob}{\mathcal P}
\newcommand{\Meas}{\mathcal M}

\newcommand{\copyK}{\Delta}
\newcommand{\delK}{\operatorname{del}}
\newcommand{\id}{\operatorname{id}}
\newcommand{\KL}{D_{\mathrm{KL}}}
\newcommand{\Exp}{\mathbb E}

\newcommand{\diverg}{\operatorname{div}}
\newcommand{\diag}{\operatorname{diag}}
\newcommand{\dd}{\,\mathrm d}

\newcommand{\N}{\mathbb N}

\newcommand{\abs}[1]{\left\lvert #1\right\rvert}
\newcommand{\norm}[1]{\left\lVert #1\right\rVert}
\newcommand{\ip}[2]{\left\langle #1,#2\right\rangle}
\newcommand{\semt}[1]{\mathbf{#1}}
\newcommand{\df}{\mathrel{:=}}

\newcommand{\QBS}{\mathbf{QBS}}

\newcommand{\Dens}{\mathbb D}
\newcommand{\Denso}{{\mathbb D}^\circ}

\newcommand{\ETrace}[1]{\mathcal E_{\Omega}[#1]}
\newcommand{\ETraceo}[1]{\mathcal E_{\Omega}^{\circ}[#1]}
\newcommand{\ETraceAt}[2]{\mathcal E_{#1}[#2]}

\newcommand{\ETraceColl}[1]{\overline{\mathcal E_{\Omega}[#1]}}
\newcommand{\ETraceoIn}[2]{\mathcal E_{#1}^{\circ}[#2]}
\newcommand{\ETraceUniv}[1]{\mathfrak E[#1]}
\newcommand{\ETraceW}[1]{{\mathcal E}^\mathbf{W}_{\Omega}[#1]}

\newcommand{\Law}{\mathbb Q}
\newcommand{\Output}{\mathbb O}
\newcommand{\ClassMap}[1]{\widehat #1}
\newcommand{\FisherMetric}{g^{\mathrm F}}

\newcommand{\curry}{\operatorname{curry}}

\newcommand{\Scal}{\operatorname{Scal}}

\newcommand{\vol}{\operatorname{vol}}

\newcommand{\evmap}{\operatorname{ev}}
\newcommand{\NatGrad}{\nabla^{\mathrm{nat}}}

\newcommand{\PSD}{\mathsf{PSD}}

\newcommand{\vkto}{\Rightarrow}

\newcommand{\ifte}[3]{\mathsf{if}\;#1\;\mathsf{then}\;#2\;\mathsf{else}\;#3}
\newcommand{\letin}[3]{\mathsf{let}\;#1=#2\;\mathsf{in}\;#3}
\newcommand{\letbind}[3]{\mathsf{let}\;#1\leftarrow #2\;\mathsf{in}\;#3}

\newcommand{\gauss}[2]{\mathsf{gauss}(#1,#2)}
\newcommand{\bern}[1]{\mathsf{flip}(#1)}

\newcommand{\dirac}[1]{\delta(#1)}

\newcommand{\return}[1]{\mathsf{return}(#1)}
\newcommand{\sample}[1]{\mathsf{sample}(#1)}
\newcommand{\score}[1]{\mathsf{score}(#1)}
\newcommand{\observe}[1]{\mathsf{observe}(#1)}

\newcommand{\Jeffreys}{\mathfrak{J}}

\newcommand{\logistic}{\sigma}

\newtheorem{theorem}{Theorem}
\newtheorem{proposition}{Proposition}
\newtheorem{definition}{Definition}

\newtheorem{corollary}{Corollary}

\newtheorem{example}{Example}
\newtheorem{lemma}{Lemma}
\newtheorem{remark}{Remark}

\newcommand{\WFR}{\operatorname{WFR}}
\newcommand{\normalise}[1]{\mathsf{normalize}(#1)}

\usepackage{pgfplots}
\usetikzlibrary{arrows.meta,calc,positioning}
\pgfplotsset{compat=1.18}

\definecolor{surfaceLight}{HTML}{DDEAF0}
\definecolor{surfaceMid}{HTML}{9DBDCC}
\definecolor{surfaceDark}{HTML}{64899A}
\definecolor{scoreCurve}{HTML}{C6463A}
\definecolor{transportCurve}{HTML}{275D91}
\definecolor{quietText}{HTML}{52616A}

\pgfmathdeclarefunction{sigmoid}{1}{%
  \pgfmathparse{1/(1+exp(-#1))}%
}

\tikzset{
  program/.style={
    align=left,
    text width=5.35cm,
    inner sep=0pt,
    font=\small
  },
  inclusion/.style={
    -{Latex[length=2.2mm,width=1.5mm]},
    line width=0.75pt,
    rounded corners=3pt
  }
}

\pgfplotsset{
  small curve/.style={
    hide axis,
    axis equal image,
    scale only axis,
    clip=false
  },
  exact curve/.style={
    no marks,
    line width=2.2pt,
    line cap=round,
    samples=64,
    smooth
  },
  colormap={program surface}{%
    color(0cm)=(surfaceDark)
    color(1cm)=(surfaceMid)
    color(2cm)=(surfaceLight)
  }
}

\title{Compositional Information Geometry of Probabilistic Programs}
\author[1]{Jacek Karwowski}
\author[2]{Frank Nielsen}
\author[1]{Sam Staton}

\affil[1]{{Department of Computer Science, University of Oxford, Oxford, UK}}
\affil[2]{{Sony Computer Science Laboratories, Inc., Tokyo, Japan}}

\date{\today}

\begin{document}

\maketitle

\begin{abstract}
    We study the compositional, intensional information geometry of probabilistic programs. For a first-order probabilistic programming language with finite random variables, we construct a contravariant lax monoidal functor that, for a program of type $X \to Y$, pulls back a (degenerate) Riemannian metric on $Y$ to $X$. It specialises to the Fisher metric in the case of standard statistical models and to ordinary pullback in the case of deterministic programs. We further justify this construction in two ways. First, we show that this functor arises as a pullback of geometry from a universal space of programs over a type $Y$ with a fixed trace space. Second, we show that it is a special case of a more general pullback construction over weighted measures. Our compositional geometry depends on the presentation of the program, and we study how it relates to the (non-compositional) geometry of the realised measures by establishing a three-level inequality relating our score-velocity metric for the program, the program's corresponding output-conditioned metric, and the standard Wasserstein–Fisher–Rao metric of the realised family. Finally, we show how the theory can be extended to accommodate continuous random variables using diffeological quasi-Borel spaces. We also discuss several concrete directions for future work.
\end{abstract}

\section{Introduction}\label{sec:intro}
\begin{figure}
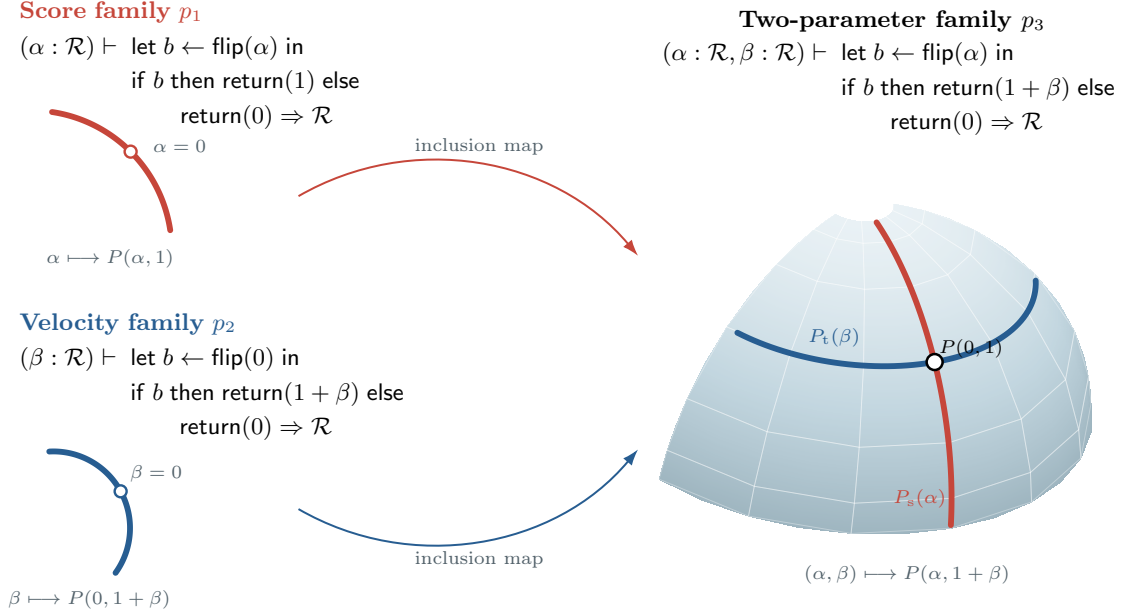

    \centering
    \include{figures/figure_sphere}
    \caption{Visualisation of the geometry of three program families: $p_1$ and $p_2$ embed into a common $p_3$. The first family is a biased coin flip, with the standard Fisher geometry on the parameter space. The second family denotes a fair coin choosing between 0 and a number $1+\beta$, varying in magnitude, with standard Euclidean geometry on the parameter space. The geometry on the two-parameter family combines the output (velocity) component and the Fisher (score) component, resulting in the geometry being locally spherical. It pulls back onto the one-dimensional families. The families $p_1, p_2$ intersect transversally in $p_3$ at $\alpha = 0$ and $\beta = 0$. Note that the Bernoulli distribution is parametrised by log-odds, so $\bern{0}$ denotes a fair coin. The plot is rescaled to standard unit-interval coordinates in $\alpha$, and clipped to a small range of $\beta$ around $0$: the spherical geometry wraps around infinitely many times around the blue coordinate.}
    \label{fig:one}
\end{figure}

Information geometry tells us that families of probability distributions are often equipped with intrinsic geometry, independent of parametrisation~\citep{amariInformationGeometryIts2016}. This gives certain such families the structure of Riemannian manifolds, making it possible to talk about the geodesics, curvature of the given family, and so on. Even large, non-parametric collections of distributions can sometimes be given geometric structure, although it tends to be less canonical~\citep{pistoneInfiniteDimensionalGeometricStructure1995,ayInformationGeometry2017b,bauer2024p}.

Probabilistic programs denote certain families of probability distributions. These families are at the same time more- and also less-general than those commonly considered in classical information geometry. More general, because programs often denote mutually singular measures (such as deterministic programs denoting moving Dirac deltas), which is an obstruction to many classical approaches, and because they subsume simple models such as exponential families, for which explicit calculation is possible. But they are also less general, because programs are highly structured: there are very specific rules of composition, and thus only very particular statistical models are denotable with programs.

When considering a program $p$ with inputs of type $X$ and (stochastic) outputs of type $Y$, written:
\begin{equation*}
X \vdash p \vkto Y
\end{equation*}
we will think of it as pulling back the metric on $Y$ to a metric on $X$ in a certain way. The magnitude of an infinitesimal change in the parameters~$X$ will depend on the corresponding magnitude of infinitesimal change in the output~$Y$, or on the change in output distribution. This already exemplifies the issue of greater-and-lesser generality: we subsume finite-dimensional models (since the output type $Y$ can be continuous, with a non-trivial metric structure), but in a controlled manner, not requiring heavy machinery of non-parametric theory. 

Thus, to programs we assign certain geometric spaces: for a program like $p$ it will be a smooth manifold on the set $X$ equipped with a degenerate Riemannian metric. When composing two programs, for example sequentially:
\[
(x: X) \vdash p \vkto Y \qquad \text{and} \qquad (y:Y) \vdash q \vkto Z \qquad \text{into}\qquad (x:X) \vdash \letbind{y}{p}{q}\vkto Z
\]
our essential desideratum (1) is to exploit this compositional structure, and for the pull-back metric to be composed through the pullbacks of $p$ and $q$. To fix the base cases, we also want (2) the geometry to be induced by the Fisher metric in case of standard statistical models, and (3) to be given by an ordinary pullback in case of deterministic programs. If one program is an instantiation of another, it should be possible to understand it as its embedded submanifold; if two programs partially overlap, they should be intersecting geometrically too. We show a visualisation of the true geometry of three programs in~\Cref{fig:one}.

It turns out that these desiderata force us to consider programs intensionally. That is, when a program induces a transformation of metrics on the output to metrics on the input, this transformation has to depend on the implementation, not only on the realised measure. Let us illustrate this with an example. Consider a program denoting a coin toss (c.f.~\Cref{fig:one}), and two deterministic programs into the unit type~$\unittype$ which ignore their argument and output a constant value:
\[
(\alpha: \RR)\vdash \bern{\alpha}\vkto \Bool \qquad \Bool\vdash \return{\unitelem} \vkto \unittype \qquad \RR\vdash \return{\unitelem} \vkto \unittype
\]
On the discrete sets of $\unittype$ and $\Bool$ only the trivial metrics $0_\unittype$ and $0_\Bool$ exist. Suppose that we have a way of pulling back the metric by $\Affine_{(\cdot)}$ from the program codomain to its domain. Now, our desiderata for compositionality and two base-case requirements from above imply that:
\[
\Fisher{\bern{\alpha}} \stackrel{(2)}{=} \Affine_{\bern{\alpha}}(0_\Bool) \stackrel{(3)}{=} \Affine_{\bern{\alpha}}(\Affine_{\return{\unitelem}}(0_{\unittype})) \stackrel{(1)}{=} \Affine_{\letbind{b}{\bern{\alpha}}{\return{\unitelem}}}(0_\unittype) \qquad \text{and} \qquad \Affine_{\return{\unitelem}}(0_\unittype) \stackrel{(3)}{=} 0_\RR
\]
But the Fisher geometry of a Bernoulli random variable is not trivial, so necessarily we have:
\[
\Affine_{\letbind{b}{\bern{\alpha}}{\return{\unitelem}}} \neq \Affine_{\return{\unitelem}}
\]
even though extensionally these programs are equal. In other words, under our assumptions (1)-(3), random choices made during the execution of the program, recorded as its trace, have to impact its geometry, even if they do not show up in the realised measure. 

Nevertheless, the extensional geometry of the realised measure and the intensional geometry of the program are intimately related: we show that when the execution information is (partially or fully) forgotten, the geometry contracts in a controlled way. To do that, we introduce a statistic $s$, which forgets some of the trace information. Thus, the phrase ``the geometry of the program'' can be said to be ambiguous: one program will have different geometries depending on whether we observe all random choices made during the execution, only its return value, or even something in between: a more complex statistic of the execution, such as the branch taken in a conditional. 

A related issue convenient to investigate in the same setup is that one program $X \vdash p \vkto Y$ can be fed many combinations of values and reused in different contexts, such as supplying a collection of i.i.d. samples, or hyperparameters. For that, we can also introduce a deterministic map $r: M \to X$ which denotes one particular statistical experiment. We then get the slogan:
\begin{center}
\emph{Geometry pulls back along $r$, composes through $p$, and contracts along $s$.}    
\end{center}

\subsection{Structure}
We start in~\Cref{sec:ppl-jets} by looking at a simple first-order generative probabilistic programming language $\PPL$ with finite random variables and algebraic Euclidean types. We consider semantics of a program $X \vdash p \vkto Y$ in $\PPL$ to consist of a set $\Omega$ and a pair of maps:
\[
\RR_{\geq 0} \xleftarrow{\quad q \quad} \Omega \times X \xrightarrow{\quad o \quad} Y
\]
where $q$ is a (parametrised) density over the trace space $\Omega$, and $o$ is a deterministic output map. To each such triple $(\Omega, q, o)$ we attach the score $S$ of $q$ and the derivative $V$ of $o$ called velocity: we call the representation $(\Omega, q, o; S,V)$ a jet trace, and we show that they can be quotiented to form a category with good properties.

\Cref{sec:algebraic} considers the algebraic formulation of the titular problem. We show how a program with inputs in $X$ and outputs in $Y$ can be thought of as pulling back the geometry of $Y$ onto $X$, compositionally. Then, the geometry of programs arises as a lax monoidal contravariant functor $\Affine$ between the category of jet traces and the category of affine cones, in particular cones of degenerate Riemannian forms on algebraic Euclidean spaces. For a jet trace $F = (\Omega, q, o; S,V)$ and a positive-semidefinite form $h$ on the codomain, it is given by:
\[
\Affine_F(h)[\xi, \eta] = \E_{q}[S_\xi S_\eta + h(V_\xi, V_\eta)]\qquad \text{or abstractly, simply as} \qquad \Affine_F(h) = \Fisher{F} + F^\sharp h
\]
that is, the sum of the Fisher metric for $F$, and the metric $h$ pulled back with $F$ and averaged.

\Cref{sec:geometric} considers a more geometric approach. This means looking at the moduli space $\ETraceo{Y}$ of programs of a given type $Y$, which, in the finite case, is $\Simplexo{\Omega}\times Y^\Omega$. We then show that the functor $\Affine$ from~\Cref{sec:algebraic} arises simply as a pullback of the canonical geometry on that moduli space $\ETraceo{Y}$, and characterise it. We also show that $\ETraceo{Y}$ can be considered a hypersurface in a product of cones (in the square-root coordinates), which allows us to explicitly compute the curvature, Levi-Civita connection, and Riemannian volume form (on regular families).

In~\Cref{sec:observe-score-normalise} we give an extension from the purely generative fragment to $\observe{} / \score{}$, i.e., non-normalised measures. We develop a compositional weighted version $\mathbb{A}$ of the functor $\Affine$, of which it is a special (normalised) case. We also explain how normalisation interacts with geometry.

In~\Cref{sec:pplr}, we arrive at the analytic formulation, which extends the base language $\PPL$ with continuous random variables, forming $\PPL_\RR$. The theory presented in the first two parts extends properly, but requires more functional-analytic care, such as additional restrictions on the tails of both degenerate-Riemannian metrics and primitives in $\PPL_\RR$, to guarantee integrability of relevant quantities. The semantics also requires separating smooth and discrete real number types. To properly discuss this, as well as to form infinite-dimensional moduli spaces $\ETrace{Y}$, we combine the theories of quasi-Borel and diffeological spaces, forming a category $\CatDQBS$ (diffeological quasi-Borel space), and prove that it carries semantics for $\PPL_\RR$.

In~\Cref{sec:many-geometries} we discuss how the geometry of programs extends in two directions: marginalising over traces (i.e., when the full trace is not observed) and pulling back on the inputs. This allows us to introduce the natural gradient on program parameter spaces (induced by the score-velocity metric). We also show that $\Affine(\ip{\cdot}{\cdot})$ is closely related to the Wasserstein-Fisher-Rao metric: specifically, that it is a (compositional) upper bound for it, tight for regular finite output-measurable atomic families.

Finally, in~\Cref{sec:future-work}, we outline many concrete ideas for future work. These include loops (via stopping-time and martingale theory), a universal length-geometric prior on the space of probabilistic programs (defining Bayesian posteriors and Bayesian non-parametric regression under a global uninformative prior), trans-dimensional Riemannian Manifold Hamiltonian Monte Carlo algorithms based on the universal posterior, connections to the global (divergence) geometry of programs, the full dualistic information geometry, and computational issues regarding the approximation of the natural gradient and variational algorithms in this setup.

\section{Related work}\label{sec:related-work}
We build this work as a bridge between information geometry and probabilistic programming, using tools from category theory and programming language semantics. There is a wealth of related literature on each of these topics, as well as some alternative approaches to program geometry developed independently, some of them recently or concurrently with this work.

\subsection{Information geometry}
Information geometry has been developed by~\citet{cencovStatisticalDecisionRules2000} and \citet{amariDifferentialGeometryCurved1982,amariInformationGeometryIts2016}. As a reference, we have mostly used the modern treatment in the book by~\citet{ayInformationGeometry2017b}, which also rigorously covers the non-parametric case, as well as other introductory tutorials~\citep{nielsenElementaryIntroductionInformation2020,nielsenManyFaces,baez2011information}. One version of non-parametric information geometry was developed earlier by~\citet{pistoneInfiniteDimensionalGeometricStructure1995}, described for the Gaussian model specifically in~\citep{pistoneInformationGeometryGaussian2018}. Another approach to the geometry of probabilistic models is through the Wasserstein metric: in~\Cref{sec:many-geometries}, we compare our metric to the Wasserstein-Fisher-Rao metric, a generalisation and unification of Wasserstein and Fisher-Rao geometry developed by~\citet{chizatInterpolatingDistanceOptimal2015}.~\citet{fujitaniInformationGeometryWarped2022} studies information geometry for warped product metrics, and finds which dual connections are compatible with a warped metric, which is complementary to our work of deriving such a metric from a compositional perspective.

The framework of information geometry has many applications beyond the common statistical model cases; for example, it has been used in optimisation (e.g., mirror descent~\citep{raskutti2015information}) or mathematical programming (e.g., semi-definite programming~\citep{ohara1999information}), in game theory~\citep{harper2009information}, in control theory (e.g., linear systems~\citep{amari1987differential}), in causal inference~\citep{janzing2012information}, etc. We hope that by placing probabilistic programming within the scope of information geometry, some of the existing richness of applications finds use in programming languages as well.

\subsection{Probabilistic and differentiable programming languages}
Probabilistic programming theory and practice are described in depth in the introductory book of~\citet{meentIntroductionProbabilisticProgramming2021}. The primary construction of~\Cref{def:trace-presentation} is directly inspired by the split into a density-output pair of~\citet{staton2020bayesian}.~\citet{lewADEVSoundAutomatic2023} also focus on the interaction between differential and probabilistic structure of probabilistic programs, and their tricks of looking at the full trace rather than raw output and the separate smooth and control types $\RRs, \RRc$ (in~\Cref{sec:pplr}) is essential for our work, but they focus on the differentiation of programs' expectations, rather than their differential (information) geometry. PAP and $\omega$PAP spaces~\citep{leeCorrectnessAutomaticDifferentiation2020, huotOmegaPAPSpaces2023} allow for discontinuities in the domain by tracking the boundaries, and would allow us to generalise our treatment further, but we have not attempted it here.~\citet{dashAffineMonadsLazy2022b} develop a language for lazy probabilistic programming, which, in combination with non-parametric approaches cited earlier, could potentially allow us to generalise our treatment with loops.

\subsection{Categorical probability and categorical information geometry}
We use the language of categorical probability theory to develop our semantics, in line with previous works on PPL semantics. In~\Cref{sec:pplr}, we develop the category $\CatDQBS$, combining quasi-Borel spaces and diffeological spaces. Quasi-Borel spaces were specifically developed as a setting for probabilistic programming semantics, in~\citep{heunenConvenientCategoryHigherOrder2017}. Diffeological spaces were used by~\citet{huotCorrectnessAutomaticDifferentiation2020a} for semantics for differentiable programming. As a reference, we used~\cite{baezConvenientCategoriesSmooth2009}. In our work, we also use the terminology of Markov or copy-discard categories, by~\citet{fritzSyntheticApproachMarkov2020b} and~\citet{choDisintegrationBayesianInversion2019}.

\citet{perroneMarkovCategoriesEntropy2024} introduced a notion of divergence-enriched categories, which is another approach to handling information geometry categorically - we briefly outline how the divergence viewpoint could be connected to our work in~\Cref{sec:future-work}.
~\citet{ciagliaCategoryTheoreticFoundationClassical2026} built a different fundamental theory of categorical information geometry, using the category of non-commutative probability spaces, which is a setting for both classical and quantum probability. Comparing our approach to theirs, in particular for the construction of natural gradient, is a subject for future work.
Compositional information geometry could in principle also be approached from the more general angle of synthetic differential geometry~\citep{kostecki2009differential}. Recent work by ~\citet{gardarssonDaggerCategoriesRiemannian2026} relates metric structures on vector spaces and vector bundles to daggers on the corresponding categories, although it stops short of applying this to information geometry specifically. Concurrent with our work,~\citet{forreSampleSmoothSpacesConvenient2026} developed Sample-Smooth Spaces as a setting for differentiable probabilistic programming, which could be an alternative to $\CatDQBS$, as we note in~\Cref{remark:dqbs-monad}.

\subsection{Diffeological spaces, stratifolds, orbifolds}
The geometry of diffeological spaces is more permissive, but more subtle, than that of smooth manifolds. As a reference, we used the introductory text by~\cite{iglesias-zemmourWhyDiffeology2025}. For the construction of the universal moduli space in~\Cref{sec:pplr}, we were inspired by an earlier work of~\cite{leDiffeologicalStatisticalModels2020}, who developed information geometry specifically in a diffeological setting. Other theories of singular geometry we considered in earlier versions of this work included stratifolds~\citep{esserModelingResolvingSingular2021} and orbifolds~\citep{harmsGeometrySampleSpaces2023}.

\subsection{Program geometry and inference over programs}
There are other approaches to (not necessarily probabilistic) program geometry.~\citet{cliftGeometryProgramSynthesis2021a} apply Singular Learning Theory to the problem of Turing machine synthesis in a Bayesian setup, and model programs themselves as singularities of analytic varieties. Recent work by~\citet{plummerObservableGeometrySingular2026} focuses on singular geometry of models independent of parametrisation.~\cite{barbarossaTropicalMathematicsLambdaCalculus2025} use tropical geometry to analyse probabilistic programs. \citet{richardson2024mixture} study a language of continuous processes, and find an interesting duality between $\observe{\cdot}$ and $\sample{\cdot}$ statements: their perspective seems complementary to ours, since their starting point is the existence of Riemannian structure.

Exponential families are a simple class of statistical models that have well-studied information geometry, but they are not closed under composition:~\citet{scherrer2017exponential} considered a free arrow-based formalism closing them under sequential composition, but did not attempt to describe the resulting geometry.

One of our starting motivations for developing geometric semantics for probabilistic programs was to be able to unify noninformative geometric priors such as Jeffreys prior with Occam's razor or minimal description length prior for standard programs, described by~\citet{grunwaldMinimumDescriptionLength2019a}. We are leaving it for future work, but the diffeological Hausdorff-Jeffrey measure introduced by~\citet{leNonparametricEstimationsDiffeological2020} could be the basis for such a prior.

\subsection{Natural gradient}
Another starting motivation for this work was to develop a natural gradient algorithm~\citep{martensNewInsightsPerspectives2020} for probabilistic programs, that is, a parametrisation-invariant gradient descent in the space of measures. Computing natural gradient for dense models is expensive, as a second-order method, but the sparse structure of programs could potentially make it more efficient. Most famously, the Adam optimiser~\citep{kingmaAdamMethodStochastic2017a} uses a rough diagonal approximation to the Fisher update. K-FAC~\citep{martensOptimizingNeuralNetworks2020} is one of the well-known algorithms to approximate natural gradient, using a sparsity assumption on the Fisher information matrix in the form of a Kronecker product decomposition.~\citet{sunRelativeFisherInformation2017a} describe another algorithm for computing relative Fisher information, exploiting a different sparsity structure, dividing a large dense system into weakly-interacting subsystems. None of these exploit a known a-priori, exact decomposition of the form we study here.

\section{Language $\PPL$ and its jet trace semantics}\label{sec:ppl-jets}
We start with a very simple first-order generative probabilistic programming language $\PPL$, with algebraic Euclidean data types and finite random variables.
\begin{align*}
\text{(Types)}\ X &\df \emptytype \mid \unittype \mid \R \mid X\times Y \mid X + Y \\
\text{(Variables)}\ v &\df x \mid y \mid z \mid \ldots \\
\text{(Contexts)}\ \Gamma &\df \emptyctx \mid \Gamma,(v:X) \\
\text{(Pure terms)}\ t &\df v \mid q \mid \unitelem  \\
&\mid t_1 + t_2 \mid t_1 \cdot t_2 \mid - t \mid f(t) \\
&\mid (t_1,t_2) \mid \pi_1 t \mid \pi_2 t \\
&\mid \inleft(t) \mid \inright(t) \mid \ifte{t}{t_1}{t_2}\\
&\mid \letin{v}{t_1}{t_2} \\
\text{(Programs)}\ p &\df \return{t} \mid \letbind{v}{p_1}{p_2}\\
&\mid \bern{t}\\
&\mid \ifte{t}{p_1}{p_2}
\end{align*}
Above, we let constants $q$ vary across rational numbers $\mathbb{Q}$, and we allow some fixed finite set of symbols $f(t) \in \Prim(\PPL)$ which will denote a collection of smooth maps $\RR^n \to \RR^
m$. Also, to simplify notation, we write $\Bool$ for the type $\unittype + \unittype$ and denote its elements $\false \df \inleft(\unitelem), \true \df \inright(\unitelem)$. 

\subsection{Typing rules}\label{ssec:ppl-typing-rules}
We use the following two judgements for pure terms and programs respectively:
\[
  \Gamma \vdash t:X  \qquad \Gamma \vdash p\vkto X
\]
Full typing rules are listed in~\Cref{appendix:typing}; they are largely standard. For now, we only have a single source of randomness, a Bernoulli random variable:
\[
\begin{minipage}[t]{0.5\linewidth}\centering
\begin{prooftree}
\AxiomC{$\Gamma \vdash x:\R$}
\RightLabel{(Bern)}
\UnaryInfC{$\Gamma \vdash \bern{x} \vkto \Bool$}
\end{prooftree}
\end{minipage}
\]
where we take $\bern{x}$ to be parametrised by log-odds $x = \log \frac{p(\true)}{p(\false)}$. Note that the map $\RR \to (0, 1)$ given by $\logistic(x) = 1/(1 + e^{-x})$ is smooth (and has polynomially-bounded tails, which will be important later in~\Cref{sec:pplr}).

\subsection{Types and pure terms}

We define the standard semantics of types and pure terms by induction: we write the semantics of a type $X$ as $\semt{X}$ and of a term $t$ as $\sem{t}$.

\begin{definition}[Algebraic Euclidean space]
    An algebraic Euclidean space is a disjoint union of Euclidean spaces of possibly varying dimensions, assumed to be equipped with the standard measurable structure and the Lebesgue measure, and the standard smooth and Riemannian structure on each component.
\end{definition}

The type $\unittype$ is interpreted as a one-element set $\{\unitelem\}$, the type $\R$ as $\RR$ with the standard Euclidean topology, standard Borel structure, and is assumed to be equipped with the Lebesgue measure $\lambda$. Sums and products are interpreted as algebraic Euclidean spaces. Pure terms $\Gamma \vdash t: X$ are interpreted as functions between algebraic Euclidean types $\sem{t} : \semt{\Gamma} \to \semt{X}$.

\subsection{Denotational semantics in Markov kernels}

We let $\CatSyn_\PPL$ be the syntax category for $\PPL$, where types are objects and programs of shape $\Gamma \vdash p \vkto X$ are morphisms between $\Gamma$ and $X$, modulo the equivalence relation of $\alpha$-conversion, $\beta/\eta$ laws for deterministic morphisms, and monadic equations for left/right unit laws and associativity of composition:
\vspace{1em}
\\
\begin{minipage}{0.48\textwidth}
\centering
\begin{prooftree}
\AxiomC{$\Gamma\vdash t:X$}
\AxiomC{$\Gamma,(x:X)\vdash p\Rightarrow Y$}
\BinaryInfC{$\Gamma\vdash\letbind{x}{\return{t}}{p} \equiv p[t/x]\vkto Y$}
\end{prooftree}
\end{minipage}
\begin{minipage}{0.48\textwidth}
\centering
\begin{prooftree}
\AxiomC{$\Gamma\vdash p\Rightarrow X$}
\UnaryInfC{$\Gamma\vdash\letbind{x}{p}{\return{x}} \equiv p\vkto X$}
\end{prooftree}
\end{minipage}
\begin{prooftree}
\AxiomC{$\Gamma\vdash p\Rightarrow X$}
\AxiomC{$\Gamma,(x:X)\vdash q\Rightarrow Y$}
\AxiomC{$\Gamma,(y:Y)\vdash r\Rightarrow Z$}
\RightLabel{$x\notin \mathrm{FV}(r)$}
\TrinaryInfC{$\Gamma\vdash\letbind{y}{(\letbind{x}{p}{q})}{r} \equiv \letbind{x}{p}{(\letbind{y}{q}{r})}\vkto Z$}
\end{prooftree}
but not $\letbind{x}{p}{q} \equiv q$ (even if $x$ is not free in $q$). We also assume the interchange law:
\begin{prooftree}
\AxiomC{$\Gamma\vdash p\Rightarrow X$}
\AxiomC{$\Gamma\vdash q\Rightarrow Y$}
\AxiomC{$\Gamma,(x:X),(y:Y)\vdash r\Rightarrow Z$}
\RightLabel{$x\notin \mathrm{FV}(q)\ \text{and}\ y\notin \mathrm{FV}(p)$}
\TrinaryInfC{$\Gamma\vdash\letbind{x}{p}{\letbind{y}{q}{r}} \equiv \letbind{y}{q}{\letbind{x}{p}{r}}\vkto Z$}
\end{prooftree}
and the $\beta$ laws for program conditionals:
\vspace{1em}
\\
\begin{minipage}{0.48\textwidth}
\centering
\begin{prooftree}
\AxiomC{$\Gamma\vdash p\Rightarrow X$}
\AxiomC{$\Gamma\vdash q\Rightarrow X$}
\BinaryInfC{$\Gamma\vdash\ifte{\true}{p}{q} \equiv p \vkto X$}
\end{prooftree}
\end{minipage}
\begin{minipage}{0.48\textwidth}
\centering
\begin{prooftree}
\AxiomC{$\Gamma\vdash p\Rightarrow X$}
\AxiomC{$\Gamma\vdash q\Rightarrow X$}
\BinaryInfC{$\Gamma\vdash\ifte{\false}{p}{q} \equiv q \vkto X$}
\end{prooftree}
\end{minipage}
\vspace{1em}
\\
We define standard denotational semantics for programs in Markov kernels in the subcategory of $\CatStoch$ consisting of algebraic Euclidean spaces and Markov kernels, as a functor $\sem{\cdot}$, where a program $\Gamma \vdash p \vkto X$ denotes a kernel $\sem{p} = K: \semt{\Gamma}\kto \semt{X}$. The semantics functor on pure terms lifts to $\CatStoch$, so there is no ambiguity with the notation.

We note one crucial property of the language, that is, smooth inputs cannot impact the values of pure discrete outputs. We will maintain this key property throughout the development of the language, even after adding more complicated, real-valued random variables in~\Cref{sec:pplr}.

\begin{proposition}[Finite values are locally constant]
\label{prop:finite-values-locally-constant}
Let \(\Gamma\vdash t:\Bool\) be a pure boolean term. Then, on each Euclidean component of $\semt\Gamma$, the denotation $\sem t$ is constant.
\end{proposition}
\begin{proof}
    There is no operation in the language turning a continuous term into a discrete one. Thus, Boolean output terms can only depend on the branch component tags of the input.
\end{proof}

Differential-geometric semantics also require smoothness, which is preserved in our language $\PPL$ as well.

\begin{proposition}\label{prop:pure-functions=are-smooth}
    For any pure term $\Gamma \vdash t: X$, the corresponding function $\sem{t}: \semt{\Gamma} \to \semt{X}$ is smooth.
\end{proposition}
\begin{proof}
The conditionals are only allowed on Boolean types:
\begin{prooftree}
\AxiomC{$\Gamma\vdash b:\Bool$}
\AxiomC{$\Gamma\vdash t_1:T$}
\AxiomC{$\Gamma\vdash t_2:T$}
\RightLabel{(If)}
\TrinaryInfC{$\Gamma\vdash \ifte{b}{t_1}{t_2}:T$}
\end{prooftree}
as a consequence of~\Cref{prop:finite-values-locally-constant} the Boolean values are locally constant, and the primitives $\Prim(\PPL)$ are assumed to only contain smooth functions.
\end{proof}

As a consequence of the above, once we move to geometric semantics, for every $\gamma\in\semt\Gamma$ and every tangent $\xi\in T_\gamma\semt\Gamma$, the branch selected by $t$ will remain locally constant along $\xi$.

\subsection{Dominated trace category}

Information geometry in the Fisher sense usually considers dominated families of distributions. When dealing with probabilistic programs, this assumption gets violated at the very start. A simple deterministic program like:
\[
(\theta: \R) \vdash \return{\theta^3}\vkto \R
\]
denotes the family of Dirac measures $\delta_{\theta^3}$, for which there is no way to assign Fisher information, because this family cannot be described by a family of density functions $f(\theta, x)$ on $\RR$ for any $\sigma$-finite measure $\lambda$ on $\RR$. Thus, the project of Fisher information for programs has to, seemingly, immediately be abandoned before it even got off the ground.

However, it turns out that the lack of a common dominating measure is very mild. As we see in the example of a moving Dirac, the dominating measure exists on the singleton trace space $\unitset$. For each fixed $x$, the singleton trace law is pushed forward by $o(x,\star)=x^3$, giving $\delta_{x^3}$. Thus, one solution is immediate: to consider a program $p: X \kto Y$ as a pair $(q, o)$, where the map $q: X \kto \Omega$ will be the stochastic (uniformly dominated) kernel onto random traces $\Omega$, and the map $o: X \times \Omega \to Y$ will deterministically push the pair $(x, \omega)$ onto the output $y = o(x, \omega)$. In this way, we arrive at the following definition.
\begin{definition}[Trace presentation]\label{def:trace-presentation}
    For algebraic Euclidean spaces $X, Y$, we say that a dominated trace presentation from $X$ to $Y$, written $F: X \to Y$, is a triple $F = (\Omega, q, o)$, where $\Omega$ is a measurable space with a $\sigma$-finite measure $\lambda$, $q: X \times \Omega \to [0, \infty) $ is a density wrt $\lambda$, such that $\int q(x, \omega)\lambda(d\omega) = 1$ and $\sqrt{q(-, \omega)}$ is smooth, and $o: X \times \Omega \to Y$ is a measurable map, smooth in the first component.
\end{definition}
In the remainder of this paper, we will use $Q$ to denote the ($x$-dependent) measure $q\lambda$, and $Q_x$ to denote the measure $q(x, -)\lambda$ for a fixed $x$.

\begin{remark}
    In the image of the $\PPL$ semantics, $q$ is locally either identically zero or strictly positive (because $\bern{t}$ is always positive, and conditional branches do not change locally). Consequently, the support of the trace law is locally constant on each smooth source component.
\end{remark}

\begin{remark}
    The requirement of $q(-,\omega)$ being smooth is not a sufficient regularity condition for smooth Fisher information, even in the finite setting. For example, take a Bernoulli random variable with bias given by $q(t,\true) = \frac{1}{1 + t^2}$. This is a smooth function on all $\RR$. But the Fisher information is:
    \[
    g_t = \E_{\omega\sim q(t)}[(\partial_t \log q(t,\omega))^2] = \frac{t^2}{1+t^2} \left(\frac{2}{t(1+t^2)}\right)^2 + \frac{1}{1+t^2} \left(-\frac{2t}{1+t^2}\right)^2 = \frac{4}{(1+t^2)^2}
    \]
    In the limit of $t \to 0$, this gives $g_t \to 4$. But the formula computed directly from $q(0,\true) = 1$ and $q(0, \false) = 0$ gives $g_t = 0$, thus Fisher information is not continuous for arbitrary smooth $q$. Smoothness of $g_t$ can be forced by requiring that the square root of $q$ is smooth, because in that case, we can reparametrise $r(t,\omega) = \sqrt{q(t,\omega)}$, i.e. $\partial_t r(t,\omega) = \frac{1}{2\sqrt{q(t,\omega)}}\partial_t q(t,\omega)$, to get:
    \[
    g_t = \int \frac{\left(\partial_t q(t, \omega)\right)^2}{q(t, \omega)}\dd \omega = 4\int (\partial_t r(t,\omega))^2\dd\omega
    \]
    Square root coordinates are very convenient for computing in information geometry: we will reuse them in other parts as well, e.g. in~\Cref{ssec:square-root-coordinates}.
\end{remark}

The most important observation in the first part of our construction is therefore: even though we need to handle programs that induce mutually-singular probability measures, their mutual singularity is very tame, only controlled by some fixed pushforward $o$. As noted previously in the semantics for types earlier, we also assume a fixed, unique reference measure for every algebraic Euclidean space constructed by structural induction: singleton measure for unit type, Lebesgue measure for $\RR$, and sums and products of measures for algebraic types, so for a given $\Omega$, we will always have the same $\lambda_\Omega$.

\begin{remark}
    We note that a restriction on the class of output maps $o$ is needed. With an arbitrary choice, we run into the randomness pushback lemma (see e.g.~\cite{fritzSyntheticApproachMarkov2020b}, for one-dimensional variables known as the inverse-cdf transform), which guarantees that any family of random variables can be implemented by first obtaining a random seed from a chosen source, e.g. $X \sim U(0,1)$, and afterward applying a parameter-dependent deterministic measurable map $f(\theta, X)$.
\end{remark}

\begin{definition}[Realised measure]
    For a trace presentation $F = (\Omega, q, o): X \to Y$, we denote its realised measure by $\Realisation(F)$, defined as the Markov kernel: 
    \[
    \Realisation(F) : X \kto Y \qquad \Realisation(F)(x, E) = \int_\Omega \mathbb{1}_E[o(x, \omega)]q(x, \omega)\lambda(\dd \omega)
    \]
    that is, the pushforward of the measure $Q$ with output map $o$.
\end{definition}

Because our basic language $\PPL$ only has boolean random variables, we can initially (at least until $\PPL_\RR$ in~\Cref{sec:pplr}) limit our analysis to finite spaces $\Omega$.

\begin{definition}[Finite trace presentation]
    We say that a finite trace presentation is a trace presentation $F = ((\Omega, \lambda), q, o)$ where the space $\Omega$ is a finite set, and $\lambda$ is the counting measure.
\end{definition}

Thus, the realisation of any finite trace presentation can be written as a sum of finitely many measures:
\[
\Realisation(F)(a, E) = \sum_{\omega \in \Omega} \mathbb{1}[o(a, \omega) \in E]\ q(a, \omega) = \sum_{\omega \in o^{-1}_a(E)} q(a, \omega)
\]

\begin{definition}[Composition of trace presentations]\label{def:composition-trace}
    Given two trace presentations:
    \[
    F = (\Omega_F, q, o): X \to Y \qquad G = (\Omega_G, r, p): Y \to Z
    \]
    we define their composition $G \circ F: X \to Z$ to be $(\Omega_F\times\Omega_G, h, s)$ where:
    \[
    h(x,(\omega_F, \omega_G)) =q(x,\omega_F)r(o(x,\omega_F),\omega_G)
    \qquad
    \text{and}
    \qquad
    s(x,(\omega_F,\omega_G)) =p(o(x,\omega_F),\omega_G)
    \]
\end{definition}

Unfortunately, to give categorical meaning to finite traces, one annoying obstacle is associativity: $\Omega_1 \times (\Omega_2 \times \Omega_3)$ is not identical to $(\Omega_1 \times \Omega_2) \times \Omega_3$; they can be identified only up to an isomorphism.
\begin{definition}[Trace isomorphism]
    For two trace presentations $(\Omega, q, o)$ and $(\Omega', q', o')$ between $X$ and $Y$, we say these are trace-isomorphic if there exists a $Q_x$-conull subset $D \subseteq \Omega$ and similarly a $q'(x, -)\lambda'$-conull subset $D' \subseteq \Omega'$ for all $x \in X$, for which we have a Borel isomorphism $h: D \to D'$ such that:
    \[
    Q'_x = h_*(Q_x) \qquad o'(x, h(\omega)) = o(x, \omega)
    \]
    for all $x \in X$ and for $Q_x$-almost-every $\omega \in D$.
\end{definition}

We need to quotient trace presentations by trace isomorphisms. An alternative, although more complicated approach, can be developed using a graded trace monad.
Also note that we intentionally do not require the map to be a Borel isomorphism on the full trace space $\Omega$. In particular, this would not respect the $\beta$ law of $\ifte{\false}{p}{q} \equiv q$. We also note that we do not include the $\eta$ law of conditionals $\ifte{b}{p}{p} \not\equiv p$, since this would potentially change the structure of the trace.

\begin{proposition}\label{prop:deternistic-are-traces}
    Every smooth deterministic map $f: X\to Y$ can be canonically written as a finite dominated trace $F = (\Omega^f, q^f, o^f): X \to Y$, for:
    \[
    \Omega^f = \unitset \qquad q^f(x, \unitsetelem) = 1 \qquad o^f(x, \unitsetelem) = f(x)
    \]
    such that we have $\Realisation(F)(x) = \delta_{f(x)}$.
\end{proposition}

\begin{definition}[Tensor product, copy and discard]\label{def:copy-discard-traces}
    We put a tensor product structure on finite dominated traces, which is Cartesian products on types, and for $F = (\Omega, q, o): X \to Y$ and $G = (\Omega', q', o'): X' \to Y'$ we put:
    \[
    (F \otimes G) = (\Omega \times \Omega', q \otimes q', o \otimes o'): X\times X'\to Y\times Y'
    \]
    where:
    \[
    (q \otimes q')((x, x'), (\omega,\omega')) = q(x,\omega)\cdot q'(x',\omega')\qquad
    (o \otimes o')((x, x'), (\omega,\omega')) = (o(x,\omega), o'(x',\omega'))
    \]
    The tensor unit is $\tensorunit = \unitset$. The copy map $\copyK: X \to X\times X$ and the delete map $!_X : X \to \tensorunit$ are given as in~\Cref{prop:deternistic-are-traces}.
\end{definition}

\begin{proposition}\label{prop:trace-isomorphism-congruence}
     Trace isomorphism is an equivalence relation, and a congruence for composition and tensor product.
\end{proposition}
\begin{proof}
    In the definition of trace isomorphism, we can take the sets $D$ and $D'$ to be the minimal conull subsets of $\Omega$ and $\Omega'$ respectively, given by the union of supports $q(x, -)$ over all $x \in X$. For three:
    \[
    F = (\Omega, q, o) \qquad G = (\Omega', q', o') \qquad H = (\Omega'', q'', o'')
    \]
    we can thus take $D \subseteq \Omega, D' \subseteq \Omega', D'' \subseteq \Omega''$, with $h: D \to D', h': D' \to D''$, which give a Borel isomorphism $h'' = h'\circ h: D \to D''$. The remaining condition is the output maps commuting:
    \[
    o''(x, h''(\omega)) = o''(x, h'(h(\omega))) = o'(x, h(\omega)) = o(x, \omega)
    \]
    where we used that both $h$ and $h'$ form trace isomorphisms. For composition, assume  we now have $F: X \to Y$ and $G: X \to Y$, and $H: Y\to Z$, then if $h: D \to D'$ is a witness of trace isomorphism between $F$ and $G$, then $h\times \id_{\Omega''}$ is a witness for $H\circ F$ and $H \circ G$. Similarly for the composition in the other direction. Finally, the tensor product is the standard Cartesian product on trace spaces, so a similar argument applies.
\end{proof}

\begin{definition}[$\CatFinTrace$]\label{def:fin-trace}
    We define the category of finite dominated trace presentations $\CatFinTrace$ to have, as objects, algebraic Euclidean spaces, and, as morphisms, finite trace presentations modulo trace isomorphisms.
\end{definition}
\begin{proof}
Composition on the equivalence classes is given as in~\Cref{def:composition-trace}, and the identity on $X$ is given as in~\Cref{prop:deternistic-are-traces}.
The category is well-defined because composition is associative thanks to the quotient by trace isomorphism, identifying $(\Omega\times\Omega')\times\Omega''$ and $\Omega\times(\Omega'\times\Omega'')$.
\end{proof}

\begin{proposition}
    Category $\CatFinTrace$ has a copy-discard structure.
\end{proposition}
\begin{proof}
    Directly from~\Cref{def:copy-discard-traces} and~\Cref{prop:trace-isomorphism-congruence}.
\end{proof}

\begin{definition}[CD functor]
    For two symmetric monoidal categories $\CatC, \CatD$, we say that a functor $F: \CatC \to \CatD$ is a CD functor, if it is a strong symmetric monoidal functor preserving the copy and discard structure.
\end{definition}

\begin{proposition}
    The realisation map $\Realisation$ defines a CD functor between $\CatFinTrace$ and $\CatStoch$.
\end{proposition}

\begin{remark}
    $\CatFinTrace$ does not have a Markov structure, because there are many morphisms $X \to 1$, arising from programs that have unused randomness. After realisation, $\CatStoch$ has a Markov structure.
\end{remark}

\subsection{Dominated jet traces}
One reason the split of $p$ into $(q, o)$ is so effective in our setting is because the scores of density maps $q$ and derivatives of output maps $o$ compose well. Indeed, we form a category of jets over dominated traces, which effectively annotate them with differentials. This does not add any information - this category will be equivalent to dominated trace spaces - but it will make working with them easier later on, and is the first step to decomposing programs' geometry.

\begin{definition}[Jet trace]\label{def:jet-trace}
Given a finite trace presentation $F = (\Omega,\lambda_\Omega,q,o): X \to Y$, we define a (holonomic 1-) jet over $F$ as a tuple:
\[
 \dot F=((\Omega,\lambda_\Omega),q,o;S,V)
\]
where:
\[
 S(x,\omega;-):T_xX\to\RR \qquad S[x, -; \xi]=\frac{\dd[(D_x q(x, -)[\xi])\lambda_\Omega]}{\dd Q_x} \in L_0^2(Q_x)
\]
is the score of $q$, given by the Radon-Nikodym derivative with respect to the measure $Q_x$, and: 
\[
 V(x,\omega;-):T_xX
      \to T_{o(x,\omega)}Y \qquad V[x,\omega; \xi]=D_x o(x, \omega)[\xi]
\]
is the derivative of the output map in the first component, called velocity.
\end{definition}

We note that only the directions of $\xi \in T_xX$ can be differentiated - this is obvious in the case of finite dominated trace presentations, but will remain true in the more general case of arbitrary (even ``smooth'') trace spaces - which will become relevant as we move towards more complicated random variables in $\PPL_\RR$, in \Cref{sec:pplr}.

\begin{definition}[Composition of jet traces]\label{def:jet-composition}
For:
\[
F=(\Omega_F,q,o;S_q,V_o):X\to Y
\qquad
G=(\Omega_G,r,p;S_r,V_p):Y\to Z
\]
we define their composition $H=G\circ F$ to be $((\Lambda, h, s); S_h, V_s)$ where the trace part $(\Lambda, h, s)$ is given as composition of trace presentations $(\Omega_G,r,p) \circ (\Omega_F,q,o)$, and the score and velocity are given by:
\begin{align*}
 S_h[x,(\omega_F,\omega_G);\xi]
 &=S_q[x,\omega_F;\xi] + S_r[o(x,\omega_F),\omega_G;V_o(x,\omega_F;\xi)]\\
 V_s[x,(\omega_F,\omega_G);\xi]
 &=V_p[o(x,\omega_F),\omega_G;V_o(x,\omega_F;\xi)]
 \end{align*}
\end{definition}

\begin{definition}[Jet trace isomorphism]
    A jet trace isomorphism between  $(\Omega, q, o; S, V)$ and $(\Omega', q', o'; S', V')$ is a trace isomorphism $h: D \to D'$ for $D \subseteq \Omega, D'\subseteq \Omega'$, such that $S(x, \omega;-) = S'(x, h(\omega);-)$ and $V(x, \omega, -) = V'(x, h(\omega), -)$ for $Q_x$-almost every $\omega \in D$.
\end{definition}

\begin{proposition}
    Every dominated trace isomorphism between $(\Omega, q, o)$ and $(\Omega', q', o')$ gives a corresponding jet trace isomorphism between $(\Omega, q, o;S,V)$  and $(\Omega', q', o';S',V')$.
\end{proposition}
\begin{proof}
    The compatibility conditions on score are automatically satisfied, so the only non-trivial condition is on velocity. Observe that if for some $x \in X$ we have $q(x, \omega) > 0$, then since $q$ is smooth, we also have $q(x', \omega) > 0$ on some open neighbourhood $x' \in U$. Thus, we must have equality of output maps $o$ and $o'$ on the whole neighbourhood, therefore, by differentiating, equality of velocity maps.
\end{proof}

\begin{definition}
    We define the category $\Jet\CatFinTrace$ to have, as objects, algebraic Euclidean spaces, and as morphisms, finite jet traces quotiented by jet trace isomorphisms.
\end{definition}

\begin{remark}\label{remark:jet-traces-equiv-to-pres}
    The assignment $\Jet(\Omega, q, o) = (\Omega, q, o; S_q, V_o)$ is a CD functor, which is an isomorphism, together with the forgetful functor $U(\Omega, q, o; S, V) = (\Omega, q, o)$. Thus, in our setup, the additional jet data $(S, V)$ is just convenient bookkeeping on top of finite dominated traces.
\end{remark}

\begin{corollary}
    The category $\Jet\CatFinTrace$ has a copy-discard structure.
\end{corollary}
In particular, $\Jet\CatFinTrace$ has the copy map $\Delta_\Gamma: \Gamma \to \Gamma \times \Gamma$, since it has the diagonal tangent map and zero score, since it is deterministic, and it has the standard velocity:
\[
S_{\Delta_\Gamma} = 0 \qquad V_{\Delta_\Gamma}(x;\xi) = (\xi, \xi)
\]
  
The delete map is just a deterministic $!_X: X \to 1$, with: \[
S_{!_X} = 0 \qquad V_{!_X} = 0
\]

\subsection{Semantics for $\PPL$ in jet traces}

We will define semantics for programs in jet traces, inductively on the program derivation. For a program $\Gamma \vdash p \vkto T$ in context $\Gamma$, we define by structural induction the map $\Trace(p)$ as:
\[
  \Trace(p)=(\Omega_p,q_p,o_p; S^p, V^p)
\]
for each case below.

\subsubsection{Return}
Return is done with the construction of~\Cref{prop:deternistic-are-traces} - for:
\begin{prooftree}
\AxiomC{$\Gamma \vdash t:T$}
\RightLabel{(Return)}
\UnaryInfC{$\Gamma \vdash \return{t}\vkto T$}
\end{prooftree}
we set:
\[
  \Omega=\{\unitsetelem\}
  \qquad
  q(\gamma,\unitsetelem)=1
  \qquad
  o(\gamma,\unitsetelem)=\sem t(\gamma)
\]
and thus:
\[
  S^{\return t}[\gamma,\unitsetelem;\xi]=0
  \qquad
  V^{\return t}[\gamma,\unitsetelem;\xi]=D\sem t_\gamma[\xi]
\]

\subsubsection{Bernoulli sampling}
For:
\begin{prooftree}
\AxiomC{$\Gamma \vdash t:\R$}
\RightLabel{(Bern)}
\UnaryInfC{$\Gamma \vdash \bern{t} \vkto \Bool$}
\end{prooftree}
we let $p(\gamma) =\logistic(\sem{t}[\gamma]): \semt{\Gamma} \to \RR$ denote the probability of the flip. Then:
\[
  \Omega=\{0, 1\},
  \qquad
  q(\gamma,i)=\begin{cases}
  \logistic(\sem{t}[\gamma]) & i = 1 \\
  1 - \logistic(\sem{t}[\gamma])& i = 0
  \end{cases}
  \qquad\qquad
  o(\gamma,i)=i.
\]
For $\xi\in T_\gamma\semt\Gamma$, the score is:
\[
  S[\gamma,i;\xi] = (i - p(\gamma))D\sem{t}_\gamma[\xi]
\]
The output is discrete, hence has zero tangent velocity:
\[
  V[\gamma,i;\xi]=0
\]

\subsubsection{Bind} \label{subsec:bind-rule}
We note that we only allow bind with context sharing, which makes the composite map more complex than simply composition of the respective semantics. Specifically, we need the map:
\[
\Gamma \xrightarrow{\Delta_\Gamma} \Gamma \times \Gamma \xrightarrow{\id_{\Gamma} \times {\Trace(p)}} \Gamma  \times X \xrightarrow{\Trace(q)} Y
\]
For:
\begin{prooftree}
\AxiomC{$\Gamma \vdash p\vkto X$}
\AxiomC{$\Gamma,(x:X) \vdash q\vkto Y$}
\RightLabel{(Bind)}
\BinaryInfC{$\Gamma \vdash \letbind{x}{p}{q}\vkto Y$}
\end{prooftree}
denoting dominated traces respectively as:
\[
  \Trace(p)=(\Omega_p,q_p,o_p)
  \qquad
  \Trace(q)=(\Omega_q,q_q,o_q)
\]
we set the trace space of the composite to $\Omega_p\times\Omega_q$. For a fixed $\omega_p\in\Omega_p$, we write $x_p=o_p(\gamma,\omega_p)$. Then, density and output are:
\[
  q_{\mathrm{bind}}(\gamma,(\omega_p,\omega_q))
  =q_p(\gamma,\omega_p)
   q_q((\gamma,x_p),\omega_q)
 \qquad
  o_{\mathrm{bind}}(\gamma,(\omega_p,\omega_q))
  =o_q((\gamma,x_p),\omega_q)
\]
For a tangent $\xi\in T_\gamma\semt\Gamma$, the upstream output moves with velocity
\[
  \dot x_{\omega_p,\xi}
  =V^p[\gamma,\omega_p;\xi]
  =D_\gamma o_p(\gamma,\omega_p)[\xi]
\]
The downstream program is therefore differentiated at the extended tangent
\[
  (\xi,\dot x_{\omega_p,\xi})
  \in T_{(\gamma,x_p)}(\semt{\Gamma}\times{\semt X})
\]
The bind score and velocities are then:
\[
  S^{\mathrm{bind}}[\gamma,(\omega_p,\omega_q);\xi]
  =S^p[\gamma,\omega_p;\xi]
   +S^q[(\gamma,x_p),\omega_q;(\xi,\dot x_{\omega_p,\xi})]
 \qquad
  V^{\mathrm{bind}}[\gamma,(\omega_p,\omega_q);\xi]
  =V^q[(\gamma,x_p),\omega_q;(\xi,\dot x_{\omega_p,\xi})]
\]

\subsubsection{If-then-else}\label{sssec:if-then-else}

For a pure boolean term:
\begin{prooftree}
\AxiomC{$\Gamma \vdash b:\Bool$}
\AxiomC{$\Gamma \vdash p_0\vkto T$}
\AxiomC{$\Gamma \vdash p_1\vkto T$}
\RightLabel{(If)}
\TrinaryInfC{$\Gamma \vdash r := \ifte{b}{p_1}{p_0}\vkto T$}
\end{prooftree}
we have:
\[
  \Omega_r=\Omega_0\sqcup\Omega_1
  \qquad 
  q_r(\gamma,(i,\omega))=
  \begin{cases}
  q_{p_i}(\gamma,\omega),&\sem b(\gamma)=i\\
  0,&\sem b(\gamma)\neq i
  \end{cases}
    \qquad  
  o_r(\gamma,(i,\omega))
  =o_{p_i}(\gamma,\omega)
\]
On the component containing $\gamma$, the selected branch $i_0=\sem b(\gamma)$ is locally constant by~\Cref{prop:finite-values-locally-constant}. Hence, along $\xi$, on the support of $q_r(\gamma,-)$, we have:
\[
  S^r[\gamma,(i_0,\omega);\xi]
  =S^{p_{i_0}}[\gamma,\omega;\xi]
  \qquad
  V^r[\gamma,(i_0,\omega);\xi]
  =V^{p_{i_0}}[\gamma,\omega;\xi]
\]
Since the score is given as the Radon-Nikodym derivative, on the non-selected branch $i\neq i_0$, the score $S^r(i,\omega;-)$ is irrelevant. For a non-pure boolean guard $\Gamma\vdash b\vkto \Bool$, we interpret it as syntactic sugar for:
\[
\Gamma\vdash\letbind{x}{b}{\ifte{x}{p_1}{p_0}}\vkto T
\]

\subsubsection{Context rules}
We also need rules for contexts, as these are used to form e.g. let-bind composition - weakening is then especially important. For:
\begin{prooftree}
\AxiomC{$\Gamma \vdash p\vkto Y$}
\RightLabel{(Weak)}
\UnaryInfC{$\Gamma,\Delta \vdash p \vkto Y$}
\end{prooftree}
we have:
\[
  q_{\mathrm{wk}}((\gamma,\delta),\omega)=q_p(\gamma,\omega)
  \qquad
  o_{\mathrm{wk}}((\gamma,\delta),\omega)=o_p(\gamma,\omega)
\]
and for tangents \((\xi,\zeta)\) and \((\eta,\tau)\):
\[
  S^{\mathrm{wk}}[(\gamma,\delta),\omega;(\xi,\zeta)]
  =S^p[\gamma,\omega;\xi]
  \qquad
  V^{\mathrm{wk}}[(\gamma,\delta),\omega;(\xi,\zeta)]
  =V^p[\gamma,\omega;\xi]
\]

Reordering variables is pullback along the corresponding coordinate permutation. Reading only some variables is weakening followed by projection, and drops the unused tangent directions.

\subsubsection{Functorial semantics}

Using all of the above, we have the theorem establishing compositional semantics of $\PPL$ in jet traces. For this, we need a small lemma about substitution.
\begin{lemma}[Substitution]\label{lemma:substitution}
    Given contexts $\Gamma, \Delta$ such that $\Delta = (x_1 :X_1, \ldots x_n: X_n)$, and a collection of pure terms $\Gamma\vdash t_i : X_i$ with semantics $\sem{t}: \semt{\Gamma}\to\semt{\Delta}$, for the program $\Delta\vdash p\vkto Y$, with semantics $\Trace(p) = (\Omega_p, q, o)$, we have substitution semantics:
    \[
    \Trace(\Gamma\vdash p[t/\Delta]\vkto Y) = (\Omega_p,(\gamma,\omega)\mapsto q(\sem{t}(\gamma),\omega),(\gamma, \omega)\mapsto o(\sem{t}(\gamma),\omega))
    \]
\end{lemma}
In other words, substitution by a term $t$ simply precomposes the argument $\gamma$ to both density and output maps with the map $\sem{t}$. The proof of the lemma is by structural induction.
\begin{theorem}\label{thm:ppl-semantics}
    The map $\Trace: \Syn_\PPL \to \Jet\CatFinTrace$ is a strong monoidal functor.
\end{theorem}
\begin{proof}
    The most important aspect to verify is that $\Trace$ respects the appropriate quotient laws in $\Syn_\PPL$. To simplify, it is enough to examine the behaviour of dominated trace-valued $\hat{\Trace} = U\circ \Trace: \Syn_\PPL \to \CatFinTrace$, per~\Cref{remark:jet-traces-equiv-to-pres}.
    The $\alpha$-conversion is automatically satisfied because it changes neither trace, density, nor output maps. Pure terms are valued in measurable maps, and then interpreted via~\Cref{prop:deternistic-are-traces}, which is extensional, so deterministic $\beta/\eta$ laws are respected.
    \begin{description}[leftmargin=0pt]

    \item[$\beta$ law for conditionals:] for a program:
    \[
    \Gamma\vdash p:=\ifte{\true}{p_\true}{p_\false} \vkto X
    \]
    and denotations:
    \[
    \hat\Trace(p_\true) = (\Omega_\true, q_\true, o_\true)\qquad \hat\Trace(p_\false) = (\Omega_\false, q_\false, o_\false)\qquad \hat\Trace(p) = (\Omega_\true\sqcup\Omega_\false, q_p, o_p)
    \]
    with the density and output maps $q_p, o_p$ as in~\Cref{sssec:if-then-else}, we construct the required isomorphism to $\hat\Trace(p_\true)$ by taking $D = D' = \Omega_\true$, $h = \id$, which trivially satisfies the density and output pullback conditions.

    \item[Left unit law:] for a program:
    \[
    \Gamma\vdash \letbind{x}{\return{t}}{p} \equiv p[t/x]\vkto Y
    \]
    assume we have:
    \[
    \hat\Trace(p) = (\Omega, q, o): \semt{\Gamma}\times\semt{X} \to Y
    \]
    From the definition of bind, this equals:
    \[
    \hat\Trace(\letbind{x}{\return{t}}{p}) = \hat\Trace{p}\circ (\id_{\Gamma} \otimes \hat\Trace(\return{t})\circ \Delta_\Gamma)
    \]
    where we have:
    \[
    \hat\Trace(p) = (\Omega, q, o)\qquad 
    \hat\Trace(\return{t}) = (\unitset, (\gamma, \unitsetelem)\mapsto 1, (\gamma, \unitsetelem)\mapsto \sem{t}[\gamma])
    \qquad
    \id_{\Gamma} = (\unitset, (\gamma, \unitsetelem)\mapsto 1, (\gamma, \unitsetelem)\mapsto \gamma)
    \]
    Unrolling the definitions:
    \[
    (\id_{\Gamma}\otimes \hat\Trace(\return{t}))\circ\Delta_\Gamma = \left(
    \unitset \times \unitset, (\gamma, (\unitsetelem, \unitsetelem)) \mapsto 1, (\gamma, (\unitsetelem, \unitsetelem)) \mapsto (\gamma, \sem{t}(\gamma))
    \right)
    \]
    Post-composing $(\Omega,q,o)\circ (\id_{\Gamma}\otimes \hat\Trace(\return{t}))\circ\Delta_\Gamma$ we get the dominated trace $(\Omega', q', o')$ given by:
    \[
     \Omega' = (\unitset \times \unitset)\times\Omega\qquad q'(\gamma, ((\unitsetelem, \unitsetelem), \omega)) = q((\gamma,\sem{t}(\gamma)),\omega)
     \qquad
     o'(\gamma, ((\unitsetelem, \unitsetelem), \omega)) = o((\gamma, \sem{t}(\gamma)), \omega)
    \]
    Thus, we take $h((\unitsetelem, \unitsetelem), \omega) = \omega$, and density and output maps follow from~\Cref{lemma:substitution}.
    
    \item For the right unit, the computation is similar, but we do not need the substitution lemma. The associativity of composition follows by the same argument we outlined in~\Cref{def:fin-trace}. 
    \item[Interchange law] for programs:
    \[
    \Gamma\vdash\letbind{x}{p}{\letbind{y}{q}{r}}\vkto Z \qquad \Gamma\vdash\letbind{y}{q}{\letbind{x}{p}{r}}\vkto Z
    \]
    we have the corresponding trace spaces: \[
    \Omega = \Omega_p\times (\Omega_q\times \Omega_r) \qquad \Omega' = \Omega_q\times (\Omega_p\times \Omega_r) \qquad 
    \]
    and we take the map $h(\omega_p, (\omega_q, \omega_r)) = (\omega_q, (\omega_p, \omega_r))$. Write $x=o_p(\gamma,\omega_p)$ and $y=o_q(\gamma,\omega_q)$, using the same context order $(\gamma,x,y)$ for $r$ on both sides. Then, because of the assumption that $x\notin \mathrm{FV}(q)$ and $y\notin \mathrm{FV}(p)$, we know that the densities factorise as:
    \begin{align*}
    q(\gamma, (\omega_p, (\omega_q, \omega_r))) &= q_p(\gamma, \omega_p)q_q(\gamma, \omega_q)q_r((\gamma, x, y), \omega_r)\\
    q'(\gamma, (\omega_q, (\omega_p, \omega_r))) &= q_q(\gamma, \omega_q)q_p(\gamma, \omega_p)q_r((\gamma, x, y), \omega_r)
    \end{align*}
    so they are equal after transporting by $h$. Outputs are equal to:
    \begin{align*}
    o(\gamma,(\omega_p,(\omega_q,\omega_r))) &= o_r((\gamma,x,y),\omega_r)\\
    o'(\gamma,(\omega_q,(\omega_p,\omega_r))) &= o_r((\gamma,x,y),\omega_r)
    \end{align*}
    which also agree after applying $h$.
    \end{description}
\end{proof}

\section{Algebraic formulation and $\Affine$}\label{sec:algebraic} 

To express the geometry of programs, we need the notion of degenerate Riemannian metrics. This is because some directions in the tangent space of the trace presentation might be null, for example those corresponding to ignored parameters.

\begin{definition}
    For a Euclidean space $X$, let $\V(X)$ be the vector space of symmetric 2-tensor fields on $X$, and let $\PSD(X)$ be the cone of positive-semidefinite metrics on $X$:
    \[
    \PSD(X) = \Gamma(\mathrm{Sym}^2\ T^*X)_{\succeq 0} \subseteq \V(X)
    \]
    For an algebraic Euclidean space, this should be understood componentwise.
\end{definition}

$\PSD(X)$ has rich structure. In particular, it supports $0$ (i.e. deterministic programs), sums (independent products), i.i.d. sampling ($nG$, i.e. multiplication by a positive $n$), hiding (cone order $G_{hidden} \preceq G_{joint}$), weakening (null directions) and the copy map ($\Delta^*(g \oplus h) = g + h$). Also, as a forward pointer, we note that when extending this construction to $\PPL_\RR$ in \Cref{sec:pplr}, the definition of $\PSD(X)$ will have to be modified to include only certain admissible (in a sense explained later) symmetric tensor fields.
In the classical case, each parametric dominated statistical model:\[
q: \Theta \to \mathcal{P}(\Omega) \qquad \text{where} \qquad \Theta \subseteq \RR^n
\]
has an assigned Fisher metric $\Fisher{q}_{ij} \in \PSD(\Theta)$, in coordinates $(\theta^i)$ on $\Theta$ via:
\[
\Fisher{q}_{ij} = \mathbb{E}_{\omega \sim q(\theta)}\left[\frac{\partial \log q(\omega|\theta)}{\partial \theta^i}\frac{\partial \log q(\omega|\theta)}{\partial \theta^j} \right]
\]

In our setup, where the density is $q: X \times \Omega \to [0, \infty)$, we can still carry out part of this construction, using the already-computed score $S(x, \omega)$.
\begin{definition}
    Given a dominated jet trace presentation $F = (\Omega, q, o; S, V)$, we define its corresponding Fisher metric $\Fisher{F} \in \PSD(X)$, by:
    \[
        \Fisher{F}_{x}[\eta, \xi] = \mathbb{E}_{\omega \sim q(x)}\left[S(x, \omega; \eta)S(x, \omega; \xi)\right]
    \]
\end{definition}
\begin{remark}
    For maps into the unit $F: X \to 1$ we do not always have $\Fisher{F} =0$. This is due to the same reason that $\CatFinTrace$ is not a Markov category: there might be some other maps $X \to 1$, with unused randomness.
\end{remark}

This is, however, only half of the construction. For a fixed $\omega$, the output map $o(-, \omega): X \to Y$ pulls back metrics in $\PSD(Y)$ to metrics in $\PSD(X)$.

\begin{definition}
    Given a dominated jet trace presentation $F = (\Omega, q, o; S, V): X \to Y$, we define its expected tangent pullback $F^\sharp : \PSD(Y) \to \PSD(X)$ by:
    \[
        F^\sharp(h)_x[\eta, \xi] = \mathbb{E}_{\omega \sim q(x)}\left[h_{o(x, \omega)}(V(x, \omega; \eta), V(x, \omega; \xi))\right]
    \]
\end{definition}

These two assignments interact with each other.

\begin{proposition}\label{prop:fisher-pullback-properties}
For maps $F: X \to Y$, $G: Y \to Z$, deterministic maps $f: X \to Y$, $g: Y \to Z$, the metrics $h_i \in \PSD(Y)$, non-negative real numbers $a_i\in \RR_{\geq 0}$, the induced Fisher metric and expected tangent pullback satisfy the following properties:
\begin{enumerate}[label={(\arabic*)}]
    \item tangent pullback is a contravariant functor, $(\id_X)^\sharp = \id_{\PSD(X)}$, and $(G \circ F)^\sharp = F^\sharp \circ G^\sharp$
    \item tangent pullback is linear in the argument, $F^\sharp\left(\sum_{i}a_i h_i\right) = \sum_{i}a_iF^\sharp h_i$
    \item $\Fisher{f} = 0$, in particular $\Fisher{\id_X} = 0$
    \item if $Y$ is discrete, then $F^\sharp h = 0$
    \item $f^\sharp = f^*$, where $f^*$ denotes the ordinary pullback
    \item $\Fisher{G \circ f} = f^*\Fisher{G}$ and $\Fisher{g \circ F} = \Fisher{F}$
    \item $\Fisher{G \circ F} = \Fisher{F} + F^\sharp \Fisher{G}$
\end{enumerate}
\begin{proof}
    For (1), the velocity of identity is identity, and $q^{\id}(x, \unitsetelem) = 1$, so $(\id^{\sharp})h_x(\eta,\xi) = h(\eta,\xi)$. If:
    \[
    F = (\Omega_F, q, o; S_q, V_o) \qquad G=(\Omega_G,r,p;S_r,V_p)
    \]
    then the output, density, velocity and score of composition $G\circ F$ are, from~\Cref{def:jet-composition}, respectively:
    \begin{align*}
    s(x,(\omega_F,\omega_G)) &=p(y,\omega_G)\\
    u(x,(\omega_F, \omega_G))&=q(x,\omega_F)r(y,\omega_G)\\
    S[x,(\omega_F,\omega_G);\xi]
    &=S_q[x,\omega_F;\xi] + S_r[y,\omega_G;V_y(\xi)]\\
    V[x,(\omega_F,\omega_G);\xi]
    &=V_p[y,\omega_G;V_y(\xi)]   
    \end{align*}
    where we denote $y = o(x,\omega_F)$ and $V_y(\xi)=V_o[x,\omega_F;\xi]$. Plugging this into the calculation:
    \begin{align*}
    (G\circ F)^\sharp(h)_x(\eta,\xi) &= \E_{\omega \sim u(x)}\left[h_{s(x, \omega)}(V(x, \omega; \eta), V(x, \omega; \xi))\right] \\
    &= \E_{\omega_F \sim q(x)}\left[
    \E_{\omega_G \sim r(y)}\left[
        h_{p(y,\omega_G)}(
            V_p[y,\omega_G;V_y(\eta)], 
            V_p[y,\omega_G;V_y(\xi)])
    \right]
    \right]\\
    &= \E_{\omega_F \sim q(x)}\left[
    (G^\sharp h)_y(V_y(\xi), V_y(\eta))
    \right]\\
    &= (F^\sharp(G^\sharp h))_x(\xi,\eta) \\
    &= ((F^\sharp \circ G^\sharp)h)_x(\xi, \eta)
    \end{align*}
    For (2), the result follows from linearity of expectations.
Then, (3) follows because the score of a deterministic map is zero, (4) follows because necessarily $h$ vanishes, in (5) the expectation has only one atom at $\delta_{f(x)}$, and (6) follows from (7), using the same notation as above for $F$ and $G$:
\begin{align*}
g^{G\circ F}_x[\eta, \xi] &= \E_{\omega \sim u(x)}\left[
    S(x, \omega;\eta)S(x, \omega;\xi)
    \right]\\
    &=\E_{\omega_F \sim q(x)}\left[
    \E_{\omega_G \sim r(y)}\left[
        (S_q[x,\omega_F;\eta] + S_r[y,\omega_G;V_y(\eta)])\cdot
        (S_q[x,\omega_F;\xi] + S_r[y,\omega_G;V_y(\xi)])
    \right]
    \right]\\
    &= \E_{\omega_F \sim q(x)}\left[S_q[x,\omega_F;\eta]S_q[x,\omega_F;\xi]\right] \\
    &\qquad\qquad+
    \E_{\omega_F \sim q(x)}\left[\E_{\omega_G \sim r(y)}\left[        S_r[y,\omega_G;V_y(\eta)]\cdot S_r[y,\omega_G;V_y(\xi)]
    \right]\right]\\
    &= g^F_x(\eta,\xi) + \E_{\omega_F \sim q(x)}\left[\E_{\omega_G \sim r(y)}\left[
    S_r[y,\omega_G;V_y(\eta)]\cdot S_r[y,\omega_G;V_y(\xi)]
    \right]\right]\\
    &= g^F_x(\eta,\xi) + (F^\sharp g^G)_x(\eta,\xi)
\end{align*}
where the cross-terms $S_q[x,\omega_F;\xi]\cdot S_r[y,\omega_G;V_y(\eta)]$ and $S_q[x,\omega_F;\eta]\cdot S_r[y,\omega_G;V_y(\xi)]$ disappeared between the second and third lines because of centering of the score:
\begin{align*}
\E_{\omega_F \sim q(x)}\left[\E_{\omega_G \sim r(y)}\left[
S_q[x,\omega_F;\xi]\cdot S_r[y,\omega_G;V_y(\eta)]
\right]\right]
&= \E_{\omega_F \sim q(x)}\left[S_q[x,\omega_F;\xi]\cdot\E_{\omega_G \sim r(y)}\left[
 S_r[y,\omega_G;V_y(\eta)]\right]\right] \\
 &=\E_{\omega_F \sim q(x)}\left[S_q[x,\omega_F;\xi]\cdot0\right] = 0
\end{align*}
\end{proof}

\end{proposition}
\begin{proposition}[$\Fisher{F}$ and $F^\sharp$ on copy-discard structure]\label{prop:fisher-on-copy-discard}
    For maps $F: X \to Y$, $G: X \to Z$ and $F': X' \to Y'$, the induced Fisher metric and tangent pullback satisfy the following properties:
    \begin{enumerate}[label={(\arabic*)}]
        \item $\Fisher{F\otimes F'} = \Fisher{F} \oplus \Fisher{F'}$
        \item For the copy map $\Delta$ followed by two independent maps:
        \[
        X \xrightarrow{\Delta_X} X \times X \xrightarrow{F \otimes G} Y \times Z
        \]
        we have $\Fisher{(F \otimes G)\circ \Delta_X} = \Delta_X^*(\Fisher{F} \oplus \Fisher{G}) = \Fisher{F} + \Fisher{G}$
        \item $(F\otimes F')^\sharp(h\oplus k)=F^\sharp h\oplus(F')^\sharp k$ for $h\in\PSD(Y)$ and $k\in\PSD(Y')$
    \end{enumerate}
\end{proposition}
\begin{proof}
    Writing $F = (\Omega, q, o;S,V)$ and $F'=(\Omega', q', o';S',V')$, statement (1) follows by independence of the joint measure $(q\otimes q')((x,x'),(\omega,\omega')) = q(x,\omega)q'(x',\omega')$. Statement (2) follows from (6) of~\Cref{prop:fisher-pullback-properties} and explicit calculation:
    \begin{align*}
    (\Delta_X^*(g^F \oplus g^G))_x(\eta, \xi) &= (g^F \oplus g^G)_{(x,x)}(d\Delta_X\eta, d\Delta_X\xi) \\
    &= (g^F \oplus g^G)_{(x,x)}((\eta,\eta), (\xi, \xi)) \\
    &= g^F_x(\eta, \xi) + g^G_x(\eta,\xi)
    \end{align*}
    Finally, (3) follows because the velocity of $F\otimes F'$ is the pair of component velocities, $h\oplus k$ has no mixed terms, and the product trace law has the two trace laws as marginals.
\end{proof}
We then have the functorial assignment of the full metric, as promised in~\Cref{sec:intro}. But first, we need to specify the target of the functor. Here, we use a simple category $\CatAffCone$ defined below, but note that this construction can be reworked (and made more canonical) by the use of semi-direct product of categories, which additionally takes care of the (somewhat awkward) non-functorial assignment $F \mapsto \Fisher{F}$ presented earlier in this section. However, the core insight is then somewhat obscured, so we proceed with the simplest version here.

\begin{definition}[Cones and positive linear maps]
A cone is a convex subset $C$ in a real vector space $V$ closed under $\RR_{\ge 0}$-multiplication. For two cones $C \subseteq V$ and $D \subseteq W$, a positive linear map between them is a linear map $L: V \to W$ such that $L(C) \subseteq D$. A positive affine map is an affine map $A: V \to W$ such that $A(C) \subseteq D$. 
\end{definition}

An affine map $A: V \to W$ can be represented as a pair $(b, L)$ where $b \in W$ and $L: V \to W$ is linear, such that $A(x) = b + Lx$.

\begin{definition}[Category $\CatAffCone$]
We define the category $\CatAffCone$ to have cones as objects and positive affine maps as morphisms. The identity is given as $(0, \id)$, and composition of $(b, L): C \to D$ and $(c, M): D \to E$ as:
\[
(c, M) \circ (b, L) = (c + Mb, M \circ L)
\]
\end{definition}

\begin{proposition}
    The category $\CatAffCone$ has monoidal structure, with the monoidal product given as Cartesian product of cones, and monoidal unit as the zero-dimensional cone. 
\end{proposition}

\begin{definition}[Functor $\Affine$]
    We define a contravariant map:
    \[
    \Affine : \Jet\CatFinTrace^\op \to \CatAffCone
    \]
    given by $\Affine(X) = \PSD(X)$ on objects, and:\[
    \Affine_F(h) = \Fisher{F} + F^\sharp h
    \]
    on morphisms.
\end{definition}
In our setup, the score contributes $\Fisher{F}$ to the metric, as in the classical information geometry, but the velocity also contributes the (pulled-back) part $F^\sharp h$, for some chosen $h \in \PSD(Y)$.

\begin{theorem}\label{thm:properties-of-affine}
The map $\Affine : \Jet\CatFinTrace^\op \to \CatAffCone$ satisfies the following properties:
    \begin{enumerate}[label={(\arabic*)}]
        \item $\Affine(F)$ is a positive affine map for every dominated jet trace $F$,
        \item $\Affine(F)$ is invariant under jet trace isomorphisms,
        \item $\Affine$ is a functor,
        \item $\Affine_{F \otimes G}(f \oplus h) = \Affine_F(f) \oplus \Affine_G(h)$,
        \item $\Affine_{\Delta_X}(f \oplus g) = f + g$,
        \item $\Affine_{!_X} = 0$,
        \item we recover Fisher information by $\Fisher{F} = \Affine_F(0_Y)$,
        \item for a deterministic $f$, we have $\Affine_f(h) = f^*h$,
        \item $\Affine$ is lax symmetric monoidal, considered with the trivial unit $\eta$ and multiplication given by:
        \[
        \mu_{X, Y}: \PSD(X) \times \PSD(Y) \to \PSD(X \times Y) \qquad \mu_{X, Y}(f, h) = f\oplus h
        \]
    \end{enumerate} 
\end{theorem}
\begin{proof}
    (1) follows because the Fisher metric is PSD, a pullback of a PSD metric is PSD, and a non-negative linear combination of PSD metrics stays PSD. For (2), we take expectations with $q$, so the result only depends on the minimal conull set $D$: consistent relabelling of $\Omega$ does not change the expectations. We show (3) by calculation, using (1), (2) and (7) of~\Cref{prop:fisher-pullback-properties}:
    \begin{align*}
    \Affine_{G \circ F}(h) &= g^{G\circ F} + (G \circ F)^\sharp(h) \\
    &= g^F + F^\sharp g^G + F^\sharp(G^\sharp h) \\
    &=  g^F + F^\sharp(g^G + G^\sharp h) \\
    &= \Affine_F(\Affine_G(h)) = (\Affine_F\circ\Affine_G)(h)    
    \end{align*}
    Property (4) follows from~\Cref{prop:fisher-on-copy-discard} by calculation:
    \begin{align*}
        \Affine_{F \otimes G}(f \oplus h) &= g^{F \otimes G} + (F\otimes G)^\sharp(f \oplus h) \\
        &= (g^F \oplus g^G) + (F^\sharp f \oplus G^\sharp h) \\
        &= (g^F + F^\sharp f) \oplus (g^G + G^\sharp h) \\
        &= \Affine_F(f) \oplus \Affine_G(h)
    \end{align*}
    (5) follows from deterministic pullback and $\Delta_X^*(f\oplus g)=f+g$. (6) follows because the delete map is deterministic, so by (3) of~\Cref{prop:fisher-pullback-properties} the Fisher part disappears, and the codomain has trivial geometry, so by (4) of the same the expected pullback also vanishes. The same pair of arguments apply to (7) and (8). Finally, for (9), naturality follows directly from (4):
    \[
    (\Affine_{F\otimes G}\circ \mu_{X',Y'})(f,h) = \Affine_{F\otimes G}(f \oplus h) = (\mu_{X,Y}\circ(\Affine_{F} \times \Affine_G))(f,h)
    \]
    Associativity, unit law, and symmetry follow by direct calculation.
\end{proof}

\begin{definition}\label{def:fisher-dirac}
    For a given positive real number $\kappa >0$, a jet trace presentation $F: X \to Y$ between algebraic Euclidean spaces $X$ and $Y$, and chosen coordinates on $Y$, we define the score-velocity metric $\FisherDirac^\kappa$ on $X$ by: \[
    \FisherDirac^\kappa = \Affine_F(\kappa\langle \cdot, \cdot \rangle)
    \]
    where $\langle \cdot, \cdot \rangle$ is the standard Euclidean product on $Y$.
\end{definition}

\begin{corollary}
    The score-velocity metric is invariant under trace isomorphisms.
\end{corollary}

We note that the score-velocity metric is not extensional on the realised kernels. That is, two different programs with the same output kernels can have different $\FisherDirac$ - this is again for the same reason as we pointed out before in the case of $\CatFinTrace$ being non-Markov. For example, we have the following two families.
\begin{example}
    Take input type $X = \RR$ and output type $Y=\RR$. One presentation of a kernel $F_0: X \to Y$ with the constant output $0$ is given by:
\[
  \quad \Omega=\unitset\quad q(\theta,\unitsetelem)=1
  \quad o(\theta,\unitsetelem)=0
\]
and has \(\FisherDirac^\kappa_{F_0}=0\). Let us take another presentation of the same kernel $F_1$ with:
\[
 \Omega=\{0,1\} \quad
  q(\theta,1)=\sigma(\theta) \quad q(\theta,0)=1-\sigma(\theta) 
  \quad o(\theta,0)=o(\theta,1)=0
\]
This gives the same deterministic map $\delta_0$ for every \(\theta\).  But its trace score in the \(\partial_\theta\) direction is \(B-\sigma(\theta)\), so
\[
  \FisherDirac^\kappa_{F_1,\theta}(\partial_\theta,
  \partial_\theta) =\sigma(\theta)(1-\sigma(\theta))>0.
\]
Thus the energy depends on the presentation, not only on the realised kernel.
\end{example}

Subsequently, we will use the unit-weight $\FisherDirac$ for $\kappa = 1$.
\begin{example}
Let us now compute some examples in programs.
\begin{itemize}
    \item For a primitive $f \in \Prim(\PPL)$, a deterministic program $x: \R^n \vdash \return{f(x)} \vkto Y$ and $\eta, \xi \in T_x\RR^n$, we have: \[
    \FisherDirac(\Trace(\return{f(x)}))[\eta, \xi] = \langle df_x[\eta], df_x[\xi]\rangle
    \]
    \item For a Bernoulli random variable $x: \R \vdash \bern{x} \vkto \Bool$, tangents $\eta, \xi \in T_x\RR$, and $p = \logistic(x)$, we have $S[x,i;\eta] = (i-p)\eta$ and velocity $V \equiv 0$, thus:
    \[
        \begin{aligned}
        \FisherDirac(\Trace(\bern{x}))_x[\eta,\xi]
        &= (\Fisher{\Trace(\bern{x})})_x[\eta,\xi]
        = \E_{i\sim\bern{x}}\big[(i-p)^2\big]\,\eta\xi\\
        &= p(1-p)\eta\xi
        = \frac{e^{-x}}{(1+e^{-x})^2}\eta\xi
        \end{aligned}
    \]
    Thus, in the log-odds coordinates, the metric is
    $p(1-p)\dd x^2$.
\item For stochastic bind $\Gamma \vdash \letbind{x}{p}{q} \vkto Y$ and $\eta, \xi \in T_\gamma\semt{\Gamma}$, assuming that $\Trace(p) = F$ and $\Trace(q) = G$, because let-bind is context sharing, we use $\widetilde F : \semt{\Gamma} \vkto \semt{\Gamma}\times \semt{X}$, given as $\widetilde{F} = (\id_{\semt{\Gamma}}\otimes F)\circ \Delta_{\semt{\Gamma}}$. Then, writing $x_p=o_p(\gamma,\omega_p)$,  $V_p[\xi] = V^p(\gamma, \omega_p, \xi)$, and $S_p[\xi] = S^p[\gamma,\omega_p;\xi]$ we have:
    \begin{align*}
    \FisherDirac(\Trace(\letbind{x}{p}{q}))[\eta, \xi] &= (\Affine_{\widetilde F} \circ \Affine_{G})(\langle \cdot, \cdot \rangle)[\eta, \xi] \\
    &= \bigl(\Fisher{G \circ \widetilde F} + (G \circ \widetilde F)^\sharp(\langle \cdot, \cdot \rangle)\bigr)_\gamma[\eta,\xi]\\
    &= \Fisher{F}[\eta, \xi] + ( \widetilde F^\sharp \Fisher{G})[\eta, \xi] + ( \widetilde F^\sharp G^\sharp)(\langle \cdot, \cdot \rangle)[\eta, \xi] \\
    &= \E_{\omega_p\sim q_p(\gamma)}\left[
            S_p[\eta] S_p[\xi]
            +
            \FisherDirac(G)_{(\gamma,x_p)}
            \left( (\eta,V_p[\eta]), (\xi,V_p[\xi])
            \right)\right]
    \end{align*}
    \item As a special case of the above, we consider deterministic pre-processing, that is, a program:
    \begin{prooftree}
    \AxiomC{$B \vdash t : X$}
    \AxiomC{$(x:X) \vdash q\vkto Y$}
    \RightLabel{}
    \BinaryInfC{$B \vdash \letbind{x}{\return{t}}{q}\vkto Y$}
    \end{prooftree}
    We then have a deterministic map $\sem{t}=f:\semt{B}\to\semt{X}$ and some jet trace $\Trace(q)=G:\semt{X}\to\semt{Y}$. From the above and taking into account context weakening, we have:
    \[
        \FisherDirac(\Trace(\letbind{x}{\return{t}}{q}))_b[\eta,\xi]
        =
        \FisherDirac(G)_{f(b)}
        \left[Df_b[\eta],Df_b[\xi]\right]
    \]
    or equivalently in matrix notation:
    \[
        \FisherDirac(\Trace(\letbind{x}{\return{t}}{q}))(b)
        =
        Df_b^T\ \FisherDirac(G)(f(b))\ Df_b.
    \]
\end{itemize}
\end{example}

\section{Geometric formulation and $\ETrace{Y}$}\label{sec:geometric} 
In the previous section, we have constructed compositional geometric semantics for program families by considering a certain functor $\Affine$ to the category of affine cones and positive affine maps, specifically valued in positive-semidefinite 2-forms - in other words, giving degenerate Riemannian geometry. There has not been any justification for this particular functor. We now wish to address that by taking a slightly different perspective on the program geometry, by considering the classifying (or moduli) spaces of programs.

The motivation is similar to what~\cite{ayInformationGeometry2017b} have done for other statistical models before, in the context of classical information geometry: to think of the geometry on a particular family $M$ on a set $\Omega$ considered as a Banach manifold, as pulled back from the global geometry on the Banach manifold of signed / non-negative / probability measures $\mathcal P(\Omega) \subset \mathcal M(\Omega) \subset \mathcal S(\Omega)$ (specifically in that work, for the so-called $k$-integrable families).

Here, we also wish to consider individual programs, such as:
\[
(\theta: \R) \vdash \return{\theta^3} \vkto \R \qquad (\theta: \R) \vdash \ifte{\bern{\theta^3}}{\return{\theta}}{\return{\theta + 1}} \vkto \R \qquad
\]
as elements or submanifolds of the global space of all possible models. This moduli space has therefore to accommodate both deterministic and stochastic families in a uniform way. It also has to have a metric, which we pull back onto the space of parameters of individual programs. It will turn out that such metrics naturally correspond to the construction of $\Affine$ we carried out before.

\subsection{The decorated trace law}

\begin{definition}\label{def:simplex}
For a given finite trace space $\Omega$, we let the closed and open simplex of probability distributions on $\Omega$ be denoted as:
\[
  \Simplex{\Omega} = 
  \left\{(q_i)_{i \in \Omega}:
      q_i \geq 0 \text{ and }
      \sum_{i}q_i = 1
  \right\} \qquad \Simplexo{\Omega} = \mathrm{relint}\left(\Simplex{\Omega}\right)
\]
\end{definition}
\begin{definition}[Classifying space]
For a fixed trace space $(\Omega,\lambda)$ and output space $Y$, define:
\[
  \ETrace{Y}= \Simplex{\Omega}\times Y^{\Omega} \qquad \ETraceo{Y}= \Simplexo{\Omega}\times Y^{\Omega}
\]
with the standard Euclidean measurable and smooth structure, together with two projections, the universal trace law:
\[
  \Law_Y:\ETrace{Y}\longrightarrow\Simplex{\Omega}
  \qquad
  \Law_Y(q,o)=q
\]
and the universal output map:
\[
  \Output_Y:\ETrace{Y}\times\Omega\longrightarrow Y
  \qquad
  \Output_Y((q,o),\omega)=o(\omega)
\]
\end{definition}

We note that $\ETrace{Y}$ does not yet identify dominated trace presentations $(q, o), (q',o')$ that only differ on zero-mass outputs, i.e. $o(x, \omega) \neq o(x',\omega')$ for $q(x, \omega) = q(x', \omega') = 0$. This cannot happen on the interior $\ETraceo{Y}$. We move to the collapsed (moduli) space only in~\Cref{sec:pplr}, after we develop the required technology to handle quotients and singular spaces.

\begin{remark}[Representability of fixed-trace presentations]
For a fixed trace space $\Omega$, for all algebraic Euclidean spaces $X,Y$ there is a natural isomorphism:
\[
  \CatSmooth\bigl(X,\ETrace{Y}\bigr)
  \cong
  \CatSmooth\bigl(X,\Simplex{\Omega}\bigr)
  \times
  \CatSmooth\bigl(X\times\Omega,Y\bigr)
\]
Under this isomorphism, a map $P:X\to\ETrace{Y}$ corresponds to a pair $(q, o)$ of a dominated law and an output map:
\[
q:X\to\Simplex{\Omega} \qquad  o:X\times\Omega\to Y
\]
\end{remark}
Note that the pair $(q, o)$ above is a dominated trace presentation if, additionally, we have that $\sqrt{q(-, \omega)}$ is smooth as well (see~\Cref{def:trace-presentation}).

The space $\ETraceo{Y}$ has a canonical smooth structure, and so we describe the tangents below. We note that essentially the same description will carry over to the continuous case in $\PPL_\RR$.

\begin{proposition}[Tangent structure of the classifying space]\label{prop:tangent-structure-etraceo}
    Let $P=(q,o):U\to\ETraceo{Y}$ be a plot. For $\xi\in T_uU$:
\[
  DP_u[\xi]=\bigl(\dot q_{u,\xi},V_{u,\xi}\bigr)
  \qquad
  V_{u,\xi}(\omega)=D_uo(u,\omega)[\xi]
\]
Since within $\Simplexo{\Omega}$ we always have a score $\dot q_{u,\xi} = S_{u,\xi}Q_u$ this can also be written as the score-velocity pair $DP_u[\xi]=(S_{u,\xi},V_{u,\xi})$. Thus, the tangent space of the classifying space at $(q(u), o(u))$ is:
\[
T_{(q(u), o(u))}\ETraceo{Y} = L^2_0(Q)\oplus\prod_{\omega\in\Omega}T_{o(\omega)}Y
\hookleftarrow \{(S_{u,\xi}, V_{u, \xi}): \xi \in T_uU\} = \im DP_u
\]
\end{proposition}
\begin{proof}
    The tangent space of a product manifold is a direct sum of tangent spaces of its factors. The tangent space of a standard open simplex: \[
    \Simplexo{n} = \left\{(x_1,\ldots, x_n): \text{for all}\ i, 0 < x_i, \sum_{i=1}^n x_i = 1\right\}
    \qquad
    \text{is}
    \qquad
    T_x\Simplexo{n} \simeq \left\{(\xi_1,\ldots, \xi_n): \sum_{i=1}^n \xi_i = 0\right\}
    \]
    see e.g.~\cite[Example 2.1]{ayInformationGeometry2017b}. In our case, this can be written as being isomorphic to $L^2_0(Q)$, since $Q$ has an everywhere positive density with respect to counting measure. Scores are indeed centered (and trivially 2-integrable in the finite setting):
    \[
    \int_\Omega S(\omega, \xi) Q(\dd\omega)  = \int_\Omega D(\log q(\omega))[\xi]q(\omega) \dd\omega = D\left(\int_\Omega q(\omega) \dd\omega\right)[\xi] = 0
    \]
\end{proof}
We note that this applies to $U$ being products of Euclidean spaces in particular, and thus extends trivially to the algebraic Euclidean spaces $X$.

\subsection{The universal metric on $\ETraceo{Y}$}\label{ssec:universal-metric}
Let $Y$ be a finite-dimensional smooth manifold (or a coproduct of manifolds), and let $h$ be a positive-semidefinite smooth $2$-form on $Y$. The density and output coordinates of $\ETraceo{Y}$ are smooth-product coordinates, but the probabilistically relevant metric is not a Cartesian product metric: its vertical part is weighted by the current trace law. This is a choice, but we argue it only makes sense this way. For example, consider a program similar to that in~\Cref{sec:intro}:
\begin{align*}
(\beta:\R)\vdash{}\;&
          \letbind{b}{\bern{1000}}{}\\[-1pt]
        &\ifte{b}{\return{\beta}}{}\\[-1pt]
        &\qquad \return{2\beta}\vkto\R
\end{align*}
If we didn't weight by the current trace, we would uniformly select the weight $1$ and $2$ on the tangent of $\beta$, even though the $\true$ branch is almost certain to happen. We explore this argument further in~\Cref{ssec:intrinsic-characterisation-g}.

\begin{definition}[Score-velocity metric]\label{def:score-velocity-metric}
For an algebraic Euclidean space $Y$ and a positive-semidefinite smooth $2$-tensor $h$ on $Y$, we define a positive-semidefinite smooth $2$-tensor on $\ETraceo{Y}$, called the score-velocity metric, at a point $(q,o)$, by:
\[
  G^{\kappa, h}_{(q,o)}
  \bigl((S,V),(T,W)\bigr)
  := \Fisher{q}(S, T) + \kappa H^h_{(q, o)}(V, W)
\]
where:
\[
\Fisher{q}(S, T) = \mathbb{E}_{\omega \sim Q}\left[S(\omega) T(\omega)\right]
\qquad
H^{h}_{(q,o)}(V,W) = \mathbb{E}_{\omega \sim Q} \left[h_{o(\omega)}\bigl(V(\omega),W(\omega)\bigr)\right]
\]
denote the score (Fisher-Rao) and transport forms respectively, and where $(S,V)$ and $(T,W)$ are two tangent pairs in $T_{(q, o)}\ETraceo{Y}$.
\end{definition}
Note that the score-velocity metric is well-defined because of the characterisation of the tangent structure we gave in~\Cref{prop:tangent-structure-etraceo}. Equivalently, in density coordinates $\dot q=S q$, $\dot q'=Tq$, $S,T\in L^2_0(Q_q)$ on the open simplex $\Simplexo{\Omega}$, we can write:
\[
  G^{\kappa, h}_{(q,o)}
  \bigl((\dot q,V),(\dot q',W)\bigr)
  =
  \int_\Omega\frac{\dot q\,\dot q'}{q}\,\dd\lambda
  +\kappa \int_\Omega q\,h(V,W)\,\dd\lambda
\]
The score (first) and transport (second) summands are then orthogonal. Since the transport tensor $H^h$ depends on $q$, $G^{\kappa, h}$ is a warped product metric rather than the product of the Fisher metric with one fixed metric on $Y^\Omega$.
As a forward pointer, we note that in the infinite case, there will arise problems of integrability: on noncompact output manifolds, smoothness of a metric does not imply that its expected pullback is finite. This will be relevant in the real-valued random variable setup in~\Cref{sec:pplr}.
\begin{proposition}
    If $h \in \PSD(Y)$ is positive-definite (i.e., Riemannian), and $\kappa > 0$, then $G^{\kappa, h}$ is also Riemannian.
\end{proposition}
\begin{proof}
    The Fisher metric on the interior of the simplex is positive-definite. Adding and averaging positive-definite metrics with positive weights also preserves positive-definiteness.
\end{proof}

Let us note two geometric invariants of the structure of $\ETraceo{Y}$ with the metric $G^{\kappa, h}$.

\begin{proposition}
For a fixed trace space $\Omega$, a smooth map $f:X\to Y$ induces a map:
\[
\ETraceo{f}: \ETraceo{X} \to \ETraceo{Y} \qquad \ETraceo{f}(q, o) = (q, f \circ o)
\]
which then acts as a pullback on metrics $h \in \PSD(Y)$ via $\ETraceo{f}^*(G^{h}) = G^{f^*h}$. In particular, output isometries induce isometries of classifying spaces.
\end{proposition}
\begin{proof}
    Take some $f: X\to Y$. Then, suppressing the common base point $(q,o)$, we have:
    \begin{align*}
    (\ETraceo{f}^*G^h)((S,V), (S', V')) &= G^h(D\ETraceo{f}[S,V], D\ETraceo{f}[S', V']) \\
    &= G^h((S,Df[V]), (S', Df[V'])) \\
    &= \Fisher{q}(S,S') + H^h(Df[V],Df[V']) \\
    &= \Fisher{q}(S,S') + H^{f^*h}(V,V') \\
    &= G^{f^*h}((S,V), (S',V'))
    \end{align*}
\end{proof}

\begin{proposition}\label{prop:permutations-omega-isometries}
    For any permutation $\pi: \Omega\to\Omega$ and an algebraic Euclidean space $Y$, the induced map between classifying spaces is an isometry:
    \[
    \ETraceoIn{\pi}{Y}: \ETraceoIn{\Omega}{Y} \to \ETraceoIn{\Omega}{Y} \qquad \ETraceoIn{\pi}{Y}(q, o) = (q \circ \pi^{-1}, o \circ \pi^{-1})
    \]
\end{proposition}
\begin{proof}
    Expectations over $\Omega$ do not depend on the permutation $\pi$, and both the output and density are precomposed with $\pi^{-1}$, so for $\omega' = \pi(\omega)$ and $o' = o \circ \pi^{-1}$, the metric basepoint stays invariant:
    \[
    h_{o'(\omega')} = h_{o(\pi^{-1}(\pi(\omega)))} = h_{o(\omega)}
    \]
\end{proof}
An injective map $\iota: \Omega\hookrightarrow\Omega'$ between trace spaces induces a map between their (closed) probability simplices:
\[
\iota_*: \Simplex{\Omega} \to \Simplex{\Omega'} \qquad (\iota_*q)(\omega') = \begin{cases}
    q(\omega) & \omega' = \iota(\omega),\ \omega\in\Omega\\
    0 & \text{otherwise}
\end{cases}
\]
Any map $o: \Omega \to Y$ then also partially defines a map $\iota_*o: \Omega' \to Y$, with the remaining choice in the coordinates in $\Omega' \backslash \iota(\Omega)$. If the metric $G^{\kappa,h}$ was extended to the boundary of the simplex (e.g. in the square-root coordinates, as in~\Cref{thm:finite-sqrt-coordinates-cone}), then we could say that the resulting induced map is an isometric embedding. This statement is a bit awkward because of the arbitrary choice of extension of the output map, which turns out not to matter. Indeed, this stems from the fact that the classifying space $\ETrace{Y}$ distinguishes points with $q$-almost-surely equal output maps. We will remove this freedom in the collapsed space, in~\Cref{def:collapsed-space}.

\subsubsection{Hessian structure on $\ETraceo{Y}$ for Euclidean $Y$}\label{ssec:hessian}

One of the ways information geometry can be derived is through the Hessian structure~\citep{shima2007geometry} prescribed on the statistical manifold. The metric is then given as the Hessian $D^2\Phi$ for some convex potential function $\Phi$. Our metric can also be described in these terms, for a Euclidean space $Y$, although we have to change coordinates, from $(q,y)$ used so far, to:
\[
\eta=(q,m) \qquad \text{where}\qquad m_\omega=q_\omega y_\omega
\]
In other words, $m_i$ is the probability-weighted output attached to a given trace $\omega$.
\begin{definition}[Hessian structure on $\ETraceo{Y}$]
We define the potential function $\Phi: \ETraceo{Y} \to \RR$ in the probability-weighted coordinates $(q, m)$ as the sum of negative entropy in $q$ and the weighted squared 2-norm of $m$:
\[
\Phi(q,m) = \sum_\omega q_\omega\log q_\omega + \frac{\kappa}{2}\sum_\omega\frac{\|m_\omega\|^2}{q_\omega}
\]
\end{definition}
\begin{proposition}
The Hessian structure $\Phi$ on $\ETraceo{Y}$ gives the score-velocity metric:
\[
D^2\Phi=G^{\kappa,\ip{\cdot}{\cdot}}
\]
\end{proposition}
\begin{proof}
Direct differentiation gives:
\[
D^2\Phi[(\dot q,\dot m)]^2
=
\sum_\omega\frac{\dot q_\omega^2}{q_\omega}
+
\kappa\sum_\omega
\frac{\|\dot m_\omega-y_\omega\dot q_\omega\|^2}{q_\omega}
\]
Since:
\[
\dot m_\omega-y_\omega\dot q_\omega=q_\omega\dot y_\omega
\]
this gives the proof.
\end{proof}

\subsection{Square-root coordinates and cone geometry}\label{ssec:square-root-coordinates}
%

\begin{figure}
  \centering
  \begin{subfigure}[c]{0.33\linewidth}
    \centering
    \begin{overpic}[height=0.265\textheight]{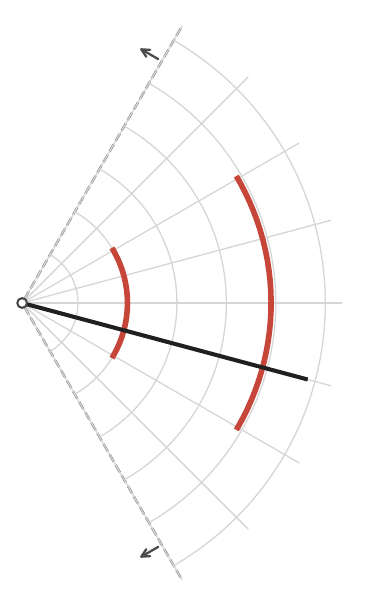}
    \end{overpic}
    \caption{Single cone in the product}
    \label{fig:cone-factor}
  \end{subfigure}
  \hfill
  \begin{subfigure}[c]{0.61\linewidth}
    \centering
    \begin{overpic}[height=0.265\textheight]{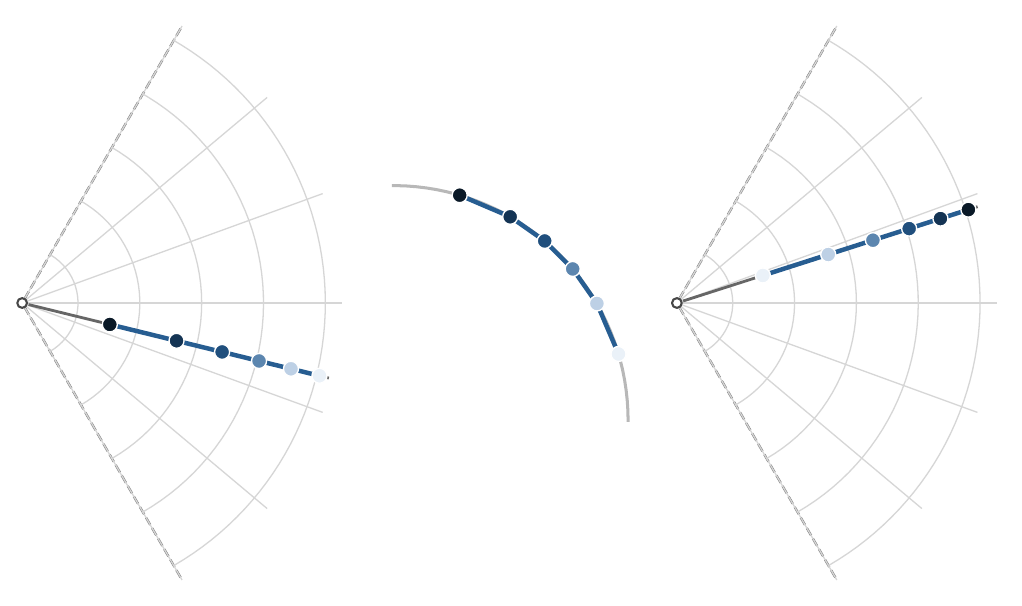}
      \put(46,48){\small normalisation}
      \put(48,42){\small $r_0^2+r_1^2=4$}
    \end{overpic}
    \caption{Two paired cones for $\Omega = \Bool$}
    \label{fig:bernoulli-cones}
  \end{subfigure}

  \caption{%
    Universal space geometry in square-root coordinates, illustrated for $Y=\mathbb{R}$ and $\kappa=1$.
    (a) A visualisation of a part of a single coordinate cone $C(Y) = (\RR_{>0}\times Y, dr^2+\frac14r^2dy^2)$ in the product $\prod_{\omega \in \Omega} C(Y)$, locally isometrically embedded in $\RR^2$. We display only the range of $y\in[-2\pi/3,2\pi/3]$ - the dashed rays are cuts of this finite window, and the whole $\RR_{> 0} \times \RR$ wraps around the origin infinitely many times. The two red arcs represent the same distance in the $Y$ coordinate, and the black line is a constant-$y$ ray.
    (b) For a Bernoulli trace the two radii are coupled by $r_0=2\sqrt{1-q}$ and $r_1=2\sqrt{q}$, so $r_0^2+r_1^2=4$ with $q\in(0,1)$. Matching shades of blue represent the same value of $q$ in the two cones and on the normalisation arc, and $y$ is held fixed.
  }
  \label{fig:cone-geometry}
\end{figure}

\begin{theorem}\label{thm:finite-sqrt-coordinates-cone}
$\ETraceo{Y}$ with the universal metric $G$ is isometric to a radius-$2$ hypersurface in the product of $|\Omega|$-many cones $C_\Omega(Y)$:
\[
  \ETraceo{Y} \simeq \left\{ 
   (r, y) \in C_\Omega(Y): \sum_{i=1}^{|\Omega|}r_i^2=4
   \right\}
   \qquad \mathrm{ where } \qquad
   C_\Omega(Y) := \prod_{i=1}^{|\Omega|}\left(\RR_{> 0}\times Y,
      \dd r_i^2+\frac{\kappa}{4}r_i^2h_i\right)
\]
\end{theorem}
\begin{proof}
Let us now assume that $\Omega = \{1, \ldots n\}$. On $\Simplexo{\Omega}$, we can write the metric simply as:  
\begin{equation*}
  G^{\kappa,h}_{(q, o)}
  =\sum_{i=1}^n\frac{\dd q_i^2}{q_i}
   +\kappa\sum_{i=1}^nq_i h_i
  \label{eq:finite-warped-metric}
\end{equation*}
This is a warped product with warping functions $f_i(q)=\sqrt{\kappa q_i}$. Using the standard trick of moving to square-root coordinates $r_i=2\sqrt{q_i}$ gives:
\[
  G^{\kappa,h}
  =\sum_{i=1}^n\left(\dd r_i^2+\frac{\kappa}{4}r_i^2h_i\right)
\]
\end{proof}

Because of the above, we can explicitly compute various geometric data for the classifying space $\ETraceo{Y}$, under the assumption that $Y$ is a Euclidean space of dimension $d$, and $h$ is a proper Riemannian metric on $Y$. These computations are all standard, but lengthy, so we omit them here.
We show a visualisation of the cone geometry in~\Cref{fig:cone-geometry}.


\begin{remark}[Levi-Civita connection on $\ETraceo{Y}$]\label{prop:levi-civita}
Denoting the ambient connection on the product of cones $C_n(Y)$ by $\bar \nabla$, and letting $R=\sum_{i=1}^{|\Omega|}r_i\partial_{r_i}$ with $\norm{R}=2$ be a normal vector field to $\ETraceo{Y}$, for $U,V$ tangent to $\ETraceo{Y}$ we have:
\[
\nabla^{\ETraceo{Y}}_U V =\bar\nabla_U V+\frac{1}{4}G^{\kappa,h}(U,V)R
\]
\end{remark}

\begin{remark}[Sectional curvature of $\ETraceo{Y}$]
A plane tangent to $Y$ has sectional curvature:
\[
  K=\frac{1}{4} + \frac{4K_h-\kappa}{\kappa r^2}
\]    
\end{remark}
From the formula, the cone is locally flat for $\dim Y\geq2$ exactly when $(Y,h)$ has constant sectional curvature $K_h=\kappa/4$. For $\dim Y=1$, the cone is always locally flat. If every pointwise cone is flat, and $Y$ is connected, the classifying space $\ETraceo{Y}$ is a space form, with constant sectional curvature $1/4$.

\begin{remark}[Scalar curvature of $\ETraceo{Y}$]\label{prop:curvature}
Scalar curvature of $\ETraceo{Y}$ is given as:
\[
  \Scal_{G^{\kappa,h}}=\sum_{i=1}^{|\Omega|}
  \frac{
      \Scal_h(y_i)-\frac\kappa4d(d-1)
  }{\kappa q_i}
  +\frac{m(m-1)}{4}
\]
for $m=|\Omega|(d+1)-1$ being the dimension of $\ETraceo{Y}$.
\end{remark}

Thus, the classifying space is rarely flat: if $n\geq3$ (that is, $|\Omega|$ is bigger than just one random choice), the Fisher simplex directions already contribute curvature. If $n\geq2$ and $d\geq1$, a plane spanned by vertical vectors supported in two different trace fibres has curvature $1/4$. The locally flat cases are therefore limited to $n=1$ with $(Y,h)$ flat, that is, deterministic programs, together with the one-dimensional case $n=2$, $d=0$, i.e. a single coin flip.

\begin{remark}
In coordinates $q_1,\ldots,q_{n-1}$ such that $q_n=1-\sum_{i=1}^{n-1}q_i$, the Riemannian volume on $\ETraceo{Y}$ is given by:
\[
  \dd\vol_{G^{\kappa,h}}
  =
  \kappa^{nd/2}
  \left(\prod_{i=1}^nq_i^{(d-1)/2}\right)
  \dd q_1\cdots\dd q_{n-1}
  \prod_{i=1}^n\dd\vol_h(y_i)
\]
\end{remark}

Note that the first component, coming from the trace, is proportional to a symmetric Dirichlet density with parameter $(d+1)/2$.  For $d=0$, this reduces to the Jeffreys density $\prod_iq_i^{-1/2}$ on the simplex. This is one of the fundamental characterisations of the Jeffreys prior.

\subsection{Pullback factorisation of $\Affine$}

\begin{proposition}\label{theorem:metric-pulls-back}
Setting $\kappa = 1$, for any algebraic Euclidean space $Y$ the metric construction defines an affine map:
\[
  \mathbb G_Y:
  \PSD(Y)
  \longrightarrow
  \PSD\bigl(\ETraceo{Y}\bigr) \qquad \mathbb G_Y(h)=\Law_Y^*\FisherMetric+H^h
\]
For a strictly positive dominated trace presentation $F:X\to Y$ with classifying map: \[
\ClassMap{F}:X\longrightarrow\ETraceo{Y}
\qquad 
  \ClassMap{F}=\left\langle q,\curry(o)\right\rangle
\]
the affine metric transformer $\Affine_F$ factors as a pullback from the universal metric:
\[\begin{tikzcd}
    {\PSD(Y)} && {\PSD(X)} \\
    & {\PSD(\ETraceo{Y})}
    \arrow["{\Affine_F}", from=1-1, to=1-3]
    \arrow["{\mathbb G_Y}"', from=1-1, to=2-2]
    \arrow["{\ClassMap{F}^*}"', from=2-2, to=1-3]
\end{tikzcd}\]
\end{proposition}
\begin{proof}
We have:
\begin{align*}
  \ClassMap{F}^*G^{1, h}_x(\xi,\eta)
  &={}
  \Exp_{Q_x}
  \left[
      S_{x,\xi}S_{x,\eta}
      +h_{o_x(\omega)}(V_{x,\xi},V_{x,\eta})
  \right]\\
  &=(\Fisher{F})_x(\xi,\eta)+(F^\sharp h)_x(\xi,\eta)
\end{align*}
\end{proof}

This is still not the full statement of the universal metric pullback, since the sample space $\Omega$ is fixed, and we only characterise maps into $\ETraceo{Y}$. The latter is not a significant obstruction, since for a program-generated presentation, the same statement holds locally after restricting to its active trace support. To make it formal (in~\Cref{prop:plotwise-metric}), we will need to consider the universal trace space and universal metric, requiring a more general theory than $\PSD$ on algebraic Euclidean spaces considered so far. Thus, we defer it until after we develop the theory of $\CatDQBS$ in~\Cref{sec:pplr}.

\subsection{Intrinsic characterization of the score-velocity metric}\label{ssec:intrinsic-characterisation-g}

As we said, the definition of the universal metric on $\ETraceo{Y}$ might initially appear somewhat arbitrary. We argue it is not. To do that, we first characterise the ``velocity'' part $H^h$ of our metric.

\begin{proposition}[Characterization of the velocity metric]
\label{prop:transport-characterization}
Fix $h\in\operatorname{PSD}(Y)$ and $\kappa \geq 0$. Suppose that, for every finite trace space $\Omega$, $H^\Omega$ is a smooth symmetric positive-semidefinite $2$-form on $\ETraceo{Y}$ satisfying:

\begin{enumerate}
    \item Independence of the score:
    \[
    H^\Omega_{(q,o)}((S,V),(T,W))
    =
    H^\Omega_{(q,o)}((0,V),(0,W))
    \]
    \item Continuity and trace invariance, i.e., there exist
    symmetric bilinear 2-forms, continuous in $p$:
    \[
    B_{p,y}:T_yY\times T_yY\to\mathbb R
    \qquad p\in[0,1]
    \qquad 
    \text{s.t.} \qquad
    H^\Omega_{(q,o)}((0,V),(0,W))
    =
    \sum_{\omega\in\Omega}
    B_{q_\omega,o_\omega}(V_\omega,W_\omega)
    \]
    \item Additivity: 
    \[
    B_{p_1+p_2,y}
    =
    B_{p_1,y}+B_{p_2,y}
    \qquad
    \text{for}\qquad p_1,p_2\geq 0, p_1+p_2\leq 1
    \]
    \item Singleton normalisation:
    \[
    B_{1,y} = \kappa h_y
    \]
\end{enumerate}

Then, we necessarily have:
\[
H^\Omega_{(q,o)}((S,V),(T,W)) = \kappa \sum_{\omega\in\Omega} q_\omega\, h_{o_\omega}(V_\omega,W_\omega) = \kappa H^h_{(q,o)}(V,W)
\]
In particular, $B_{0,y}=0$.
\end{proposition}
\begin{proof}
For $y\in Y$ and fixed $\xi,\eta\in T_yY$, we denote:
\[
b(p)=B_{p,y}(\xi,\eta)
\]
By (3), we have additivity $b(p_1+p_2)=b(p_1)+b(p_2)$,  so by continuity from (2), $b(p)=p b(1)$ on $[0,1]$. Singleton normalization (4) therefore gives us $B_{p,y}=p\kappa h_y$. Substitution into the trace-local decomposition (2) proves the claim.
\end{proof}

Note that these assumptions are relatively strong. The trace invariance and additivity are reminiscent of the argument made in e.g. Jaynes' characterisation of entropy~\citep{jaynesInformationTheoryStatistical1957a}, or in Cencov's characterisation of Fisher information~\citep{cencovStatisticalDecisionRules2000}. Finding a more elegant characterisation of this metric, presumably involving its compositional properties, is one of the future directions we discuss in~\Cref{sec:future-work}.

\begin{proposition}[Characterization of the universal metric] \label{prop:universal-metric-characterization} 
Denoting both parts of the tangent space of $T\mathcal E^\circ_\Omega[Y]=\mathcal H\oplus\mathcal V$ as:
$$
\mathcal H_{(q,o)}=T_q\Simplexo{\Omega}\qquad
\mathcal V_{(q,o)}=\prod_{\omega \in \Omega}T_{o(\omega)}Y
$$
for fixed $\kappa\geq0$ and $h \in \PSD(Y)$, the unique smooth positive-semidefinite symmetric tensor $G$ satisfying:
$$
G|_{\mathcal H\times\mathcal H}=\Law_Y^*\FisherMetric
\qquad
G|_{\mathcal V\times\mathcal V}=\kappa H^h
\qquad
G(\mathcal H,\mathcal V)=0
$$
is $G = G^{\kappa,h}$.
\end{proposition}
\begin{proof}
    Existence follows from~\Cref{def:score-velocity-metric}, so it remains to prove uniqueness. Suppose there is some $G$ satisfying the hypotheses, that is, we have:
    \[
    G((S,V), (S',V')) = G((S,0),(S',0)) +  G((S,0),(0,V')) + G((0,V),(S',0)) + G((0,V),(0,V'))
    \]
    using the characterisation of the tangent space in~\Cref{prop:tangent-structure-etraceo} and bilinearity. The cross-terms disappear by orthogonality hypothesis:
    \[
    G((S,V), (S',V')) = G((S,0),(S',0)) + G((0,V),(0,V'))
    \]
    and score and velocity parts are also fixed by hypotheses, giving the thesis.
\end{proof}

For the non-degenerate case, we have an even more straightforward characterisation.

\begin{corollary}[Characterization of the universal metric, Riemannian case] \label{prop:universal-metric-characterization-riem} Given a Riemannian metric $h \in \PSD(Y)$ and $\kappa >0$, \(G^{\kappa,h}\) is the unique Riemannian metric on $\ETraceo{Y}$, such that the projection: \[
\Law_Y: \ETraceo{Y}\to\Simplexo{\Omega}
\] is a Riemannian submersion onto the open Fisher simplex, with horizontal distribution given by the density tangents and fibre metric \(\kappa H^h\). 
\end{corollary}



The degenerate Riemannian description above applies in the interior $\Simplexo{\Omega}$. The boundary is naturally described by collapsing output values attached to zero-mass traces. We need notions of a quotient by $Q$-almost-sure equality and stratified spaces to define it on the whole $\ETrace{Y}$. This is one of the uses for the machinery of $\CatDQBS$ developed in~\Cref{sec:pplr}.

\section{Weighted programs: $\observe$, $\score$ and normalisation}\label{sec:observe-score-normalise}

So far, we have discussed only the generative version of probabilistic modelling. But standard probabilistic programming languages usually provide access to $\score{\cdot}$ or $\observe{\cdot}$. Semantically, these allow conditioning on data, making writing Bayesian models easier. The operational perspective is that a call to $\score{w}$ multiplies the current execution weight by a (non-negative) number $\exp(w)$, and an observation $\observe{x \text{ from } \mathcal D}$ from distribution $\mathcal D$ multiplies it by a likelihood $p_{\mathcal D}(x)$, implemented as e.g. a call to $\score{\log p_{\mathcal D}(x)}$. Thus, now we extend the language $\PPL$ with a $\score$ primitive to form $\PPLscore$:
\[
\begin{minipage}[t]{0.5\linewidth}\centering
\begin{prooftree}
\AxiomC{$\Gamma \vdash w:\R$}
\RightLabel{(score)}
\UnaryInfC{$\Gamma \vdash \score{w} \vkto \unittype$}
\end{prooftree}
\end{minipage}
\]
This moves the denotations from probability measures to more general finite measures. We point out that because weights are always strictly positive, this preserves the key property of locally constant support. Also, as a forward pointer, this is also the natural domain of Wasserstein-Fisher-Rao geometry: transport moves mass in the output space, while reaction creates or removes mass, so much of the theory survives. But first, we have the following updated definition of the trace presentations.
\begin{definition}[Finite weighted trace presentation]
Let $X$ and $Y$ be algebraic Euclidean spaces. A finite weighted trace presentation $F=(\Omega,q,o):X\to Y$ is a trace presentation, except that $q:X\times\Omega\longrightarrow[0,\infty)$ is no longer required to be a probability measure, only a finite measure. Its total mass is:
\[
Z_F(x)=\sum_{\omega\in\Omega} q(x,\omega) < \infty
\]
\end{definition}
The usual finite trace presentation is the special case of $Z_F=1$. Composition of trace presentations is unchanged except that the component weights need not sum to one. Jet traces $(\Omega, q, o; S, V)$ also have an unchanged definition, apart from the fact that the Radon-Nikodym derivative defining $S$ is no longer $L_0^2(Q_x)$, only $L^2(Q_x)$. The definition of jet trace isomorphisms stays unchanged.

\begin{definition}
    We define the category $\Jet\CatFinTraceWeighted$ to have, as objects, algebraic Euclidean spaces, and as morphisms, finite weighted jet traces quotiented by jet trace isomorphisms.
\end{definition}

\begin{proposition}
    The category $\Jet\CatFinTraceWeighted$ has a copy-discard structure, and the inclusion $\Jet\CatFinTrace \hookrightarrow \Jet\CatFinTraceWeighted$ is a CD functor.
\end{proposition}

We extend the semantics functor $\Trace: \Syn_{\PPL} \to \Jet\CatFinTrace$ for $\PPL$ to semantics $\Tracescore$ for $\PPLscore$, by interpreting $\score{w}$ as exponential weighting. We set $\Tracescore\left(\Gamma \vdash \score{w}\vkto \unittype\right)$ as the following finite weighted jet trace:
\[
\Omega = \unitset \qquad  q(\gamma,\unitsetelem)=\exp(\sem{w}(\gamma)) \qquad o(\gamma,\unitsetelem)=\unitsetelem
\]
\[
 S[\gamma,\unitsetelem;\xi]=D\sem{w}(\gamma)[\xi]
 \qquad V=0
\]

\begin{proposition}
    The semantics map defined above: \[
    \Tracescore: \Syn_{\PPLscore} \to \Jet\CatFinTraceWeighted
    \]
    is a strong monoidal functor. Moreover, semantics for the normalised fragment $\PPL$ factors through:
    \[\begin{tikzcd}
      {\Syn_{\PPLscore}} && {\Jet\CatFinTraceWeighted} \\
      {\Syn_{\PPL}} && {\Jet\CatFinTrace}
      \arrow["{\Tracescore}", from=1-1, to=1-3]
      \arrow[hook, from=2-1, to=1-1]
      \arrow["\Trace", from=2-1, to=2-3]
      \arrow[hook, from=2-3, to=1-3]
    \end{tikzcd}\]
\end{proposition}

\subsection{Weighted affine functor $\AffineW$}
To develop a weighted version of the pullback functor $\AffineW$, it is useful to step back and take a slightly different perspective from what we did in~\Cref{sec:algebraic}. Instead of starting from the ``global'' perspective of Fisher information $\Fisher{F}$ and extending it with the output metric $F^\sharp$, we now take a local, operational perspective of tracing the change along each possible execution branch of the program. Intuitively, the energy of infinitesimal change of the parameter in the tangent direction $\xi$ should correspond to an expected pullback of this change along each trace. Let us first formalise the (weighted) execution map.
\begin{definition}
For a weighted jet trace $F = (\Omega, q, o; S,V):X\to Y$, for a fixed trace $\omega \in \Omega$ we define the weighted execution map:
\[
W_{F,\omega}: \RR\times X \to \RR \times Y
\qquad
(c,x) \mapsto \bigl(c+\log q(x, \omega), o(x, \omega)\bigr).
\]
where we use the convention $\log 0 = 0$.
\end{definition}
For fixed randomness, this map is essentially how operational semantics carries both the weight and current value throughout the program execution. We work in $\log$-likelihood coordinates, because we want the weights to be additive. For our purposes, the map $W_{F,\omega}$ will only matter on possible trajectories, so the choice of $\log 0$ is unimportant - we could as well define it up to $Q_x$-a.s. equality.

Continuing the discussion from the initial paragraph, the geometry of the codomain of a map $F: X \to Y$ is in principle encoded as some $h \in \PSD(Y)$. But because we want the change of the likelihood to also have non-zero energy (in particular, discrete probabilistic models should still be described by Fisher geometry), the metric is, in principle, some $B \in \PSD(\RR \times Y)$, where the additional first coordinate is the total weight. In principle, $B$ itself could depend on the $\RR$ coordinate too, but it is slightly unnatural for the shape of geometry to depend on (non-local) information about the weight accumulated so far. Thus, we make the choice of looking at metrics independent of the first coordinate, which are the following.
\begin{definition}[Weight-independent metrics]
    For a given algebraic Euclidean space $X$, we define the cone of weight-independent positive-semidefinite 2-forms on $\RR$-extended space as:
    \[
    \PSDW(\RR \times X) = \{B \in \PSD(\RR \times X) \mid B_{c,x}\ \text{independent of}\ c\}
    \]
\end{definition}
We note that such $B$ is not quite equivalent to what we considered in~\Cref{sec:algebraic}, since it has non-vanishing cross-terms between $V$ and $S$. There is a good reason for this. For weighted kernels, the ``weighted expected score'' term:
\[
\sum_{\omega\in\Omega}q(x,\omega)S_{x,\xi}(\omega) = DZ_F(x)[\xi]
\]
is no longer zero (which it is in probability kernels, because $Z_F\equiv1$). If two maps $F$ and $G$ have scores $S_F$ and $S_G$, then the score of $F \circ G$ is $S_F+S_G$, and its square contains a genuine mixed term $2S_FS_G$. This term will, after further composition, mix with $V$ terms. This is unavoidable in the non-normalised case. The old affine functor $\Affine$ exploited the fact that the expectation of the score vanishes, and so, almost by luck, could work on pure $\PSD(Y) \to \PSD(X)$, because $S$ and $V$ parts pulled back separately.

Once we settle on the relevant approach being a trace-by-trace averaged pullback of the output $B$, the following definition of the transformation is obvious.
\begin{definition}
We define the contravariant weighted map $\AffineW$ by:
\[
\AffineW_F : \PSDW(\RR \times Y) \to \PSDW(\RR \times X)
\qquad
\AffineW_F(B)_x = \sum_{\omega\in\Omega} q(x,\omega) DW_{F,\omega,x}^{\,*} B_{o(x,\omega)}
\]
\end{definition}

And indeed, we can show that it has nice properties. First, it is functorial.
\begin{proposition}
For weighted jet trace presentations $F, G$, we have:
\[
\AffineW_{G\circ F}
=
\AffineW_F\circ\AffineW_G
\qquad
\AffineW(\id) = \id
\]
so the map $\AffineW: \Jet\CatFinTraceWeighted^{op} \to \CatAffCone$ is a contravariant functor. 
\end{proposition}

\begin{proof}
Scores add and velocities compose, hence:
\[
DW_{G \circ F} = DW_{G}\circ DW_{F}
\]
The composite trace weight is the product of the two component weights. Substituting these identities into the defining equation for $\AffineW$ and rearranging the finite sums gives the result.
\end{proof}
For an algebraic Euclidean space $Y$, $B \in \PSDW(\RR \times Y)$ can be decomposed as:
\[
B_y
=
\begin{bmatrix}
 z(y)&\alpha_y\\
 \alpha_y&h_y
\end{bmatrix}
\]
where $z(y)$ is a non-negative scalar field, $\alpha$ is a 1-form, and $h \in \PSD(Y)$.
Writing:
\[
\mathbb A_F(z,\alpha,h)=(z',\alpha',h')
\]
we can explicitly compute the output as:
\begin{align*}
z'(x)&=\sum_\omega q_x(\omega)z(o_x(\omega))\\
\alpha'_x(\xi)&=
\sum_\omega
 q_x(\omega)
\left[
 z(o_x(\omega))S_{x,\omega}[\xi]
 +
 \alpha_{o_x(\omega)}(V_{x,\omega}[\xi])
\right] \\
h'_x(\xi,\eta)
&=
\sum_\omega
 q_x(\omega)
\Big[
 z(o_x(\omega))S_{x,\omega}[\xi]S_{x,\omega}[\eta]
 +
 S_{x,\omega}[\xi]\alpha_{o_x(\omega)}(V_{x,\omega}[\eta])
\\
&\qquad + S_{x,\omega}[\eta]\alpha_{o_x(\omega)}(V_{x,\omega}[\xi])
 +
 h_{o_x(\omega)}(V_{x,\omega}[\xi],V_{x,\omega}[\eta])
\Big]
\end{align*}
The mixed terms are therefore forced by the derivative of the continuation mass; they are not additional arbitrary geometric data. To fully justify the preceding argument, let us look at how it behaves on the normalised slice.
\begin{definition}
    For every algebraic Euclidean space $X$, we define a block-diagonal embedding $j: \PSD(X) \to \PSD(\RR \times X)$ as:
    \[
    j_X(h) = 
    \begin{bmatrix}
1&0\\
0&h
\end{bmatrix}
\]
\end{definition}
\begin{remark}\label{remark:unnormalised-block-matrix}
When starting from a block-diagonal form, we have:
\[
\mathbb A_F\left(j(\kappa h)\right)
=
\begin{bmatrix}
Z_F&DZ_F\\
DZ_F& G^{F,\kappa,h}
\end{bmatrix}
\]
If $F: X \to Y$ is normalised, i.e. $Z_F\equiv1$, then $DZ_F=0$, and the lower-right block is precisely:
\[
G^{F,\kappa, h} = \Fisher{F}+\kappa F^\sharp h
=
\Affine_F(\kappa h)
\]
Thus the affine transformer of~\Cref{sec:algebraic} is the unit-mass slice of the weighted block transformer, in particular:
\[
\AffineW_F(j_Y(h)) = j_X(\Affine_F(h))
\]
\end{remark}
Indeed, as we explained, it is the non-vanishing of the expected score that produces the cross-terms $\alpha = DZ_F$. If we start without them, with the initial $\alpha = 0$,  they never reappear in the normalised setting. This is what allowed the simplified definition of $\Affine$ before. It is also another (the third) way of deriving it.

\subsection{Normalisation}
Since our language $\PPLscore$ is naturally valued in weighted measures and there is no first-class type of probability distributions, it does not need an explicit $\normalise{}$ primitive. However, it is still interesting to discuss how normalisation interacts with the language. As we show, it cannot be compositional, but it does have a concrete geometric interpretation.

\begin{definition}[Normalisation of jet traces]
Let $F=(\Omega,q,o; S,V): X \to Y$ be a weighted jet trace and suppose that $0<Z_F(x)<\infty$. We define its normalisation $\normaliseF{F}: X \to Y$ to be:
\[
\normaliseF{F} = \left(\Omega, q^\mathrm{norm}, o; S^{\mathrm{norm}}, V\right)
\]
where:
\[
q^\mathrm{norm}(x, \omega) = \frac{q(x,\omega)}{Z_F(x)}
\qquad \text{and} \qquad
S^{\mathrm{norm}}[x,\omega;\xi] =S[x,\omega;\xi] - D\log Z_F(x)[\xi]
\]
\end{definition}
\begin{proposition}
The map $\normaliseF$ satisfies the following properties:
\begin{itemize}
    \item $\normaliseF{F}$ is a (standard) jet trace for any weighted jet trace $F$, 
    \item $\normaliseF{\normaliseF{F}} = \normaliseF{F}$,
    \item $\normaliseF{(F \otimes G)} = \normaliseF{F} \otimes \normaliseF{G}$,
    \item $\normaliseF(\Trace(\letbind{\_}{\score w}{p})) = \normaliseF(\Trace(p))$ for a constant weight $w \in \RR$
    \item $\normaliseF{(G \circ F)} = G \circ \normaliseF{F}$, if $G$ is a standard jet trace.
\end{itemize}
\end{proposition}
\begin{proof}
    The first four properties follow straightforwardly from the definition of the $\normaliseF$ map. For the final property, let $F=(\Omega_F,q,o;S,V)$ and let $G=(\Omega_G,r,p;T,W)$ be a standard (normalised) jet trace. Since $\sum_{\omega_G} r(y,\omega_G)=1$ for every $y$, the normalisation constant for the composition is:
    \begin{align*}
    Z_{G\circ F}(x) &= \sum_{\omega_F,\omega_G} q(x,\omega_F)r(o(x,\omega_F),\omega_G) \\
    &= \sum_{\omega_F} q(x,\omega_F)\sum_{\omega_G}r(o(x,\omega_F),\omega_G) \\
    &= \sum_{\omega_F} q(x,\omega_F) = Z_F(x)
    \end{align*}
    Hence its normalised density is:
    \[
    \frac{q(x,\omega_F)r(o(x,\omega_F),\omega_G)}{Z_F(x)} = q_{\normaliseF{F}}(x,\omega_F)r(o(x,\omega_F),\omega_G)
    \]
    which is exactly the density of $G\circ \normaliseF{F}$. Using the composition rule for scores $S_{G\circ F}=S_F+S_G \circ V_F$ normalization subtracts $D\log Z_F$, so also:
    \[
    S_{N(G\circ F)} =(S_F-D\log Z_F)+S_G\circ V_F =S_{G\circ \normaliseF{F}}
    \]
    The output map and velocity are unchanged by normalization. Therefore, we get $N(G\circ F)=G\circ NF$.
\end{proof}
\begin{remark}
We note that $\normaliseF: \Jet \CatFinTraceWeighted\to \Jet\CatFinTrace$ is not a functor, because it does not respect compositionality. Let us take:
\[
F = \sem{\bern{0}}: \unitset \to \Bool
\qquad 
G = (\unitset, \{\true \mapsto 1, \false \mapsto 2\}, \id): \Bool \to \Bool
\]
Then, $\normaliseF{G}  \circ \normaliseF{F}$ is a fair coin, but $\normaliseF{(G \circ F)}$ is a $1/3$-biased coin. This is because the normalisation in $\normaliseF{G}$ is done for each input separately, while in the composed case $G \circ F$ one branch affects the other.
\end{remark}
We have pointed out that $\Affine$ is a special case of $\AffineW$. Thus, it is natural to ask how $\AffineW(F)$ is related to $\Affine(\normaliseF{F})$. We have the following characterisation.
\begin{proposition}
     For a weighted jet trace $F$ with $0<Z_F<\infty$ and $\AffineW_F(j(\kappa h)) = (Z_F, DZ_F, G^{F,\kappa, h})$, we have:
\[
\Affine_{\normaliseF{F}}(\kappa h)
=
\frac{1}{Z_F}G^{F,\kappa, h}
-
D\log Z_F\otimes D\log Z_F
=
\frac{1}{Z_F} G^{F,\kappa, h}
-
\frac{D Z_F\otimes D Z_F}{Z_F^2}
\]
\end{proposition}
\begin{proof}
    Follows from~\Cref{remark:unnormalised-block-matrix}.
\end{proof}
Equivalently, the above can be written as $Z_F^{-1}$ times the lower-right Schur complement of the matrix in~\Cref{remark:unnormalised-block-matrix}. We note that normalisation is a partial, non-linear operation.
\begin{remark}
Applying normalisation $\normaliseF$ to the program:
\[
\begin{aligned}
t: \R \vdash &\letbind{y}{p}{\\
&\letbind{\_}{\score{t\cdot f(y)}}{\\
&\return{y}}}\vkto Y   
\end{aligned}
\qquad \text{implements the exponential family}\qquad 
\mu_t(y) = \frac{e^{tf(y)}\mu_p(y)}{\mathbb{E}_{\mu_p}[e^{tf(Y)}]}
\]
On the other hand, an ordinary conditional:
\[
\begin{aligned}
t: \R \vdash & \letbind{b}{\bern{t}}{\\
&\ifte{b}{\return{y_1}\\
&\qquad}{\return{y_2}}} \vkto Y
\end{aligned}
\qquad 
\text{implements the mixture family}\qquad \mu_t = \sigma(t)\delta_{y_1} + (1-\sigma(t))\delta_{y_2}
\]
In information geometry, mixture coordinates and exponential coordinates are dual to each other. It is interesting to see in what sense the language constructs of $\mathsf{score}$ and $\mathsf{if}$ could be said to be dual. We do not have a good answer, and leave that for future work, but note that one interesting angle is through~\citep{richardson2024mixture}.
\end{remark}


\subsection{Classifying spaces for $\PPLscore$}
We can repeat the construction of $\ETraceo{Y}$ for $\PPLscore$, while replacing $\Simplexo{\Omega}$ by $\RRplus^{|\Omega|}$. The details remain broadly the same, but the classifying space is then no longer only a radius-2 hypersurface; rather, it is simply the whole product of cones.
\begin{definition}
    We define the classifying space for $\PPLscore$ as:
    \[
    \ETraceW{Y} = \RRplus^{|\Omega|} \times Y^\Omega
    \]
    carrying the metric:
    \[
    G^{\kappa, h}((S, V), (T, W)) = \sum_{\omega \in \Omega} q(\omega)S(\omega)T(\omega) + \kappa\sum_{\omega \in \Omega}q(\omega)h(V(\omega),W(\omega))
    \]
\end{definition}
We note that this metric is exactly the same as in the normalised case in~\Cref{def:score-velocity-metric}: only the domain is enlarged from $\Simplexo{\Omega}$ to $\RRplus^{|\Omega|}$.

\begin{proposition}\label{thm:orthogonal-projection}
$\ETraceW{Y}$ with the universal metric $G$ is isometric to $C_\Omega(Y)$ from~\Cref{thm:finite-sqrt-coordinates-cone}. The normalisation map $(q, o) \mapsto (q / Z, o)$ is then a radial projection onto $\ETraceo{Y}$.
\end{proposition}

\section{Analytic formulation and $\PPL_\RR$}\label{sec:pplr}
We can now present an extension of $\PPL$ to a more expressive language, allowing for real-valued random variables. Specifically, we add a new base type, $\Rc$, a Gaussian distribution introduction rule:
\[
\begin{minipage}[t]{0.5\linewidth}\centering
\begin{prooftree}
\AxiomC{$\Gamma \vdash x:\Rs$}
\RightLabel{(Gauss)}
\UnaryInfC{$\Gamma \vdash \gauss{x}{1} \vkto \Rc$}
\end{prooftree}
\end{minipage}
\]
The new type $\Rc$ will host \emph{control} values, that is, real values which nevertheless can be cast to $\Bool$, and we will write $\Rs$ for the ``standard'', previously used type of \emph{smooth} $\R$ to avoid confusion. We have two new primitives, $\coerce{x}$ and $\isnneg{x}$ with the following introduction rules:
\[
\begin{minipage}[t]{0.5\linewidth}\centering
\begin{prooftree}
\AxiomC{$\Gamma \vdash x:\Rc$}
\RightLabel{(Coerce)}
\UnaryInfC{$\Gamma \vdash \coerce{x} : \Rs$}
\end{prooftree}
\end{minipage}
\qquad
\begin{minipage}[t]{0.5\linewidth}\centering
\begin{prooftree}
\AxiomC{$\Gamma \vdash x:\Rc$}
\RightLabel{(Is-non-negative)}
\UnaryInfC{$\Gamma \vdash \isnneg{x} : \Bool$}
\end{prooftree}
\end{minipage}
\]

Thus, real values of type $\Rc$ are more permissive than those of $\Rs$. In particular, we can now split $\textsf{if}$'s on real values. However, there is intentionally no rule for turning $\Rs$ into $\Rc$, which is key to our smooth semantics - we have to treat $\Rc$ as having discrete topology, and thus trivial smooth structure, but standard measurable structure, isomorphic to $\RR$. This is essentially forced, as we show in~\Cref{prop:control-diffeology-is-forced}. Nevertheless, programs having intermediate values in $\Rc$ are still smooth. To show this, we first need to properly define the semantics.

\begin{remark}
    Let us point out the need for $\theta: \Rs \vdash \gauss{\theta}{1} \vkto \Rc$ instead of a zero-parameter $\textsf{gauss}() := \gauss{0}{1}$. Of course, in terms of stochastic dynamics, $\gauss{\theta}{1}$ is equal in distribution to $\gauss{0}{1} + \theta$. But while the former consumes $\Rs$ and outputs $\Rc$, the latter can only consume $\theta: \Rc$, which would be more restrictive.
\end{remark}

\subsection{Diffeological and quasi-Borel spaces}
To build semantics for $\PPL_\RR$, we have several key requirements. To keep track of $\Rc$ and $\Rs$, we need a theory that allows for arbitrary, but compositional, separate measurable and smooth structures, as opposed to defining them jointly (as standard smooth structure and Borel measurable structure). To build moduli spaces $\ETrace{Y}$ for uncountable $\Omega$'s with non-trivial measurable structure, we need function spaces. To extend metrics to the boundary of (infinite-dimensional) $\Simplex{\Omega}$, we will need well-behaved quotients. All these requirements make formalisms of diffeological spaces and quasi-Borel spaces attractive here - indeed, both of them have been used before as a convenient way to build semantics for differentiable and probabilistic programs. We restate known results about diffeological and QBS spaces in~\Cref{appendix:dqbs}, and proceed with combining them below in a lightweight, straightforward manner.

\begin{definition}[$\CatDQBS$ category]
A diffeological quasi-Borel space is a triple:
\[
    (X,\mathcal D_X,\Meas_X)
\]
consisting of a diffeology $\mathcal D_X$ and a quasi-Borel structure $\Meas_X$ on the same set $X$, such that every plot $(p:U\to X)\in\mathcal D_X$ is a quasi-Borel morphism when $U$ is equipped with its standard Borel quasi-Borel structure. A map between diffeological quasi-Borel spaces:
\[
    F:(X,\mathcal D_X,\Meas_X) \to (Y,\mathcal D_Y,\Meas_Y)
\]
is a set map $X \to Y$ which is both diffeologically smooth and a quasi-Borel morphism. We denote the resulting category by $\CatDQBS$.
\end{definition}

We note that the compatibility condition is one-way: every smooth plot is measurable, but a random element need not be a smooth plot. This is intentional, since random elements are by design not required to be smooth - this is obvious in the context of discrete sample spaces $\Omega$ considered so far (c.f.~\Cref{def:trace-presentation}), but will remain true as we extend to real-valued random variables later.

\begin{theorem}[Categorical structure of $\CatDQBS$]\label{thm:dqbs-complete-cocomplete-ccc}
The category $\CatDQBS$ is complete, cocomplete, and Cartesian closed.
Consequently, its Cartesian structure gives canonical copy and discard maps.
\end{theorem}
\begin{proof}
    For limits, we take the standard limit of the underlying sets, and give it initial quasi-Borel and initial diffeological structures. For colimits, we also take the standard colimit of the underlying sets, and give it final diffeological structure and final quasi-Borel structure. Both of these preserve the compatibility condition for $\CatDQBS$. For the exponential $Y^X$, we take the underlying set:
    \[
    Y^X = \Hom_\CatDQBS(X, Y) \subseteq \Hom_\QBS(X, Y)
    \]
    with the subspace quasi-Borel structure. Its diffeological structure is given by plots $p: U \to Y^X$ such that the uncurried $\hat p: U \times X \to Y$ is both smooth and measurable. Evaluation map:
    \[
    \evmap: Y^X \times X \to Y
    \]
    is measurable because we took the domain with subspace QBS structure, and it is smooth because for each plot $p$, the composition $\evmap\circ (p\times\id_X) = \hat p$ by the universal property of $\evmap$, and $\hat p$ was required to be smooth. Finally, standard diagonal and terminal maps of the underlying sets are $\CatDQBS$ morphisms, and give it copy-discard structure.
\end{proof}

\begin{remark}[Probability monad on $\CatDQBS$]\label{remark:dqbs-monad}
It is difficult to lift the quasi-Borel probability monad to an endofunctor on $\CatDQBS$, because of issues with specifying diffeology on $\Prob(X)$. And in any case, for any diffeology, the expectation map would fail to be smooth: taking the Cauchy distribution $\mu$ on $\RR$ as a constant plot, and pairing it with the bounded smooth function $f(t,x)=\cos(tx)$, the expectation:
\[
    (\mu, f) \mapsto \int_\RR \cos(tx)\mu(\dd x)=e^{-\abs{t}}
\]
is not differentiable at $t=0$. Thus, we keep interpreting programs as dominated trace presentations, rather than use Kleisli morphisms. 
\end{remark}
Incidentally, the same example has been used in the concurrently released work of~\citet[Theorem 6.15]{forreSampleSmoothSpacesConvenient2026}, who develops a different category combining smooth and probabilistic settings, called $\mathbf{SSS}$, for Sample-Smooth Spaces. However, even though $\mathbf{SSS}$ has a probability monad, it still fails to have expectations as smooth morphisms, and moreover, it is incompatible with ordinary smooth plots into $\mathcal{P}(\Bool)$, and having nonconstant plots into Bernoulli distributions is important for our development.

\begin{definition}[Final diffeology]\label{def:final-diffeology}
Let $(X_i)_{i\in I}$ be diffeological spaces and let $e_i:X_i\longrightarrow X$ be maps whose images cover the set $X$. The final diffeology generated by the maps $e_i$ is defined as follows. A map $p:U\to X$ is a plot if, for every $u\in U$, there are a neighbourhood $V\subseteq U$ of $u$, an index $i$, and a plot $\widetilde p:V\to X_i$ such that:
\[
    p|_V=e_i\circ\widetilde p
\]
\end{definition}

In other words, a plot of $X$ is a family which can be lifted locally through
one of the generating maps. This diffeology is characterised by the following
universal property, stated in~\Cref{appendix:dqbs}.

\begin{definition}[Internal tangent space]\label{def:internal-tangent-space}
Let $(X, \mathcal D, \mathcal M)$ be an object in $\CatDQBS$, and let $x\in X$. For every pointed plot:
\[
    p:(U,0)\longrightarrow(X,x)
\]
and every $\xi\in T_0U$, introduce a symbol $p_*\xi$ for the lift of $\xi$. The internal tangent space $T_xX$ is the vector space generated by these symbols, modulo linearity in $\xi$ and the relations:
\[
    (p\circ f)_*\eta=p_*\bigl(Df_0[\eta]\bigr)
\]
for every smooth pointed map $f:(V,0)\to(U,0)$. A smooth map $F:X\to Y$ induces:
\[
    DF_x:T_xX\longrightarrow T_{F(x)}Y,
    \qquad
    DF_x(p_*\xi)=(F\circ p)_*\xi
\]
\end{definition}

For a manifold this recovers the ordinary tangent space. For a set with the
discrete diffeology, every pointed plot is locally constant, and hence:
\[
    T_xX_{\mathrm{disc}}=0.
\]
For a final diffeology, the internal tangent space is generated by tangents of local lifts. A general tangent may be a finite sum of such generators and need not be represented by one lift.

Since the tangent bundle over any manifold has a structure of a smooth manifold as well, for any algebraic Euclidean $X$, $TX$ is also a diffeological space, and thus (considered with a standard Borel structure) an object in $\CatDQBS$.

\subsection{Trace spaces in $\CatDQBS$}

To start defining the semantics in $\CatDQBS$, we first need to do it on types - and this means picking semantics for the $\Rc$ generator, as everything else will be forced by induction and the already-defined finite fragment $\PPL$. It turns out that this choice is forced, as soon as we have the cast $\isnneg{x}$ on real random values, and require conditionals $\mathsf{if-then-else}$ to be smooth. Let us denote the semantics of $\Rc$ as $\RRc \in \CatDQBS$, and then fix its measurable and random structure by putting some:
\[
    \RRc = (\RR,\mathcal D_c, \mathcal M_c),
\]

We first observe that $\mathcal M_c$ must be $\Mstd$. First, we want the identity-on-points coercion $\coerce{}: \Rc \to \Rs$ to be allowed, thus we have $\mathcal M_c \subseteq \Mstd$. Second, the Gaussian semantics dominated trace map will have output map $o(\omega) = \omega$, thus forcing $\Mstd \subseteq \mathcal M_c$.

Therefore, the remaining choice is in $\mathcal D_c$. We also note that we do not choose a topology for $\RRc$ separately: whenever we refer
to the topology of \(\RRc\), we will mean the D-topology induced by
\(\mathcal D_c\) (see~\Cref{appendix:dqbs} for the definition).

We interpret \(\Bool\) as in $\PPL$, that is, with its finite discrete diffeology and its standard finite quasi-Borel structure. We call \(\mathcal D_c\) compatible with the deterministic fragment when every well-typed pure term in $\PPL_\RR$ denotes a morphism in \(\CatDQBS\).

\begin{proposition}[The control diffeology is forced]\label{prop:control-diffeology-is-forced}
If \(\mathcal D_c\) is compatible with the deterministic fragment, then
\(\mathcal D_c\) must be the discrete diffeology. Consequently, the
D-topology of \(\RRc\) is the discrete topology. Conversely, the discrete diffeology $\Ddisc$ on $\RRc$ is compatible with the control-real primitives. It is therefore the unique compatible diffeology on the
carrier \(\RR\) with standard quasi-Borel structure.
\end{proposition}

\begin{proof}
Equality of control reals is definable in the deterministic fragment.
Indeed, consider the pure term:
\[
x:\RRc,\ y:\RRc
\vdash
\mathsf{eq}(x,y)
\df
\ifte{\isnneg{x-y}}
     {\isnneg{y-x}}
     {\false}
:
\Bool
\]
Its denotation:
\[
    \sem{\mathsf{eq}}
    :
    \RRc\times\RRc
    \longrightarrow
    \Bool
\]
is smooth by assumption, and satisfies:
\[
    \sem{\mathsf{eq}}(x,y)=\true
    \quad\Longleftrightarrow\quad
    x=y
\]

Let $p:U\to\RRc$ be an arbitrary plot, let $u_0\in U$, and put
$a=p(u_0)$. The constant map:
\[
    \overline a:U\longrightarrow\RRc
    \qquad
    \overline a(u)=a
\]
is a plot. Hence the composite:
\[
    U
    \xrightarrow{(p,\overline a)}
    \RRc\times\RRc
    \xrightarrow{\sem{\mathsf{eq}}}
    \Bool
\]
is a plot of $\Bool$. Since $\Bool$ carries the discrete
diffeology, this composite is locally constant. Its value at $u_0$ is $\true$, so there is a neighbourhood $V$ of $u_0$ such that:
\[
    \sem{\mathsf{eq}}(p(u),a)=\true
\]
for every $u\in V$. Therefore $p(u)=a$ on $V$.

Thus every plot of $\RRc$ is locally constant. Every diffeology
contains all locally constant maps, so $\mathcal D_c$ is exactly the discrete diffeology. It follows immediately that every subset of $\RR$ is D-open.

Conversely, a finite product of discrete diffeological spaces is
discrete, and every set map out of a discrete diffeological space is smooth. The control-real arithmetic, the hard guard, and the coercion map are Borel measurable, so with the standard quasi-Borel structures they are morphisms in $\CatDQBS$.
\end{proof}

\begin{corollary}[$\CatDQBS$ structure on $\Rc$]
The control-real type is necessarily interpreted as:
\[
    \RRc
    =
    (\RR,\mathcal D_{\mathrm{disc}},
          \mathcal M_{\mathrm{std}})
\]
In particular, we have $T_x\RRc=0$ for every $x\in\RRc$.
\end{corollary}

\begin{corollary}\label{cor:coerce-is-smooth}
The identity-on-points $\semcoerce:\RRc\to\RRs$ is a morphism in \(\CatDQBS\), but the reverse identity \(\RRs\to\RRc\) is not smooth.
\end{corollary}



We thus have an extension of the key~\Cref{prop:finite-values-locally-constant} pointed out in~\Cref{sec:ppl-jets}.

\begin{corollary}[Smooth-control separation]\label{cor:smooth-control-separation}
After fixing all control coordinates and coproduct tags, every pure term with codomain \(\RRc\) or \(\Bool\) is locally constant (in the remaining smooth coordinates).
\end{corollary}

Because of the above, we call any such space a trace space, as in the definition above.

\begin{definition}[Trace space as an object in $\CatDQBS$]\label{def:trace-object}
    A trace space is an algebraic Euclidean space $(\Omega, \lambda)$ equipped with the standard quasi-Borel structure and discrete diffeology:
    \[
      \Omega_\lambda = (|\Omega|, \Ddisc, \Mstd)
    \]
\end{definition}
As a matter of notation, we suppress $\lambda$ and write simply $\Omega = \Omega_\lambda$.

\subsection{Polynomially bounded maps}

Extending the construction of~\Cref{sec:ppl-jets}, we now want to build semantics of $\PPL_\RR$ in 1-jet traces over (no longer necessarily finite) trace presentations. Thus, we first need a notion of trace presentation $(\Omega_\lambda, q, o): X \to Y$ between $\CatDQBS$ objects $X, Y$ where $\Omega_\lambda$ is a trace object in $\CatDQBS$, $q: X \times \Omega \to [0, \infty)$ is a (smooth) density with respect to $\lambda$, and $o: X \times \Omega \to Y$ is a (smooth) output map.
``Smooth'' here means simply that the map is a morphism in $\CatDQBS(X, Y)$ - where all morphisms are required to be smooth in the sense of diffeologies in $X$, $Y$ and $\Omega$. Because we put discrete diffeology on trace objects, this definition generalises~\Cref{def:trace-presentation}. Jet traces can then remain broadly the same as before in~\Cref{def:jet-trace}, but now the domains of the maps: \[
S(x,\omega;-):T_xX\to\RRs \qquad
\qquad
 V(x,\omega;-):T_xX
      \to T_{o(x,\omega)}Y
\]
are interpreted as internal tangent spaces in $\CatDQBS$. Moreover, because the trace space $\Omega$ has discrete diffeology, we can interpret them as having the domain of $T_{x, \omega}(X\times \Omega)$, because the other part of the tangent is trivial. Composition of jets is given by the same formulas, and jet trace isomorphisms remain the same.

This is indeed how we proceed, but there is still one more subtlety with this approach - namely, integrability issues. This will force us to move from the category $\CatDQBS$ to a certain concrete subcategory over it. We explain these in the section below.

\subsubsection{Integrability of score}\label{sssec:integrability-of-score}

One problematic aspect of extending the language to continuous spaces is the integrability of the score. That is, for a given jet trace $F = (\Omega, q, o; S, V)$, the score $S(x, \omega;\xi)$ can fail to be $L^1(\Omega, q(x))$, which prevents us from writing the identity:
\[
 \mathbb{E}_{\omega\sim q(x)}[S(x, \omega; \xi)]=0
\]
This identity was the key to composition of 1-jet traces in~\Cref{sec:algebraic}. And even if the identity holds, the score $S$ can still fail to be square-integrable, which is required for the existence of the Fisher-Rao metric $g^F$. Thus, some more restrictions are needed on trace presentations.

\begin{example}[A program separating $L^1$ and $L^2$ integrability]
\label{example:moment-counterexample}
Let us fix some hyper-parameter $\alpha>0$, and consider the following statistical model:\[
 Z\sim N(0,1)
 \qquad
 Y\mid Z\sim N\bigl(\theta e^{\alpha Z^2},1\bigr)
\]
This model could be described by a program of the shape:
\begin{align*}
\theta: \Rs \vdash & \letbind{z}{\gauss{0}{1}}{\\
&\letbind{y}{\gauss{\theta\cdot\exp(\alpha\cdot \coerce{z}^2)}{1}}{\\
&\return{\coerce{y}} \vkto \Rs}
}
\end{align*}
provided we have the primitives $\exp(x)$ and $z^2$ in $\Prim$. The trace space is $\Omega = \RR \times \RR$; we denote its elements by $(z,y) \in \Omega$. Then, the score (for a fixed $\alpha$) is:
\begin{align*}
 S_\alpha(\theta, (z, y))&= D_\theta \log p(z,y) \\
 &=D_\theta \log p(z) + D_\theta \log p(y|z) \\
&= D_\theta \left(-\frac{(y - \theta e^{\alpha z^2})^2}{2}\right) \\
&=  y{e^{\alpha z^2}} - \theta {e^{2\alpha z^2}}
\end{align*}
Evaluating at $\theta=0$:
\[
S_\alpha(0, (z,y)) = y e^{\alpha z^2}
\]
Since $Z$ and $Y$ are independent standard Gaussians at $\theta=0$:
\begin{align*}
 \Exp|S_\alpha|<\infty
 &\Longleftrightarrow
 \Exp e^{\alpha Z^2}<\infty
 \Longleftrightarrow \alpha<\tfrac12\\
 \Exp[S_\alpha^2]<\infty
 &\Longleftrightarrow
 \Exp e^{2\alpha Z^2}<\infty
 \Longleftrightarrow \alpha<\tfrac14
\end{align*}
Hence:
\begin{enumerate}
\item for $0<\alpha<1/4$, the score is in $L^2$
\item for $1/4\leq\alpha<1/2$, the score is in $L^1$ but not $L^2$;
\item for $\alpha\geq1/2$, the score is not even integrable.
\end{enumerate}
\end{example}

\begin{proof}
The Gaussian identity
$\Exp e^{tZ^2}=(1-2t)^{-1/2}$ holds exactly for $t<1/2$.
The trace claims follow immediately.  Differentiating the output density for
$\alpha<1/2$ gives the output score in the following remark.
\end{proof}

We note that the example above uses only a finite program: what fails is not unbounded computation (as in recursion or a non-terminating loop) but rather an uncontrolled growth of tails in smooth primitives applied to an unbounded latent value. We also note that the above example shows that considering programs intensionally (i.e. before hiding) can limit expressivity.
\begin{remark}\label{remark:hiding-not-free}
In the example of~\Cref{example:moment-counterexample}, for $1/4\leq\alpha<1/2$, hiding $Z$ nevertheless gives an output derivative with score:
\[
S_\alpha(0, y) = y\Exp e^{\alpha Z^2}
\]
which is in $L^2(N(0,1))$.  Thus marginalisation can turn a non-2-integrable score into an integrable one. In this way, treating trace presentations as primary objects does limit the expressiveness of the theory to non-zero degree, compared to working with the realised measures directly.
\end{remark}

Thus, some restrictions on the primitives are definitely needed, as we have already foreshadowed in the definition of the language $\PPL$, in~\Cref{ssec:ppl-typing-rules}, by noting that the logistic map $\sigma(x)$ is polynomially bounded. This is exactly the condition we will put on elements of $\Prim(\PPL)$ and $\Prim(\PPL_\RR)$, in~\Cref{def:polynomially-bounded}.

\subsection{Jet semantics for $\PPL_\RR$}
We define the syntax category $\Syn_{\PPL_\RR}$ for $\PPL_\RR$ to be quotiented by the same laws as $\Syn_\PPL$.

\begin{definition}[Jet trace over $\CatDQBS$]
A dominated trace presentation over $\CatDQBS$ spaces $X, Y$ consists of $F = (\Omega, q, o)$, where $\Omega$ is a trace object, density $q:X \times \Omega \to [0, \infty)$ is a $\CatDQBS$ morphism such that $\sqrt{q(-,\omega)}$ is smooth, $\int q(x,\omega)\dd\omega = 1$, and $o: X\times \Omega \to Y$ is a $\CatDQBS$ morphism.
\end{definition}

A jet trace remains unchanged from~\Cref{def:jet-trace}, but now uses dominated trace presentations over $\CatDQBS$ spaces. Every jet trace over algebraic Euclidean spaces is also a jet trace over corresponding $\CatDQBS$ spaces. We can thus extend $\Trace$ semantics of $\PPL$ to $\Trace_\RR$ semantics for $\PPL_\RR$. Pure terms have two new constructors. Coercion is interpreted as identity on points:
\[
\sem{\coerce{\cdot}} = \semcoerce : \RRc \to \RRs \qquad \semcoerce(x) = x
\]

The non-negativity check is interpreted as:
\[
\sem{\isnneg{\cdot}}: \RRc \to \Bool \qquad \sem{\isnneg{x}} = \semisnneg{x}
\]

These are well-defined, as a direct consequence of~\Cref{cor:coerce-is-smooth} and~\Cref{prop:pure-functions=are-smooth}. The whole construction preserves the key property of~\Cref{cor:smooth-control-separation}, that if $\Gamma\vdash c:\Rc$ is a pure term, and two environments $\Gamma,\Gamma'$ agree on all control coordinates and all coproduct tags, then they give the same value to $c$, regardless of their smooth-real coordinates.

For a program $\Gamma \vdash \gauss{t}{1} \vkto \Rc$, we take the trace space to be $\Omega = \RRc$, with the density and output functions:
\[
\qquad q(\gamma,\omega)=\varphi(\omega-\sem{t}(\gamma)) \qquad o(\gamma,\omega)=\omega
\]
and score and velocity:
\[
 S(\gamma,\omega;\xi) =(\omega-m(\gamma))Dm_\gamma[\xi] 
 \qquad V=0
\]
where $m(\gamma)=\sem{t}(\gamma)$ and $\varphi$ is the standard normal density function.

\begin{proposition}\label{prop:pplr-congruence}
    Jet trace isomorphisms over $\CatDQBS$ still form an equivalence relation, a congruence for jet trace composition and tensor product.
\end{proposition}
\begin{proof}
    First, we prove that this is an equivalence relation. Reflexivity and symmetry are obvious; for transitivity, we no longer can find the unique smallest conull set, as we did in the proof of the corresponding~\Cref{prop:trace-isomorphism-congruence} for $\PPL$, but if there are two maps $h: D \to D'$ and $h': E \to E'$, then we can take $C = D'\cap E$ as the conull set in the middle. The rest of the proof remains the same.
\end{proof}
\begin{definition}\label{prop:jet-dqbs-well-defined}
    We define a category $\Jet\CatDQBS_\Trace$ to be generated by $\Trace_\RR(\Syn_{\PPL_\RR})$, under standard jet trace composition, where jet traces are quotiented by jet trace isomorphisms.
\end{definition}
\begin{proof}
    This category is well-defined. Associativity works due to quotients, as a consequence of~\Cref{prop:pplr-congruence}, and the composition preserves smoothness of the $\sqrt{q(-,\omega)}$ map, since the square root of composite density is given as the product $
    \sqrt{q_{G\circ F}(x,(\omega_F,\omega_G))} =\sqrt{q_F(x,\omega_F)} \cdot \sqrt{q_G(o(x,\omega_F),\omega_G)}$.
\end{proof}

\begin{theorem}
    The semantics map defined above: \[
    \Trace_\RR: \Syn_{\PPL_\RR} \to \Jet\CatDQBS_\Trace
    \]
    is a strong monoidal functor. Moreover, assuming that $\PPL$ primitives are all polynomially bounded, semantics for the finite fragment $\PPL$ factors through:
\[\begin{tikzcd}
  {\Syn_{\PPL_\RR}} && {\Jet\CatDQBS_\Trace} \\
  & {\Syn_{\PPL}}
  \arrow["{\Trace_\RR}", from=1-1, to=1-3]
  \arrow[hook, from=2-2, to=1-1]
  \arrow["\Trace"', from=2-2, to=1-3]
\end{tikzcd}\]
where we interpret $\Trace$ as putting $\CatDQBS$ structure on types.
\end{theorem}
\begin{proof}
    This follows directly in the same way as~\Cref{thm:ppl-semantics}, using~\Cref{prop:pplr-congruence} and~\Cref{prop:jet-dqbs-well-defined}.
\end{proof}

\subsection{Functor $\Affine$ and score-velocity metric $\FisherDirac$ in $\PPL_\RR$}

The construction of the degenerate-Riemannian metrics, Fisher metrics and expected tangent pullbacks was already done in $\Jet\CatFinTrace$, and carries through to $\Jet\CatDQBS_\Trace$ for $\PPL_\RR$. Thus, we still have the (unchanged) formula for $\Affine$:
\[
\Affine_F(h) = \Fisher{F} + F^\sharp h
\]
However, its domain changes, because we once again run into functional-analytic issues with integrability. On a (noncompact) output manifold, smoothness of $h$ does not imply that its expected pullback is finite. 

First, thanks to the restriction on the set of primitives $\Prim$, we have a key property of polynomially-bounded tails for maps in $\Jet\CatDQBS_\Trace$. Informally, each map in the category has all derivatives controlled polynomially (in the smooth coordinates). We make this precise below. Second, the polynomial-growth restrictions used for program primitives have to also be followed by similar restrictions on metrics, which we define below.

\begin{remark}
    The objects of $\Jet\CatDQBS_\Trace$ are generated from $\emptyset, \unitset, \RRc, \RRs$ by finite sums and products. For an object $X$, we can thus write it as a disjoint union over finitely many components $k$: \[
    X \simeq \bigsqcup_k X_k = \bigsqcup_k \RRc^{c_k}\times \RRs^{s_k}
    \]
    for some dimensions $c_k, s_k \in \NN$.
\end{remark}

\begin{definition}[Polynomially bounded function]\label{def:polynomially-bounded}
    We say that a smooth function $f: \RRs^{n_s} \times\RRc^{n_c} \to \RRs^m$ is polynomially bounded, if it is smooth in the smooth coordinates, jointly measurable, and for every multi-index $\alpha$ there exist constants $C > 0$ and $N \in \NN$, such that:
    \[
    \norm{D_s^\alpha f(s, c)} \leq C(1 + \norm{s}+\norm{c})^N
    \]
    where the differentiation $D_s^\alpha f$ is done only in smooth coordinates. The same definition applies to a domain being a finite disjoint union of such components, separately on each component.
\end{definition}

\begin{remark}[Permitted primitive classes]\label{remark:bounded-tails-primitives}
The functions $\sin$, $\cos$, $\chi(x) = x + \sqrt{1 + x^2}$, the logistic function $\sigma(x) = \frac{1}{1 + e^{-x}}$ on both $\Rs$ and $\Rc$ are polynomially bounded, and thus can be included in $\Prim_{\PPL_\RR}$. Functions $\log$ and $\exp$ are not polynomially bounded.
\end{remark}

We note that this also means that extending the language $\PPL_\RR$ to unnormalised measures would mean changing semantics of $\score{w}$ from multiplying the weight by $\exp(w)$ to some other, polynomially bounded, function.

\begin{remark}\label{remark:primitives-sufficient-necessary}
The restriction on primitives is sufficient, but not necessary, and forbidding the non-polynomially-bounded primitives comes at a cost. This is another price we pay for compositionality, on top of the already discussed~\Cref{remark:hiding-not-free}.
\end{remark}

\begin{definition}[Polynomially bounded metrics]
For a $\Jet\CatDQBSL$ object $Y$, we define the cone of polynomially bounded metrics on $Y \simeq \bigsqcup Y_k$, written $\PSDpoly(Y)$, as a set of metrics of the form $h = \bigsqcup_k h_k$, such that $h_i(s, c)$ is a positive-semidefinite metric on the component $Y_k$, smooth in the smooth coordinates $s$ and jointly measurable in both coordinates, and its matrix components $(H_i(s, c))_{ij}$ are polynomially bounded.
\end{definition}
The decomposition of $Y$ into $Y_i$ is not unique, but the above definition does not depend on which of them (up to reassociation and permutation) is chosen. To prove the main theorem of this section, we need a lemma about Gaussian averaging of polynomially-bounded functions.
\begin{lemma}\label{lemma:gaussian-average}
For an object $X$ in $\Jet\CatDQBS_\Trace$, for any polynomially bounded morphism $a: X \times \RRc \to \RRs$ and another polynomially bounded morphism $m: X\to\RRs$, their Gaussian averaging $(a\star m): X\to \RRs$ given as:
\[
    (a\star m)(s,c) = \int_{\RR} a(s,c,z)\varphi(z - m(s,c))\dd z
\]
is also polynomially bounded.
\end{lemma}
\begin{proof}
    We can write $a$ in components as $a_i(s,c,z): \RRs^{n_s}\times\RRc^{n_c}\times \RRc\to \RRs$. We can also write $m$ as $m_i(s, c): \RRs^{n_s}\times\RRc^{n_c} \to \RRs$. We now have to prove that:
    \[
    (a\star m)_i(s,c) = \int_{\RR} a(s,c,z)_i\varphi(z - m_i(s,c))\dd z
    \]
    is also polynomially bounded, that is, jointly measurable, smooth in the smooth coordinates $s$, and has all derivatives in $s$ bounded in the relevant sense under some constants $C, N$. Let us take some fixed component $i$ throughout, suppressed from notation from now on, and define $F(s,c,z) = a(s,c,z)\varphi(z - m(s,c))$.

    We now prove by induction on $\alpha$ (ordered lexicographically) the following fact: for any $\alpha$, we have:
    \[
    \partial^\alpha F = W_\alpha(s,c,z)\varphi(z - m(s,c))
    \]
    for some $W_\alpha$ which is polynomially bounded. The base case $\alpha =0$ is trivial, $W_0 = a$. Now, assume that the above holds for $\alpha$, and we want to show it for $\alpha + e_j$.  By the chain rule, we have:
    \begin{align*}
    \partial^{\alpha + e_j}F &= \partial_j\partial^\alpha F \\
    &= \partial_j (W_\alpha(s,c,z)\varphi(z - m(s,c))) \\
    &= (\partial_j W_\alpha(s,c,z))\varphi(z - m(s,c)) + W_\alpha (z - m(s,c))(\partial_j m(s,c))\varphi(z - m(s,c)) \\
    &= \bigl(\partial_j W_\alpha(s,c,z) + W_\alpha(s,c,z)(z - m(s,c))\partial_j m(s,c)\bigr)\varphi(z - m(s,c))
    \end{align*}
    where we used the fact that for the standard Gaussian density, we have:
    \[
    \varphi'(z) = -z\varphi(z)
    \]
    This gives $W_{\alpha + e_j} = \partial_j W_\alpha(s,c,z) + W_\alpha(s,c,z)(z - m(s,c))\partial_j m(s,c)$. Being polynomially bounded is closed under addition, multiplication, and taking derivatives in the smooth coordinates, and $m$ is also polynomially bounded by assumption, finishing the induction.
    
    Assuming that $\norm{W_\alpha(s,c,z)} \leq C(1 + \norm{s} + \norm{c} + \norm{z})^N$, we can bound the relevant integral by:
    \begin{align*}
    \norm{\partial^\alpha \int_\RR F(s,c,z)\dd z} &\leq \norm{\int_\RR W_\alpha(s,c,z)\varphi(z - m(s,c))\dd z} \\
    &\leq \int_\RR \norm{W_\alpha(s,c,z)\varphi(z - m(s,c))}\dd z\\
    &\leq \int_\RR \norm{C(1 + \norm{s} +\norm{c} + \norm{z})^N\varphi(z - m(s,c))} \dd z\\
    &= C\E_{Z\sim \mathcal{N}(0,1)}\left[(1+\norm{s}+\norm{c}+\norm{Z+m(s,c)})^N\right]\\
    &\leq C'(1+\norm{s}+\norm{c}+\norm{m(s,c)})^{N}
    \end{align*}
    where the last inequality is from standard Gaussian moments. This is polynomially bounded because $m$ is, and differentiation under the integral is justified by local domination.
\end{proof}

\begin{theorem}\label{thm:affine-dqbs}
    For any program-induced jet trace $F = (\Omega, q,o;S,V)$ in $\Jet\CatDQBS_\Trace$, its score is centered $\E_Q[S] = 0$. The conclusions of~\Cref{thm:properties-of-affine} remain true for the functor: \[
    \Affine : \Jet\CatDQBSL^\op \to \CatAffCone
    \]
    given by $\Affine(X) = \PSDpoly(X)$ on objects and $\Affine_F(h) = \Fisher{F} + F^\sharp h$ on morphisms.
\end{theorem}
\begin{proof}
    We have to show that for the program-generated jet traces $F = (\Omega,q,o;S,V): X\to Y$, for every polynomially bounded metric $h$ on $Y$, the formula:
    \[
    \Affine_F(h)_x(\xi, \eta) = \E_{\omega\sim Q_x}\Big[S_{x,\omega}[\xi]S_{x,\omega}[\eta] + h_{o(x,\omega)}(V_{x,\omega}[\xi], V_{x,\omega}[\eta])\Big]
    \]
    is finite, jointly measurable, smooth in (the relevant coordinates of) $x$, and bounded under the appropriate constants $C,N$. The proof is by structural induction over the program generating $F$, but first note that, because of~\Cref{cor:smooth-control-separation}, we can do it on a component-by-component basis: with the control coordinates and trace fixed, the output coproduct tag and control outputs are locally constant in the smooth coordinates.

    All primitives $\Prim_{\PPL_\RR}$ are assumed to be polynomially bounded in the sense of~\Cref{def:polynomially-bounded}. Together with the smooth-control separation from~\Cref{cor:smooth-control-separation}, this proves that $\Affine_{f}(h) = f^*h$  is also polynomially bounded for a deterministic jet trace $f$, proving the base case of $\return{t}$. Score is zero, so it is trivially centered.
    
    Conditionals $\ifte{b}{p_1}{p_2}$ collapse to either $p_1$ or $p_2$, to which we apply the inductive hypothesis. Smoothness is satisfied because the branch does not depend on the smooth coordinates. For the common bound, we take the supremum over bounds in $p_1$ and $p_2$.

    For composition $G \circ F$, where $F: X \to Y$ and $G: Y\to Z$ we can reuse the already-done calculation in the proof of~\Cref{thm:properties-of-affine}, showing $\Affine_{G\circ F}= \Affine_F \circ \Affine_G$. By inductive hypothesis, we know that for all $h\in\PSDpoly(Z)$, we have $\Affine_G(h)\in \PSDpoly(Y)$, and for all $h\in\PSDpoly(Y)$, $\Affine_F(h) \in \PSDpoly(X)$, and both vary smoothly. The calculation remains valid because everything remains finite and smooth by the induction hypothesis: the only thing to check in the previous argument is that we are still allowed to cancel the cross-terms in the score $\E_{(\omega_F, \omega_G)}[S_F S_G] = \E_{\omega_F}[S_F\E_{\omega_G}[S_G|\omega_F]] = \E_{\omega_F}[0] = 0$, from the inductive assumption.

    Then, we consider the case of a Gaussian random variable. Because the
    composition in our language is done by context-sharing, instead of $\Trace(\gauss{t}{1}): X\to\RRc$, it is easier to take the jet trace $F_m: X\to X\times \RRc$, which maps $F_m(x) = (x, z)$ for $z \sim \mathcal{N}(m(x), 1)$ for $m(x) = \sem{t}[x]$. Then, we can apply deterministic projection to $F_m$ to get back $\Trace(\gauss{t}{1})$.
    
    Let us pick some $h \in \PSDpoly(X\times \RRc)$, which is represented in the smooth coordinates by a symmetric positive-semidefinite $H(s,c,z)$, all of whose entries are polynomially bounded. For this jet trace, we have the score and velocity given by:
    \[
    S_\xi = (z - m)D(m)[\xi] \qquad V_\xi = (\xi, 0) \qquad h(V_\xi, V_\eta) = \xi^T H(s,c,z) \eta
    \]
    Therefore, we have the coordinates of the output matrix:
    \[
    \Affine_{F_m}(H)(\partial_i, \partial_j)(s,c) = \int_{\RR} ((z - m)^2(\partial_i m) (\partial_j m) + H_{ij}(s,c,z))\varphi(z - m) \dd z = (\partial_i m) (\partial_j m) + (H_{ij}\star m)(s,c)
    \]
    or, in other words:
    \[
    \Affine_{F_m}(H)= \dd m \otimes \dd m + (H_{ij}\star m)
    \]
    The first term is obviously smooth in smooth coordinates because $m$ is, and polynomially bounded by inductive assumption. For the second term, this follows from~\Cref{lemma:gaussian-average}. Both are positive-semidefinite, so indeed we have $\Affine_{F_m}(h)\in\PSDpoly(X)$. The centering of the score is automatic.
    
    Finally, for the Bernoulli case $\bern{t}$, the calculation is very similar to the Gaussian case above, for $m(x)=\sem{t}(x)$ and $p(x) = \sigma(m(x))$ giving:
    \[
    \Affine_{F_p}(H)(s,c) = p(1-p)\dd m\otimes \dd m + p H(s,c,1) + (1-p)H(s,c,0)
    \]
    which is polynomially bounded in the same way. Score is still centered, as in the finite language $\PPL$.

    All other properties of $\Affine$, such as it being affine, symmetric, invariant under trace isomorphisms and so on, follow by the same argument as in the finite case.
\end{proof}

Similarly to~\Cref{sec:algebraic}, we extend the score-velocity metric $G^{\kappa,h}$ and $\FisherDirac$ to $\PPL_\RR$.

\subsection{Moduli spaces and geometric semantics}

We now want to repeat the construction of $\ETrace{Y}$ from~\Cref{sec:geometric}, for the much bigger - indeed, infinite-dimensional - spaces of $\PPL_\RR$. To form this space, we have to move from the small target to the full $\CatDQBS$.

\subsubsection{Exponentials $Y^\Omega$}
As a consequence of~\Cref{thm:dqbs-complete-cocomplete-ccc}, $\CatDQBS$ has exponentials, so we can in particular form the space of output maps $Y^\Omega$, with the functional quasi-Borel and functional diffeological structures. We will need a few properties of it, which we state below.

\begin{proposition}[Exponentials for $\Omega$ in $\CatDQBS$]
For a given algebraic Euclidean space $Y$ and a trace space $\Omega$, the only requirement for a map to be in $\CatDQBS(\Omega, Y)$ is to be a quasi-Borel morphism. Thus, for $Y$ being an algebraic Euclidean space, $\CatDQBS(\Omega,Y)$ consists exactly of measurable maps $\Omega \to Y$. For a finite or countable $\Omega$, these are all such functions.
\end{proposition}

A map $q:\RR\to Y^{\Omega}$ is a quasi-Borel random element exactly when its uncurrying:
\[
  \widetilde q:\RR\times\Omega\longrightarrow Y
  \qquad
  \widetilde q(r,\omega)=q(r)(\omega)
\]
is a quasi-Borel morphism.  A map $o:U\to Y^{\Omega}$ from a Euclidean domain is a plot exactly when:
\[
  \widetilde o:U\times\Omega\longrightarrow Y
  \qquad
  \widetilde o(u,\omega)=o(u)(\omega)
\]
is a morphism in $\CatDQBS$, that is, it is jointly measurable in both components and smooth in the first component. This generalises our previous treatment of trace presentations in~\Cref{sec:ppl-jets}, where we also required the output map to be smooth in the first component and measurable in both components.

\begin{proposition}[Canonical velocity of a mapping-space tangent]
\label{prop:canonical-velocity}
Let $\Omega$ be a trace space, and let $Y$ be an algebraic-Euclidean space. For every $o:\Omega\to Y$, evaluation induces a linear isomorphism:
\[
\nu_o:T_o\bigl(Y^\Omega\bigr)
  \simeq
  \left\{z : \Omega \to TY \mid z(\omega) \in T_{o(\omega)}Y\right\}
\]
where the right hand side has subset diffeology and subspace quasi-Borel structure in the function space $TY^\Omega$, where $TY$ is the sum of tangent bundles over all components of $Y$.
\end{proposition}

For finite and countable trace spaces, the measurable structure on $\Omega$ is discrete, and thus we have a simple isomorphism:
\[
\nu_o:T_o\bigl(Y^\Omega\bigr)
  \simeq
  \prod_{\omega\in\Omega}T_{o(\omega)}Y
\]

\subsubsection{2-forms on $\CatDQBS$ and $\ETrace{Y}$}
Ultimately, our goal is to redefine the score-transport 2-form $G$ on the moduli space. In diffeological spaces, we have the notion of a plotwise form.

\begin{definition}[Plotwise positive form]
A plotwise positive-semidefinite $2$-form on a diffeological space $X$ assigns to every plot $p:U\to X$ a symmetric positive-semidefinite tensor $g_p$ on $U$, subject to:
\[
    g_{p\circ f}=f^*g_p
\]
for every smooth $f:V\to U$. We let $\PSDplot(X)$ denote the set of all such plotwise forms on $X$.
\end{definition}

However, because we moved to $\CatDQBS$, we still have the problem of non-integrability. The same formula might give an integrable result in one coordinate system, but non-integrable in another. Therefore, we have to restrict the class of plots. Fortunately, diffeological spaces allow us to do that easily.

\begin{definition}[Dominated probability laws]
For a given trace space $\Omega$, we define the set of probability distributions on $\Omega$ as:
\[
  \Dens_\lambda(\Omega) = 
  \left\{q \in L^1(\Omega, \lambda):
      q \geq 0\ \lambda\text{-a.s.}\ \text{ and }
      \int_\Omega q(\omega)\lambda(\dd \omega) = 1
  \right\}
  \qquad
  \Denso_\lambda(\Omega) = 
  \left\{q \in \Dens_\lambda(\Omega) : 
      q > 0\ \lambda\text{-a.s.}
  \right\}
\]
It is an object of $\CatDQBS$ with the subset diffeology and subspace quasi-Borel structure inherited from $L^1(\Omega,\lambda)$.
\end{definition}
We note that the above definition strictly generalises~\Cref{def:simplex}.

\begin{definition}[$\ETrace{Y}$ and $\ETraceo{Y}$]
We define the classifying spaces:
\[
\ETrace{Y} = \Dens(\Omega) \times Y^\Omega \qquad \ETraceo{Y} = \Denso(\Omega) \times Y^\Omega
\]
equipped with the product quasi-Borel structure. Its diffeology $\mathcal D$ consists of $\CatDQBS$ maps $(q, o): U \to \ETrace{Y}$ such that for every $u \in U$, \(\xi\in T_uU\), the map $u \mapsto \sqrt{q(u)}$ valued in $L^2(\lambda)$ is smooth, we have a Radon-Nikodym derivative $Dq_u[\xi]=S_{u,\xi}q_u$ with $S_{u,\xi}\in L^2(Q_u)$, and for every $h \in \PSDpoly(Y)$, the tensor: \[
  G^{\kappa, h}_{(q,o)}
  \bigl((S,V),(T,W)\bigr)
  =
  \int_\Omega S(\omega) T(\omega)\,Q_q(\dd\omega)
  +\kappa\int_\Omega h_{o(\omega)}\bigl(V(\omega),W(\omega)\bigr)\ Q_q(\dd\omega)
\]
is finite and smooth on $U$.
\end{definition}

\begin{proposition}
    The diffeologies on $\ETrace{Y}$ and $\ETraceo{Y}$ are well-defined.
\end{proposition}
\begin{proof}
    Constant maps have zero score and velocity, thus satisfy the assumptions. For reparametrisation $f: W\to U$, tensor $G^{\kappa, h}$ is transformed by a pullback $G^{\kappa, h}_{q(f(u)), o(f(u))} = f^*G^{\kappa, h}$, so it remains finite and smooth, and 2-integrability and smoothness of $\sqrt{q}$ are also preserved under reparametrisation.
\end{proof}
\begin{proposition}
    For every $h \in \PSDpoly(Y)$, $G^{\kappa, h}$ defines a plotwise 2-form on $\ETrace{Y}$.
\end{proposition}
\begin{proof}
A bilinear expression defined from the pointwise $\dot q_{u,\xi}, V_{u,\xi}$ is consistent under reparametrisation. Indeed, for a plot $p: U\to \ETrace{Y}$ and smooth $f:W\to U$:
\[
  D(p\circ f)_w[\xi]=Dp_{f(w)}[Df_w[\xi]]
\]
and the same identity holds pointwise after applying every evaluation map. Therefore any integral formula built bilinearly from these derivatives defines a plotwise tensor on $\ETrace{Y}$.
\end{proof}

We note there is an important distinction between a plotwise form and a global form on the full internal tangent space. In a diffeological space $X$, the plotwise form is only required to be defined for tangents coming from each individual plot separately $a: U \to X$ (and on top of that, being invariant under reparametrisations). On the other hand, the internal tangent space contains symbols $a_* \xi, a'_*\eta$ for all plots $a: U \to X$, $a': U' \to X$ and $\xi \in T_uU, \eta\in T_{u'}U'$. In standard manifolds, these two coincide - in general, not necessarily. However, in our setup, we indeed have the stronger property as well.

\begin{theorem}\label{thm:internal-tangent-space-metric}
    For every $h \in \PSDpoly(Y), \kappa \geq 0$, $G^{\kappa, h}$ defines a positive-semidefinite form on the internal tangent space $T_{(q,o)}\ETrace{Y}$.
\end{theorem}
\begin{proof}
    For every fixed plot $p: U \to \ETrace{Y}$ at the point $u$ such that $p(u) = (q,o)$, we have a linear map:
    \[
    J_p: T_uU \to L_0^2(Q)\oplus T_oY^\Omega \qquad J_p(\xi) = (S_\xi, V_\xi)
    \]
    These maps respect reparametrisation relations in the internal tangent space, and therefore descend to a linear map on the internal tangent space:
    \[
    J: T_{(q,o)}\ETrace{Y} \to L_0^2(Q)\oplus T_oY^\Omega
    \]
    Thus $G^{\kappa, h}$ on $T_{(q,o)}\ETrace{Y}$ can be defined by:
    \[
    G^{\kappa, h}(\xi,\eta) = G^{\kappa, h}(J(\xi),J(\eta))
    \]
\end{proof}

\begin{theorem}[Programs are geometric plots]
Every program $X \vdash p \vkto Y$ in $\PPL_\RR$ with semantics $\Trace(p) = F$ defines a geometric plot $\widehat{F}: X \to \ETrace{Y}$, where $\widehat{F}$ denotes the classifying map $\widehat F = \langle q, \curry(o)\rangle$ as in~\Cref{theorem:metric-pulls-back}.
\end{theorem}
\begin{proof}
The proof is directly from the well-definedness of the score-velocity metric~\Cref{thm:affine-dqbs}.
\end{proof}

\subsubsection{Relation to diffeological information geometry}

The projection $\Law_Y:\ETrace{Y}\to\Dens_\lambda(\Omega)$ forgets the output map.  Every plot $P:U\to\ETrace{Y}$ therefore determines a dominated diffeological statistical model:
\[
  U\xrightarrow{\Law_Y\circ P}\Dens(\Omega)
  \xrightarrow{\iota_\lambda}\Prob(\Omega)
\]
where $\iota_\lambda:\Dens_\lambda(\Omega)\to \Prob(\Omega)$ turns densities into probability measures. On directions for which the logarithmic tangent lies in $L^2(Q)$, the diffeological Fisher tensor of~\cite{ayInformationGeometry2017b} is:
\[
  \FisherMetric_{q,u}(\xi,\eta)
  =\int_\Omega S_{u,\xi}(\omega)S_{u,\eta}(\omega)\,Q_u(\dd\omega)
\]
Thus the density component of the construction is exactly the canonical diffeological Fisher geometry. The decorated space $\ETrace{Y}$ has, in addition, the parameter-dependent output map, thus requires a choice of output metric.

\subsubsection{Pullback factorisation and square-root coordinates for $\PPL_\RR$}
We now want to generalise~\Cref{theorem:metric-pulls-back} to the real-valued setting. One issue is that there is no size structure on the (large) space $\ETraceo{Y}$ - a naive notion of size would not be compatible with different charts (since size is not preserved under diffeomorphism, which is the only compatibility condition in $\CatDQBS$). One way out would be to introduce a more principled way to handle polynomially bounded maps between sized spaces. We describe this in~\Cref{sec:future-work}
and leave it for future work, and for simplicity, here we opt to just restrict the space of metrics to exactly those that pull back to admissible ones on the codomain.
    
\begin{proposition}\label{prop:factorisation-pplr}
We have an affine map:
\[
  \mathbb G_Y: \PSDpoly(Y) \longrightarrow \PSDpoly\bigl(\ETrace{Y}\bigr)
  \qquad
  \mathbb G_Y(h)=\Law_Y^*\FisherMetric+ H^h = G^{1,h}
\]
For a morphism $F:X\to Y$ in $\Jet\CatDQBS_\Trace$ with classifying map $\ClassMap{F}:X\longrightarrow\ETrace{Y}$ we have the factorisation through metrics on $(\ETrace{Y})$:
\[\begin{tikzcd}
  {\PSDpoly(Y)} && {\PSDpoly(X)} \\
  & {\PSDpoly(\ETrace{Y})}
  \arrow["{\Affine_F}", from=1-1, to=1-3]
  \arrow["{\mathbb G_Y}"', from=1-1, to=2-2]
  \arrow["{\ClassMap{F}^*}"', from=2-2, to=1-3]
\end{tikzcd}\]
where:
\[
    \PSDpoly(\ETrace{Y}) = \{h \in \PSDplot(\ETrace{Y}) :\forall X \forall F \in\Jet\CatDQBS_\Trace(X, Y), \ClassMap{F}^*h \in \PSDpoly(X)\}
\]
\end{proposition}

Like in the finite case, we can put the square-root coordinates $r(\omega):=2\sqrt{q(\omega)}$ on the density part, and get the product-of-cones formulation, subject to an additional condition of $\lambda$-almost-sure equality of $q$'s. However, formulating exactly the analogous extensions of Riemannian geometry, such as the Levi-Civita connection or curvature from~\Cref{prop:levi-civita} and~\Cref{prop:curvature} is non-trivial and we postpone it to future work. There is also no canonical infinite-dimensional Riemannian volume form, so analogously building Riemannian volume seems difficult as well.

\subsection{Collapsed moduli space $\ETraceColl{Y}$}
We have already pointed out in~\Cref{ssec:universal-metric} that the definition of the classifying space $\ETrace{Y}$ distinguishes points whose output maps only differ on zero-probability outcomes. For example, for:
\[
\Omega = \{1,2,3\} \qquad Y = \Bool
\]
these two points:
\[
\begin{aligned}
q_1 &= \left\{\frac{1}{2}, \frac{1}{2}, 0\right\} \\
o_1 &= \{1 \mapsto \true, 2\mapsto \false, 3\mapsto \false\}
\end{aligned}
\qquad
\begin{aligned}
q_2 &= \left\{\frac{1}{2}, \frac{1}{2}, 0\right\} \\
o_2 &= \{1 \mapsto \true, 2\mapsto \false, 3\mapsto \true\}
\end{aligned}
\]
are not identified. This is contrary to the quotient assumed in the jet trace definition (in both the finite and infinite cases), where we consider output maps only up to $Q_x$-almost-sure equality. We fix it by applying a similar quotient to $\ETrace{Y}$, as we do below, at a price of ending up with more complex structure.

\begin{definition}\label{def:collapsed-space}
We define the collapsed moduli space as:
\[
\ETraceColl{Y} = \ETrace{Y}/{\sim}
\]
where the equivalence relation is $(q, o) \sim (q', o')$ if  $q = q' \ \lambda$-a.s., and $o = o' \ q\lambda$-a.s. We consider the moduli space as an object in $\CatDQBS$ with the quotient quasi-Borel and quotient diffeological structures.
\end{definition}

\begin{proposition}[Stratification of $\ETraceColl{Y}$]
    For a finite or countable $\Omega$, the moduli space $\ETraceColl{Y}$ can be written, as a set, as a disjoint union of the form:
    \[
    \ETraceColl{Y} = \bigsqcup_{J \subseteq \Omega} \ETraceoIn{J}{Y} =  \bigsqcup_{J \subseteq \Omega} \Simplexo{J} \times Y^J
    \]
    Here $\Simplexo{J}=\{(q_j)_{j\in J}:q_j>0,\ \sum_{j\in J}q_j=1\}$, also for countably infinite $J$.
\end{proposition}

\begin{proposition}[Plotwise metric on quotient diffeology]
    Suppose that a space $X$ has the quotient (final) diffeology generated by $e_i:X_i\to X$ and that each $X_i$ carries a positive form $g_i$. These forms descend to a plotwise form on $X$ exactly when:
    \[
        e_i\circ r=e_j\circ s
        \quad\Longrightarrow\quad
        r^*g_i=s^*g_j
    \]
    for all smooth maps $r:U\to X_i$ and $s:U\to X_j$.
\end{proposition}

Our previous~\Cref{thm:finite-sqrt-coordinates-cone} for finite or infinite $\Omega$ can be understood stratum-wise. Note that a program $X \vdash p \vkto Y$ defines a plot into a single stratum of $\ETraceColl{Y}$ on each connected component of $X$.

\begin{proposition}
    For a program $X \vdash p \vkto Y$, let its jet trace be $F = (\Omega, q, o; S, V): X \to Y$, and $x \in X$. Then, for any connected neighbourhood (in the $D$-topology) $U$ of $x$, all trace laws $Q_{x'}$ for $x' \in U$ are mutually absolutely continuous. This means the classifying map factors through some measurable stratum $J_U \subseteq \Omega$:
    \[
    \ClassMap{F}|_U: U \to \ETraceoIn{J_U}{Y} \to \ETraceColl{Y}
    \]
\end{proposition}

Under this identification $G^{\kappa,h}$ stratifies: for a fixed support \(J\subseteq\Omega\), the corresponding stratum is $\Simplexo{J}\times Y^J$ with the Riemannian metric $G_J^{\kappa,h}$. In particular, an output value attached to a trace of mass zero carries no geometric information.

\begin{corollary}\label{prop:plotwise-metric}
    For $h\in \PSDpoly(Y)$ and any trace object $\Omega$, $G^{\kappa, h}$ defines a plotwise metric on $\ETraceColl{Y}$.
\end{corollary}

\subsection{Universal moduli spaces}
Reducing the structure of the collapsed space $\overline{\ETrace{Y}}$ even further, we now turn our attention to $\Omega$. The trace space is, in a way, unimportant for our considerations: all results hold under arbitrary permutations of $\Omega$, since a particular seed $\omega$ does not (and should not) carry any semantic information. Permutations of $\Omega$ are thus additional data that we can quotient out, obtaining the following definition of the universal moduli space, consisting of only the ``legitimately distinct'' measures.

\subsubsection{Universal trace spaces for $\PPL$}

\begin{definition}[Universal trace space for finite random variables]
    We define the universal trace space to be a pair $(\OmegaUL, \Lambda)$ where:
    \[
    \OmegaUL = \bigsqcup_\Omega |\Omega| \qquad \Lambda = \sum_{\Omega} \lambda_\Omega
    \]
    where $\Omega$ varies across all possible algebraic types generated by $\unittype, \emptytype$, and $\Lambda$ is the counting measure on $\OmegaUL$.
\end{definition}

We note that $\OmegaUL$ as a set is countable, and the measure $\Lambda$ is $\sigma$-finite. This definition can be extended, preserving $\sigma$-finiteness, to accommodate real random variables, by generating $\OmegaUL$ on all algebraic Euclidean spaces, as done below.

\begin{definition}[$\ETraceUniv{Y}$]\label{def:universal}
    We define the universal moduli space on the output type $Y$ as:
    \[
\ETraceUniv{Y} := 
\left\{
(q,o)\ :
q\in\ell^1_+(\mathbb N),
\sum_iq_i=1,
\#\operatorname{supp}(q)<\infty,
o:\mathbb N\to Y
\right\}
/ \sim
\]
where $(q, o)\sim (q', o')$ for finite permutations and $q$-a.e. equality of $o$, equipped with the final $\CatDQBS$ structure.
\end{definition}

\begin{proposition}[Colimit characterisation of $\ETraceUniv{Y}$]
Let $J = \mathbf{FinInj}$ be the diagram category consisting of finite sets and injections. We then have:
\[
\ETraceUniv{Y} = \operatorname{colim}_{J} \overline{\ETrace{Y}}
\]
\end{proposition}
\begin{proof}
An injection $i:\Omega\hookrightarrow\Omega'$ induces:
\[
i_!:\overline{\ETraceAt{\Omega}{Y}}\longrightarrow \overline{\ETraceAt{\Omega'}{Y}}
\]
by transporting $q$ and $o$ along $i$ and assigning zero mass outside its image (the output map outside the support no longer needs an arbitrary extension because of the quotient). The category $\CatDQBS$ is cocomplete, so the colimit exists.
\end{proof}
 
For each individual $\ETrace{Y}$, we have the (generally not injective) cocone map into universal space:
\[
i_\Omega:\ETrace{Y}\longrightarrow\ETraceUniv{Y}
\]
For a given dominated trace $F: X \to Y$ with the classifying map $\ClassMap{F}_\Omega:X \to \ETrace{Y}$, we obtain the global map:
\[
\widehat F =  i_\Omega\circ\widehat F_\Omega: X\longrightarrow \ETraceUniv{Y}
\]
\begin{proposition}
For each $h\in \PSD(Y)$, we have a global plotwise form $\mathbb G_Y^{\kappa,h}$ on  $\ETraceUniv{Y}$, given by:
\[
i_\Omega^*\mathbb G_Y^{\kappa,h} = G_{\Omega,Y}^{\kappa,h}
\]
and therefore:
\[
\Affine_F(h) = \widehat F^{*}\mathbb G_Y^{1,h}
\]
\end{proposition}
\begin{proof}
    As we noted in~\Cref{prop:permutations-omega-isometries}, permutations of $\Omega$ act by isometries. Zero-padding also preserves the form and the intrinsic metric of the support stratum: the new zero-mass coordinates contribute neither Fisher nor transport energy. Consequently the fixed-trace forms are compatible with the colimit maps.
\end{proof}
\begin{figure}
    \centering
\[\begin{tikzcd}
    {\widehat F_\Omega^*G^{\kappa,h}_\Omega=(\iota_\Omega\circ\widehat F_\Omega)^*G^{\kappa,h}_\Sigma=\widehat F^*\mathbb G^{\kappa,h}} & X &&&&& \\
    {G^{\kappa,h}_\Omega=\iota_\Omega^*G^{\kappa,h}_\Sigma=(i_\Sigma\circ\iota_\Omega)^*\mathbb G^{\kappa,h}} & {\ETraceAt{\Omega}{Y}} \\
    {G^{\kappa,h}_\Sigma=i_\Sigma^*\mathbb G^{\kappa,h}} & {\ETraceAt{\Sigma}{Y}} &&& {\ETraceUniv{Y}} && {\mathbb G^{\kappa,h}} \\
    {G^{\kappa,h}_{\Omega'}=\iota_{\Omega'}^*G^{\kappa,h}_\Sigma=(i_\Sigma\circ\iota_{\Omega'})^*\mathbb G^{\kappa,h}} & {\ETraceAt{\Omega'}{Y}} \\
    {(\widehat F'_{\Omega'})^*G^{\kappa,h}_{\Omega'}=(\iota_{\Omega'}\circ\widehat F'_{\Omega'})^*G^{\kappa,h}_\Sigma=(\widehat F')^*\mathbb G^{\kappa,h}} & {X'}
    \arrow["on"{description}, draw=none, from=1-1, to=1-2]
    \arrow["{\widehat F_\Omega}"', from=1-2, to=2-2]
    \arrow["{\widehat F}", from=1-2, to=3-5]
    \arrow["on"{description}, draw=none, from=2-1, to=2-2]
    \arrow["{\iota_\Omega}"', hook', from=2-2, to=3-2]
    \arrow["{i_\Omega}", from=2-2, to=3-5]
    \arrow["on"{description}, draw=none, from=3-1, to=3-2]
    \arrow["{i_\Sigma}", from=3-2, to=3-5]
    \arrow["with"{description}, draw=none, from=3-5, to=3-7]
    \arrow["on"{description}, draw=none, from=4-1, to=4-2]
    \arrow["{\iota_{\Omega'}}", hook, from=4-2, to=3-2]
    \arrow["{i_{\Omega'}}", from=4-2, to=3-5]
    \arrow["on"{description}, draw=none, from=5-1, to=5-2]
    \arrow["{\widehat F'}"', from=5-2, to=3-5]
    \arrow["{\widehat F'_{\Omega'}}", from=5-2, to=4-2]
\end{tikzcd}\]
    \caption{Universal moduli space $\ETraceUniv{Y}$ with the universal metric $\mathbb{G}^{\kappa,h}$ and associated pullback metrics. Here $\Sigma=\Omega\sqcup\Omega'$, and $\iota_\Omega,\iota_{\Omega'}$ are induced by the two trace-space inclusions.}
    \label{fig:universal-moduli-space}
\end{figure}
\begin{remark}
Note that for the universal space, we no longer get a metric on the internal tangent space: ~\Cref{thm:internal-tangent-space-metric} does not extend, even in the finite case. For example, let us take the following program:
\[
    (t: \Rs) \vdash \letbind{b}{\bern{t}}{\return{\unitsetelem}}\vkto \unittype
\]
with the corresponding jet trace of:
\[
\Omega = \Bool \qquad q(t) = (\sigma(t), 1 - \sigma(t)) \qquad o \equiv \unitelem \qquad S[b,t;\partial_t] = b - \sigma(t)
\]
We have a diffeomorphism of the domain $\gamma : \RR \to \RR$ given by $\gamma(t) = -t$, which induces a permutation of trace space, by swapping the two elements of $\Omega$. By~\Cref{def:universal}, $\ClassMap{F}\circ \gamma$ and $\ClassMap{F}$ are identified, as maps into $\ETraceUniv{\unittype}$.
Now, using~\Cref{def:final-diffeology} and~\Cref{def:internal-tangent-space}, we must have:
\[
\begin{aligned}
D(\ClassMap{F})_0[\partial_t]
&= D(\ClassMap{F}\circ \gamma)_0[\partial_t] \\
&= D(\ClassMap{F})_0[D\gamma_0[\partial_t]]
= -D(\ClassMap{F})_0[\partial_t]
\end{aligned}
\]
as elements in a vector space, so necessarily this is a null vector $D\ClassMap{F}[\partial_t] = 0$. Thus, any 2-form $H$ on an internal tangent space would have to assign $H(D\ClassMap{F}[\partial_t], D\ClassMap{F}[\partial_t]) = 0$. But on the contrary, we have:
\[
(\ClassMap{F}^*\mathbb G^{\kappa, h})_0(\partial_t, \partial_t) = \sigma(0)(1 - \sigma(0)) = \frac{1}{4} \neq 0
\]
thus, $\mathbb G^{\kappa, h}$ could not arise from any norm on the internal tangent space.
\end{remark}

Nevertheless, we still have a valid plotwise metric, and we can now give a proper meaning to~\Cref{fig:one}. For programs $p_1, p_2, p_3$, if:
\[
\widehat p_3:\Theta_3\longrightarrow\ETraceUniv{Y}
\]
is the two-parameter family, and $i_1,i_2$ are the two coordinate inclusions, then:
\[
\widehat p_1=\widehat p_3\circ i_1,
\qquad
\widehat p_2=\widehat p_3\circ i_2.
\]

The intersection of these families is simply equality of points in the global moduli object $\ETraceUniv{\RR}$. This also gives a precise meaning to other statements, such as that one program family is a submodel of another, or two program families intersect transversally, which we noted in~\Cref{fig:one}.

\subsubsection{Universal trace spaces for $\PPL_\RR$}
Building a universal trace space for $\PPL_\RR$ is more complex. The first part of the construction, i.e., the definition of $\Omega_\PPL$ and $\Lambda$, survives, as we can give:
\[
    \OmegaULR = \bigsqcup_\Omega |\Omega| \qquad \Lambda = \sum_\Omega \lambda_\Omega
\]
where $\Omega$ varies across all possible algebraic types generated by $\unittype, \emptytype, \RRc$, and $\Lambda$ is the sum of their respective associated ($\sigma$-finite) measures, so  $\Lambda$ remains $\sigma$-finite. However, quotienting $(q, o)$ by finite permutations cannot simply be replaced by Borel isomorphisms. If we did that, then the space of mean-parameterised Gaussians $\mathcal{N}(\theta, 1)$ arising from a program:
\[
\theta : \Rs \vdash \letbind{\_}{\gauss{\theta}{1}}{\return{\star}} \vkto \unittype
\]
would be identified with the single-point $\mathcal{N}(0, 1)$. We leave fully working out the appropriate universal moduli space in that case for future work.


\section{Many geometries of a program}\label{sec:many-geometries}

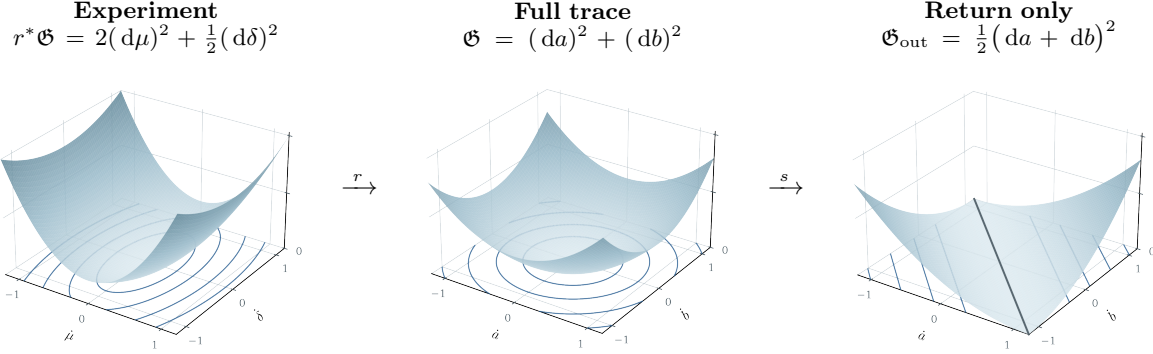
\begin{figure}
\centering
\input{figures/contraction-composition-pullback}
\caption{%
Three geometries of a real-random variable program $p$, visualized by a surface with
Hessian $G$.  The experiment map $r(\mu,\delta)=\bigl(\mu+\tfrac12\delta,\,\mu-\tfrac12\delta\bigr)$ pulls the full-trace metric back to the
elliptical bowl on the left.  Hiding the trace via $s=!_{\Omega}$ leaves the return law $Y\sim\mathcal N(a+b,2)$ and contracts $\FisherDirac$ to $\Gout$, where the direction $a-b$ becomes null.
}
\label{fig:many-geometries-program}
\end{figure}

So far, we have worked with\ intensional geometry: two different programs $X \vdash p_1 \vkto Y$ and $X \vdash p_2 \vkto Y$ denoting the same probability kernel $\sem{p_1} = \sem{p_2} : X \kto Y$ will potentially have different interpretations in the dominated trace or geometric sense. The unused randomness, which we noted already as a problem blocking Markov structure on dominated traces, is one of the culprits.
In this section, we examine what happens when we relax the assumption about observing the whole trace, that is, what is the extensional geometry of programs. Making now more precise the claim from the introduction, we introduce a statistic:
\[
\pi :\Omega \to \Omega'
\]
which potentially forgets some of the trace information $\Omega$. This disambiguates the phrase ``the geometry of the program'': the map $\pi$ controls whether we observe the full trace, no trace at all (i.e. only the return value), or something in between. However, only the full-trace geometry is compositional.

As we noted, the other issue convenient to investigate in the same setup is supplying common context, i.e., a map $r: M \to X$ which denotes one particular statistical experiment. We will do so in~\Cref{ssec:natgrad}. The full setup formally becomes:
\[
M
\xrightarrow{r}
X
\overset{p}{\rightsquigarrow}
(\Omega, Y)
\xrightarrow{\pi \times \id}
(\Omega', Y)
\]

\subsection{Forgetting the trace}
Let us fix some jet trace presentation $F=(\Omega,q,o;S,V):X\longrightarrow Y$, $h \in \PSD(Y)$, and \(\kappa>0\). For a given tangent vector \(\xi\in T_xX\), we can write the score and the velocity as random variables:
\[
  S_\xi(\omega):=S(x,\omega;\xi)
  \qquad
  V_\xi(\omega):=V(x,\omega;\xi)
\]
Let us now assume that we have a statistic on $(\Omega, Y)$, a measurable map $\pi:\Omega\longrightarrow\Omega'$. We say that the observed execution variable is then:
\[
Z_x^\pi(\omega):=\bigl(\pi(\omega),Y_x(\omega)\bigr)\in\Omega'\times Y
\]
Conditional expectations with respect to $Z_x^\pi$ define projected jets:
\[
S_\xi^\pi
=
\mathbb E_{Q_x}
\left[
S_\xi\mid Z_x^\pi
\right]
\qquad
V_\xi^\pi
=
\mathbb E_{Q_x}
\left[
V_\xi\mid Z_x^\pi
\right]
\]

\begin{definition}[Projected score-velocity]\label{def:projected-score-velocity}
    For $F$ and $\pi$ as above, we define the projected score-velocity as:
    \[
    \bigl(G^{\kappa,h}_{F\mid \pi}\bigr)_x(\xi,\eta) = \E_{Q_x}\Big[ S_\xi^\pi S_\eta^\pi+ \kappa h_{Y_x} \bigl(V_\xi^\pi,V_\eta^\pi\bigr) \Big]
    \]
    \end{definition}

Note that the projected score-velocity is possibly no longer smooth, as the following example illustrates.
\begin{example}\label{example:moving-mass-simult}
Consider a program:
\begin{align*}
t: \R \vdash & \letbind{b}{\bern{0}}{\\
&\ifte{b}{\return{t}}{\return{-t}} \vkto \R
}
\end{align*}
and a map $\pi: \Omega \to \unitset$ which forgets the trace. Take $\kappa=1$ and $h=\dd y^2$. The randomness is not parametrised, so $S = 0$, and: \[
V_{\true} = \id  \qquad V_{\false} = -\id
\]
so for a tangent $\xi\partial_t\in T_t\RR$ we have:
\[
(G^{1,h}_F)_t(\xi\partial_t,\xi\partial_t)
=\frac{1}{2}\xi^2+\frac{1}{2}(-\xi)^2=\xi^2
\]
or in other words, $G^{1,h}_F=\dd t^2$. For every $t\neq0$, the output identifies the branch, so the velocity is measurable wrt the output, and $(G^{1,h}_{F\mid\pi})_t=(\dd t^2)_t$. But for $t=0$, the conditional expectation collapses both branches, setting $\E[V_\xi\mid Y_0]=0$, and thus $(G^{1,h}_{F\mid\pi})_0=0$. This is a discontinuous metric in $t$.
\end{example}

Nevertheless, we can still write a pointwise bound of (not necessarily smooth) 2-forms.

\begin{theorem}[Trace data-processing inequality]\label{thm:projected-data-processing}
Projected score-velocity bounds score-velocity from below:
\[
G^{\kappa,h}_{F\mid \pi} \preceq G^{\kappa,h}_{F}
\]
\end{theorem}
\begin{proof}
We have $\E[S_\xi - S^\pi_\xi | Z^\pi] = 0$, and therefore:
\[
\E[S_\xi^2 - (S^\pi_\xi)^2 | Z^\pi]  = \E[(S_\xi - S^\pi_\xi)^2 | Z^\pi] \geq 0
\]
Similarly, $\E[h(V_\xi - V^\pi_\xi, V^\pi_\xi) | Z^\pi] = 0$. We can thus write:
\begin{align*}
\E[h(V_\xi, V_\xi)| Z^\pi] - \E[h(V^\pi_\xi, V^\pi_\xi)| Z^\pi]
&= \E[h((V_\xi - V^\pi_\xi) + V^\pi_\xi, (V_\xi - V^\pi_\xi) + V^\pi_\xi)| Z^\pi] - \E[h(V^\pi_\xi, V^\pi_\xi)| Z^\pi] \\
&= \E[h(V_\xi - V^\pi_\xi, V_\xi - V^\pi_\xi)| Z^\pi] 
+ 2\E[h(V_\xi - V^\pi_\xi, V^\pi_\xi)| Z^\pi]\\
&=\E[h(V_\xi - V^\pi_\xi, V_\xi - V^\pi_\xi)| Z^\pi] \geq 0
\end{align*}
Plugging these both into the calculation:
\begin{align*}
    (G^{\kappa,h}_{F} - G^{\kappa,h}_{F\mid \pi})_x(\xi, \xi) &=\E\left[S_\xi^2+\kappa h(V_\xi,V_\xi)\right]
    -\E\Big[(S_\xi^\pi)^2+ \kappa h(V_\xi^\pi, V_\xi^\pi)\Big]\\
    &=\E[(S_\xi - S^\pi_\xi)^2]
    +\kappa\E\left[h(V_\xi-V^\pi_\xi, V_\xi-V^\pi_\xi)\right]
    \geq0
\end{align*}
we get the thesis.

\end{proof}

\begin{definition}[$\Gout$]\label{def:g-out}
    For a jet trace $F: X\to Y$, a $\kappa > 0$ and $h \in \PSD(Y)$, we define the output-conditioned score-velocity metric $\Gout^{F,\kappa, h}$ as $G^{\kappa,h}_{F|!_\Omega}$, where $!_\Omega: \Omega\to\unitset$ is the delete map.
\end{definition}
In other words, the $\Gout$ metric is the projected score-velocity when the statistic is forgetting the whole trace. This makes the execution variable simply the output $Z^!_x(\omega) = (\unitsetelem, Y_x(\omega))$.


\subsection{Distribution theory and the Wasserstein-Fisher-Rao geometry}
We characterised the geometry resulting from forgetting part of the trace. However, even if all of the trace is forgotten, this does not make the resulting metric presentation independent: $\Gout$ is still intensional. There might be several different programs that realise the same family of measures, and their implementations differ in the tradeoff between score and velocity parts. For example, let us look at two programs:
\[
\begin{aligned}
\theta: \Rs \vdash & \letbind{z}{\gauss{\theta / 2}{1}}{\\
&\return{2\cdot\coerce{z}} \vkto \Rs}
\end{aligned}
\qquad
\qquad
\begin{aligned}
\theta: \Rs \vdash & \letbind{z}{\gauss{0}{1}}{\\
&\return{\theta + 2\cdot\coerce{z}} \vkto \Rs}
\end{aligned}
\]
They implement the same statistical model, but their $\Gout^{1,\ip{\cdot}{\cdot}}$ metrics are, respectively:
\[
\Gout=\frac{1}{4}\,d\theta^2
\qquad
\qquad
\Gout=d\theta^2
\]
To get a truly presentation-independent metric, we would need to minimise over, at least, all possible such presentations. This is the role of Wasserstein-Fisher-Rao theory. To explain it, we first need to briefly outline the theory of distributions.

\subsubsection{Distribution theory}
Let us look at a family of distributions $\theta\mapsto\mu_\theta$ for $\theta\in\RR$. Without any assumptions about a common dominating measure, we want to differentiate it with respect to $\theta$. For example, we might have the Dirac family we already talked about before:
\[
    \theta\mapsto \delta_{f(\theta)}
\]
A derivative of Dirac delta is not a probability measure, but it can be described using the more general language of distributions. A distribution is simply a (continuous) functional map on a certain space of functions, which we introduce below. The material in this subsection is standard, and to a large extent adapted from~\cite{kristensenB43DistributionTheory}.

\begin{definition}[Distribution]
    Given an open subset $U \subseteq \RR^n$, a scalar-valued distribution is a continuous linear functional of type $\Ctest(U) \to \RR$, where $\Ctest(U)$ are test functions, i.e., smooth functions on $U$ with compact support. For $d \in \NN$, a vector-valued distribution is a distribution of a type $\Ctest(U,\RR^d)\to\RR$.
\end{definition}

Test functions have a canonical topology, but we do not need to discuss its details here. Intuitively, the continuity condition says that, on each fixed compact set, the value of a distribution is controlled by finitely many derivatives of the test function.  As notation, we write $\ip{F}{f}$ to denote a distribution $F$ acting on a function $f$.

\begin{example}
The following are examples of distributions.
\begin{enumerate}
    \item Every locally finite measure $\mu$ on $U$ is a distribution, where the functional is given by integration:
    \[
    \ip{\mu}{f} = \int_U f(x) \mu(\dd x)
    \]
    \item In particular, Dirac delta $\dirac{x}$ acts on functions as evaluation at $x$.
    \item If $\mu$ is a measure and $u: U\to \RR^k$ is a vector-valued $\mu$-integrable function, then $u\mu$ is a vector-valued distribution, which acts on vector test functions by inner product:
    \[
    \ip{u\mu}{f} = \int_U  f(x)\cdot u(x) \mu(\dd x)
    \]
\end{enumerate}
\end{example}
Note that it is not necessary to restrict to probability measures only, or even to finite measures, since test functions vanish outside a compact domain, making them integrable wrt all locally finite measures.

\begin{definition}[Derivative of a distribution]
    For a distribution $F$ on $\Ctest(U)$, we define its $i$-th distributional derivative $\partial_i F$ as:
    \[
    \ip{\partial_i F}{f} = -\ip{F}{\partial_i f}
    \]
\end{definition}
This definition is chosen by analogy to integration by parts:
\[
\int_a^b (\partial_i F)f(x)\dd x = (Ff)(x)\Big|_a^b - \int_a^b F(\partial_i f)(x)\dd x = -\int_a^b F(\partial_i f)(x)\dd x
\]
where the first component is zero because $f$ vanishes outside compact support.
\begin{definition}[Divergence]
    For a vector-valued distribution $F = (F^1,\ldots F^d)$ on $\Ctest(U, \RR^d)$, we define its divergence as a scalar distribution:
    \[
    \diverg(F) = \sum_{i=1}^d \partial_i F^i
    \]
\end{definition}

The divergence acts similarly to a standard derivative, by ``transferring'' the divergence to its argument. For a scalar-valued test function $f$, we have:
\[
\ip{\diverg(F)}{f} = -\ip{F}{\nabla f}
\]

\begin{definition}\label{def:curve-of-measures-distributional-tangent}
    For a parametrised family of measures $\mu_\theta$ over $\Theta$, we say that $D\mu_\theta$ is a distributional derivative of $\mu_\theta$, if for every $\xi \in T_\theta\Theta$ we have that:
    \[
    \ip{D\mu_\theta[\xi]}{\phi}
    =D_\theta\left(\int_Y\phi\dd\mu_\theta\right)[\xi].
    \]
\end{definition}

The distributional derivative might not always exist, or in general be complicated. What we are interested in here is a particular representation of such a derivative.

\begin{definition}[Reaction-transport representation]
Let $\mu$ be a finite non-negative measure on $Y$. A pair $(a,u)$ is a
reaction-transport representation of a distributional tangent $\dot\mu$ if:
\[
    \dot\mu=a\mu-\diverg(u\mu)
\]
as distributions. Equivalently, for every $\phi\in\Ctest(Y)$,
\[
    \ip{\dot\mu}{\phi}
    =\int_Y a(y)\phi(y)\mu(\dd y)
     +\int_Y d\phi_y(u(y))\mu(\dd y).
\]
The scalar field $a$ is the reaction rate and the vector field $u$ is the transport velocity.
\end{definition}
\begin{example}[Derivative of parametrised Dirac deltas]
Let $f:I\to U\subseteq\RR^d$ be smooth, and let us calculate $\dot\delta_{f(t)}$. For every test function $\phi$, we have:
\begin{align*}
    \ip{\dot\delta_{f(t)}}{\phi}
    &=\frac{\dd}{\dd t}\phi(f(t))\\
    &=\nabla\phi(f(t))\cdot f'(t)\\
    &=\ip{-\diverg\bigl(f'(t)\delta_{f(t)}\bigr)}{\phi}
\end{align*}
\end{example}

\begin{remark}\label{remark:score-derivative}
    Suppose that a (finite) measure $\mu_\theta$ has density $\mu_\theta = p_\theta\lambda$ with respect to a fixed base measure $\lambda$. In that situation, assuming we can differentiate under the integral, we can write its derivative simply as:
    \[
    \ip{D\mu_\theta[\xi]}{\phi} = D\left[\int_Y p_\theta(y) \phi(y)\lambda(\dd y) \right][\xi]
    = \int_Y Dp_\theta(y)[\xi] \phi(y)\lambda(\dd y) = \ip{(Dp_\theta[\xi])\lambda}{\phi}
    \]
    If, on top of that, $Dp_\theta$ is absolutely continuous wrt $\mu_\theta$, we can multiply and divide by $p_\theta$ and write:
    \[
    D\mu_\theta[\xi] = Dp_\theta[\xi]\lambda = \frac{Dp_\theta[\xi]}{p_\theta}p_\theta\lambda = S[\theta;\xi]\mu_\theta
    \]
    where $S[\theta;\xi]$ is the score of $\mu_\theta$. Thus, such measures have a simple representation consisting only of the reaction term given by the score.
\end{remark}

\subsubsection{Wasserstein-Fisher-Rao metric}
For distributions that have a reaction-transport representation,~\citet{chizatInterpolatingDistanceOptimal2015} define the so-called Wasserstein-Fisher-Rao metric.

\begin{definition}[Local $\WFR$ quadratic form]\label{def:wfr}
Let $\mu$ be a finite scalar-valued non-negative measure on $Y$ and let $\dot\mu$ be a distributional tangent. Define:
\[
    \norm{\dot\mu}_{\WFR,\kappa,h}^2
    \df
    \inf_{a,u}
    \left\{
        \int_Y\bigl(a(y)^2+\kappa h_y(u(y),u(y))\bigr)\mu(\dd y)
        :
        \dot\mu=a\mu-\diverg(u\mu)
    \right\}
\]
The infimum ranges over
$a\in L^2(\mu)$ and $u\in L^2(\mu;TY)$. By polarisation, this gives the metric $\GWFR$.
\end{definition}


The infimum is exactly the answer to the problem posed at the beginning of this section: it is important, because a continuous measure-valued tangent can admit many different reaction-transport representations. The $\WFR$ form chooses the one with the least energy. Also note that an arbitrary distributional tangent need not admit a reaction-transport representation - the $\WFR$ then assigns value $+\infty$. However, in the case of programs, the tangent always exists, as we explain in the following section. Thus, even though technically on some tangents $\GWFR$ can be undefined, we never encounter this below.

If $\mu_t$ is a curve of probability measures, every admissible representation satisfies the centering of the reaction term:
\[
    \int_Y a_t\dd\mu_t=0
\]
But this does not imply $a_t=0$: reaction may create and remove mass at different points while preserving total mass. We will return to this when discussing the extension to non-normalised measures in~\Cref{ssec:wfr-unnormalised}.

\subsubsection{Distributional tangents for probabilistic programs}

The distributional formalism gives a direct passage from a jet trace to the derivative of its realised output law. We state the calculation on one Euclidean output component; coproducts are handled componentwise.

\begin{proposition}\label{prop:tangent-decomposition-score-velocity}
Let $F = (\Omega, q, o; S, V) : X \to \RR^d$ be a jet trace (either in $\Jet\CatFinTrace$ or in $\Jet\CatDQBS_\Trace$), and write $\mu_x = (o_x)_*(Q_x)$ for the pushed-forward output measure. Then, $D\mu_x[\xi]$ is a distribution, and for $w_{\xi}(y)=\E[S_{\xi}\mid Y_x=y]$ and $u_{\xi}(y)=\E[V_{\xi}\mid Y_x=y]$ we have a distributional tangent:
\[
D\mu_x[\xi]
    =w_{\xi}\mu_x-\diverg(u_{\xi}\mu_x)
\]
\end{proposition}

\begin{proof}
For every $\phi\in\Ctest(Y)$, we have:
\[
    \ip{D\mu_x[\xi]}{\phi}
    =\E\left[
        \phi(Y_x)S_{\xi}
        +\nabla\phi(Y_x)\cdot V_{\xi}
    \right]
\]
This is from the definition of the pushforward:
\[
    \int_Y\phi(y)\mu_x(\dd y)
    =\int_\Omega
      \phi(o_x(\omega))q_x(\omega)\lambda(\dd\omega).
\]
Differentiate the right-hand side. The product rule gives the score term and
the chain rule gives the velocity term. Taking conditional expectations given $Y_x$ gives
\[
    \ip{D\mu_x[\xi]}{\phi}
    =\int_Y\phi(y)w_{\xi}(y)\mu_x(\dd y)
     +\int_Y u_{\xi}(y)\cdot\nabla\phi(y)\mu_x(\dd y)
\]
which is exactly the weak reaction-transport equation. If $\supp(\phi)\subseteq K$, then:
\[
    \abs{\ip{D\mu_x[\xi]}{\phi}}
    \leq
    \E[\abs{S_{\xi}}]\sup_K\abs{\phi}
    +\E[\norm{V_{\xi}}]\sup_K\norm{d\phi}
\]
which is bounded, so indeed $D\mu_x[\xi]$ is a distribution.
\end{proof}

Note that the conditional expectations we have used for $w_{\xi}(y)$ and $u_{\xi}(y)$ are an instance of our earlier construction in~\Cref{def:projected-score-velocity}, for the statistic $\pi$ being simply the delete map $\delK_\Omega$.

\begin{theorem}\label{cor:jet-trace-inequalities}
    For a jet trace $F = (\Omega, q, o; S, V) : X \to \RR^d$ and $\mu_x = (o_x)_*(Q_x)$, we have the inequalities between positive (not necessarily smooth) 2-forms:
    \[
    \GWFR^{\kappa,h} \preceq \Gout^{F,\kappa,h} \preceq  G^{F,\kappa,h}
    \]
\end{theorem}
\begin{proof}
The first term in the hierarchy is the extensional local $\WFR$ energy. The middle term is the score-velocity lift projected onto the output. The final term is presentation-level score-velocity energy.

The first inequality follows from the definition of the $\WFR$ as the infimum over representations of the distributional tangent. The second inequality is a special case of~\Cref{thm:projected-data-processing}.
\end{proof}
As we show in the next section, on a regular finite atomic stratum, the first inequality is an equality. But first, let us expand on the relationship between realised distribution score and the conditioned score, which we already noted do not have to be equal.

\begin{remark}
Suppose that for $F = (\Omega, q,o;S,V):\Theta \to Y$ we have a realised measure $\Realisation(F) = \mu_\theta$ that has density $p_\theta$, and $D\mu_\theta[\xi]=S_\xi^{\mathrm{out}}(\theta)\mu_\theta$, as in~\Cref{remark:score-derivative}. From the distributional tangent equation of~\Cref{prop:tangent-decomposition-score-velocity}, dividing both sides by $p_\theta$:
\[
    S^{\mathrm{out}}_\xi(\theta)
    =w_\xi-\frac{\diverg(p_\theta u_\xi)}{p_\theta}
\]
where $w_\xi(y)=\E[S_\xi\mid Y_\theta=y]$ is the output-conditioned score, and $u_\xi(y)=\E[V_\xi\mid Y_\theta=y]$ is the corresponding velocity. Thus, the output score is, generally, not only the conditional expectation of the trace score. The formula $S^{\mathrm{out}}_\xi(\theta)=\E[S_\xi\mid Y_\theta]$ works only in special cases, such as when conditional velocity vanishes, which happens in particular for discrete domains $Y$.
\end{remark}

\subsubsection{Regular finite atomic strata}

\begin{definition}[Regular finite atomic stratum]
\label{def:regular-finite-atomic-stratum}
Let $U\subseteq\RR^k$ be open, and $n \in \mathbb{N}$. We say that a family of probability measures $\mu_\theta$ lies in a regular finite atomic stratum on $U$ if there is a smooth map $m: U\to \Simplexo{n}$ and smooth atom-location functions $x_j:U\to Y$ for $j=1,\ldots,n$ such that:
\[
  \mu_\theta=\sum_{j=1}^n m^j(\theta)\delta_{x_j(\theta)}
\]
and, for every $\theta\in U$, $x_i(\theta)\neq x_j(\theta)$ whenever $i\neq j$.
\end{definition}
Thus the number of visible atoms is constant, and no two of them ever collide on $U$.

\begin{proposition}[Finite atomic formula]
\label{prop:finite-atomic-wfr}
Let $\mu_\theta$ lie in a regular finite atomic stratum on $U$, given by mass function $m$ and location function $x$. Then, for any $\xi\in T_\theta U$, its distributional tangent has a unique reaction-transport representation on the support of $\mu_\theta$ given by:
\[
    a_{\xi}(x_j(\theta))
    =\frac{Dm_j(\theta)[\xi]}{m_j(\theta)},
    \qquad
    u_{\xi}(x_j(\theta))
    =Dx_j(\theta)[\xi].
\]
Consequently:
\[
    \norm{D\mu_\theta[\xi]}_{\WFR,\kappa,h}^2
    =\sum_{j=1}^n
      \frac{Dm_j(\theta)[\xi]^2}{m_j(\theta)}
     +\kappa\sum_{j=1}^n
      m_j(\theta)
      h_{x_j(\theta)}\bigl(Dx_j(\theta)[\xi],Dx_j(\theta)[\xi]\bigr).
\]
\end{proposition}
\begin{proof}

For $\phi\in\Ctest(Y)$, differentiating both sides of:
\[
    \int_Y\phi\dd\mu_\theta
    =\sum_{j=1}^n m_j(\theta)\phi(x_j(\theta)).
\]
gives:
\[
    D\mu_\theta[\xi]
    =\sum_{j=1}^n Dm_j(\theta)[\xi]\delta_{x_j(\theta)}
     -\diverg\left(
        \sum_{j=1}^n
        m_j(\theta)Dx_j(\theta)[\xi]\delta_{x_j(\theta)}
      \right)
\]
and the displayed reaction and transport terms. Since the atoms are distinct, one may choose a test function supported near a single atom and prescribe independently its value and first derivative at that atom. Hence the Dirac masses and their first distributional derivatives at different atoms are linearly independent. The values of $a_{\xi}$ and $u_{\xi}$ on the support are therefore forced. Substitution into the definition of the $\WFR$ form gives the formula.
\end{proof}
By polarisation, for the regular finite atomic families, the $\WFR$ form is then derived as:
\[
    \GWFR(\xi,\eta)_\theta
    =\sum_{j=1}^n
      \frac{Dm_j(\theta)[\xi]Dm_j(\theta)[\eta]}{m_j(\theta)}
    +\kappa\sum_{j=1}^n
      m_j(\theta)
      h_{x_j(\theta)}\bigl(Dx_j(\theta)[\xi],Dx_j(\theta)[\eta]\bigr)
\]
This looks exactly like the score-velocity metric. Indeed, we have the following.
\begin{corollary}
For any jet trace $F$ realising a regular finite atomic family $\mu$, $\Realisation(F) = \mu$, the inequality chain from~\Cref{cor:jet-trace-inequalities} becomes:
\[
\GWFR^{\kappa,h} = \Gout^{F,\kappa,h} \preceq G^{F,\kappa,h}
\]
\end{corollary}
It might still be the case that the last inequality is strict, for example if $F$ has unused randomness. But if that doesn't happen, we get a nice uniqueness result.
\begin{corollary}\label{cor:uniqueness-velocity}
    If two jet traces $F_1, F_2$ as above have the same realised output measure, and both of them are output-measurable, then the above corollary necessarily forces:
    \[
    G^{F_1,\kappa,h} = \Gout^{F_1,\kappa,h} = \Gout^{F_2,\kappa,h} = G^{F_2,\kappa,h}
    \]
\end{corollary}

However, we note that this is one of a few observations that only apply in the finite case. In the continuous case, it is no longer true, as we will show in~\Cref{prop:output-not-forced-continuous}.

\begin{figure}
    \centering
    \input{figures/cor5-prop30-geometries}
    \caption{The metric contractions of~\Cref{cor:jet-trace-inequalities} and~\Cref{prop:finite-atomic-wfr}.
    The program families are (1) regular finite atomic and output-measurable, (2) regular finite atomic but not output-measurable, and (3) neither regular finite atomic nor output-measurable. Within each row, all three panels use the same parameter range and scale. Red varies $\alpha$, blue varies $\beta$, and the white point marks $(\alpha,\beta)=(0,0)$. In the bottom row, $s=1+\beta$, and in all rows we assume $1 + \beta >0$. In columns, we use $\GWFR$ as defined in~\Cref{def:wfr}, $\Gout$ as defined in~\Cref{def:g-out}, and $\FisherDirac$ as defined in~\Cref{def:fisher-dirac}, that is, we take $h = \langle \cdot,\cdot \rangle$ and $\kappa = 1$.}
    \label{fig:many-geometries}
\end{figure}

\subsection{WFR geometry of unnormalised measures}\label{ssec:wfr-unnormalised}
We now aim to rederive the inequalities of~\Cref{cor:jet-trace-inequalities} between WFR, output, and standard forms, for unnormalised measures. As a first observation, we note that Wasserstein-Fisher-Rao geometry permits changing mass via the reaction term $a$ in the tangent decomposition $(a, u)$ of the measure tangent $\dot{\mu}$. Neither~\Cref{def:curve-of-measures-distributional-tangent} nor~\Cref{def:wfr} required the normalisation assumption. But, in the normalised case, we additionally have $\int_Y a(y) \mu(\dd y) = 0$, which corresponds, in particular, to the centering of score (score-velocity is one representation of the distributional tangent, as we pointed out in~\Cref{prop:tangent-decomposition-score-velocity}).

\begin{proposition}
Given a kernel $\mu: X \kto Y$ between Euclidean spaces $X, Y$ and some $Z: X\to \RR_+$ defining a weighted kernel $\widetilde\mu_x=Z_x\mu_x$, a chosen $\xi \in T_xX$, and a weighted distributional tangent $D\widetilde\mu[\xi] = a_\xi\widetilde\mu- \diverg(u_\xi\widetilde\mu)$, under suitable integrability and differentiability assumptions, we have:
\[
D\log Z_x[\xi]
=
\int_Y a_\xi(y)\mu_x(\dd y)
\]
\end{proposition}
\begin{proof}
Write:
\[
DZ_x[\xi] = D_x\left(\int_Y1\,d\widetilde\mu_x\right)[\xi] = \int_Y a_\xi \dd\widetilde\mu_x = Z_x\int_Y a_\xi\mu_x(\dd y)
\]
and divide by $Z_x>0$.
\end{proof}
This property allows us to relate WFR form for $\mu$ and $\tilde\mu$. Writing $\bar a=\int a\dd\mu_x$, differentiation of $\mu_x=Z_x^{-1}\widetilde\mu_x$ gives
\[
D\mu[\xi]
=
(a_\xi-\bar a_\xi)\mu
-
\diverg(u_\xi\mu)
\]
so for every distributional tangent representation $(a, u)$ of $D\widetilde\mu$, the pair $(a - \bar a, u)$ is a valid distributional tangent representation for the normalised $\mu$: normalisation removes the mean reaction and leaves the transport field unchanged. We also note that the decomposition:
\[
a = \bar a + (a - \bar a)
\]
is orthogonal in $L^2(\mu_x)$. Thus, we can write:
\[
\int_Y a^2+\kappa h(u,u)\dd \mu = 
\bar a^2 + \int_Y (a - \bar a)^2+\kappa h(u,u)\dd \mu
\]
and taking infimum over both sides and multiplying by $Z$, we get:
\[
\norm{D\widetilde \mu[\xi]}_{\WFR,\kappa,h}^2 = Z(\bar a^2 + \norm{D\mu[\xi]}_{\WFR,\kappa,h}^2)
\]
It turns out that removing the expectation of the reaction term and scaling by $Z$ can be done uniformly across all metrics we have studied so far. Namely, we have the following theorem.
\begin{theorem}
Given a weighted jet trace $F: X \to Y$ denoting the realised kernel $\widetilde\mu_x$, with $\xi \in T_xX$ and $m_\xi = D \log Z_x[\xi]$, we have (suppressing $x, \kappa, h$ dependence from the notation):
\[
\begin{array}{ccccc} \norm{D\widetilde\mu[\xi]}_{\WFR}^2 &\leq&\Gout^{F}(\xi,\xi) &\leq& G^{F}(\xi,\xi) \\
\rotatebox{90}{$=$}&&\rotatebox{90}{$=$}&&\rotatebox{90}{$=$} \\
Z_F\!\left(m_\xi^2+\norm{D\mu[\xi]}_{\WFR}^2\right) &\leq& Z_F(m_\xi^2+\Gout^{\normaliseF F}(\xi,\xi)) &\leq& Z_F(m_\xi^2+G^{\normaliseF F}(\xi,\xi)) \end{array}
\]
where $\mu = \widetilde\mu / Z_F$ is the normalised measure realised by $\normaliseF F$.
\end{theorem}

\subsection{WFR geometry for $\PPL_\RR$}

In~\Cref{cor:uniqueness-velocity}, we have shown that for jet traces realising regular finite atomic families which are output-measurable, we have a uniquely determined velocity and score. This property can fail for families with continuous output distributions, as the following example in $\PPL_\RR$ shows.
\begin{remark}\label{prop:output-not-forced-continuous}
    Let us take the trace $\Omega = \Rc^2$ and input and output spaces $\Theta = \Rs$, $Y = \Rs^2$, and let $q(z)$ be the (independent of the parameter $\theta$) density of a standard 2-dimensional Gaussian $\mathcal{N}(0,I_2)$. Then, consider two jet traces $F_1 = (q, o_1), F_2 = (q, o_2)$ with output maps:
    \[
    o_1(\theta, z_1, z_2) = (z_1, z_2)
    \qquad
    o_2(\theta, z_1, z_2)  = \begin{bmatrix}
        \cos \theta & \sin \theta \\ 
        -\sin \theta & \cos \theta
    \end{bmatrix}
     \begin{bmatrix}
        z_1 \\
        z_2
    \end{bmatrix}
    \]
    By definition, we have equality of realised measures, and also equality of trace densities (and thus scores). Also, both jet traces are output-measurable. But nevertheless, the velocities differ:
    \[
    V_1 \equiv 0 \qquad V_2 = \partial_\theta o_2 \neq 0
    \]
\end{remark}

\subsection{Natural gradient}\label{ssec:natgrad}
An experiment is a deterministic way to feed source environments to one or more copies of a program. Let $F: X \to Y$ be a dominated trace presentation. An \(n\)-observation experiment with a state space \(M\) is a smooth map:
\[
  r=(r_1,\ldots,r_n):M\to X^n
\]
Let us define $F^r = F^{\otimes n} \circ r: M \to Y^n$, and let $Y^n$ be equipped with a product tensor $h^{n}$, where $h \in \PSD(Y)$. Then, $F^r$ induces the product kernel:
\[
  \Realisation(F^r)_m(\dd y_1\cdots\dd y_n)
  =\prod_{i=1}^n \Realisation(F)_{r_i(m)}(\dd y_i).
\]

\begin{theorem}[Experiment geometry]
For $\xi\in T_mM$ and $\omega=(\omega_1,\ldots,\omega_n)$, the score and velocity of $F^r$ are:
\[
S^{F^r}_{m,\xi}(\omega)=\sum_{i=1}^n S^F_{r_i(m),Dr_i{}_m[\xi]}(\omega_i)
\qquad
V^{F^r}_{m,\xi}(\omega)=\bigl(V^F_{r_i(m),Dr_i{}_m[\xi]}(\omega_i)\bigr)_{i=1}^n
\]
Consequently, the induced score-velocity form on $M$ is:
\[
G^r =\Affine_{F^r}(\kappa h^{\oplus n}),
\qquad
G^r_m(\xi,\eta)
=\sum_{i=1}^n
(G_F^{\kappa,h})_{r_i(m)}
\bigl(Dr_i{}_m[\xi],Dr_i{}_m[\eta]\bigr)
\]
\end{theorem}

Thus fixed covariates, learned covariates, shared hyperparameters, computed inputs, are all handled uniformly, by the same formula.

\subsubsection{Riemannian and degenerate Riemannian gradients}
In standard Riemannian geometry, a Riemannian form $g: TM\times TM \to \RR$ induces a pair of maps, called musical isomorphisms, between the tangent and cotangent bundles. Let us fix some $m\in M$ and write $V = T_mM$. We have:
\[
g^\sharp : V^* \to V \qquad g^\flat: V \to V^*
\]
defined by the equations:
\[
g(g^\sharp(\omega), \eta) = \omega(\eta) \qquad g^\flat(\xi)(\eta) = g(\xi,\eta)
\]
for a covector $\omega \in V^*$ and tangent vectors $\xi, \eta \in V$. Musical isomorphisms are inverses of each other. The natural gradient of a smooth scalar function $\ell:M \to \RR$ is then defined as:
\[
\NatGrad\ell(m) = g^\sharp(D\ell_m)
\]
The same definition can be used in the case of a degenerate-Riemannian metric $g \in \PSD(M)$, but the musical maps are no longer isomorphisms. The flat $g^\flat(\xi) = g(\xi, -)$ can still be defined, but it possibly is no longer injective or surjective, and the sharp is a map only on the image of $g^\flat$. Thus, we have an isomorphism:
\[
g^\flat_m: T_mM / \ker g^\flat \to \text{im} g^\flat
\]
For a positive-semidefinite symmetric form, we have the characterisation of the kernel:
\[
\ker g^\flat = \{\eta \in V : g(\eta, \xi) = 0 \text{ for all } \xi \in V\} =  \{\eta \in V : g(\eta, \eta) = 0\}
\]
so we can talk about the kernel of $g$ and $g^\flat$ interchangeably. Thus, we have the following definition.
\begin{definition}
Let $\ell:M\to\RR$ be some smooth objective (or loss) function. The natural gradient of $\ell$ with respect to the score-velocity form $G^r$ at $m$ is an element of the quotient space $T_mM/\ker G^r_m$ given by:
\[
\NatGrad\ell(m) = [(G^r_m)^\sharp(D\ell_m)]
\]
\end{definition}

Natural gradient exists exactly when $D\ell_m$ annihilates $\ker G^r_m$. In local Euclidean coordinates, we can compute its (Moore-Penrose) representative with:
\[
\NatGrad\ell(m)=G^r(m)^\dagger\nabla\ell(m)
\]
Thus the program supplies the metric, while the optimisation problem supplies the covector. In particular, $\ell$ does not need to be a log-likelihood. However, we note that the resulting vector field might fail to be globally smooth.

At the two extremes, if $F=f$ is deterministic, then $G_F^{\kappa,h}=\kappa f^*h$. If $Y$ is finite, then $TY=0$ and $G_F^{\kappa,h}=\Fisher{F}$, so the construction reduces to the usual (trace) Fisher natural gradient. In that case, for observations $D=(y_1,\ldots,y_n) \in Y^n$, the log-likelihood covector is simply:
\[
\alpha_D(\xi)=D\ell_D(m)[\xi]
=\sum_{i=1}^n
S^{\mathrm{out}}_{r_i(m),Dr_i{}_m[\xi]}(y_i)
\]
and the kernel condition is then satisfied automatically.

Note that for real outputs, likelihood is more complicated. This is because $S$ is only defined up to an equivalence class, so we have to pick a concrete representative to evaluate it at a point. But in the finite-dimensional Bernoulli regression example discussed in~\Cref{appendix:bernoulli-example}, likelihood evaluation is unambiguous.

\section{Future work}\label{sec:future-work}

This work has been motivated by the vision of programs as geometric objects: transforming geometry, collected into moduli spaces, intersecting or pulling back onto each other. Infinite-dimensional information geometry is famously non-canonical and computationally intractable, but by restricting attention to certain well-behaved, still rich families of distributions, we aimed to expand its scope and applicability. The main motivating goal was to put a global, uninformative, parametrisation-invariant, compositionally constructed prior on programs, formalising Bayesian non-parametric regression over a rich class of models, extending the minimal description length paradigm with smooth parameters, and then sampling from a posterior using genuinely geometric, efficient MCMC algorithms such as Riemannian Manifold Hamiltonian Monte Carlo. Although closer to realisation, reaching that point and fully realising that vision still requires solving a number of problems. We outline some of them below.

\subsection{Loops}\label{ssec:loops}
One idea is to extend our language with loops. The key insight here is that the score is a martingale: if the loop termination condition is written as a stopping time $\tau$, then the second moment $\E[S_\tau(\xi)^2]$ can be written as the expected predictable quadratic variation $\E[\langle S(\xi)\rangle_\tau]$ (if the expected bracket is infinite, we could put the value $+\infty$, or define it only partially). Under appropriate differentiation and integrability hypotheses, the metric of an almost surely terminating loop then should take the form of:
\[
G^h(\xi,\eta)
=
\E\left[
\left\langle S(\xi),S(\eta)\right\rangle_\tau
\right]
+
\E\left[
h_{Y_\tau}
\bigl(V_\tau(\xi),V_\tau(\eta)\bigr)
\right].
\]
The first term would be controlled by the predictable bracket, but the second would require a separate tail estimate for the output velocity and for the growth of $h$. If loops do not terminate with positive probability, this would force us to the theory of subprobability measures, whose scores are no longer centered - essentially the same problem arises when adding loops to the language $\PPLscore$. One could also simply condition on termination, or add a partial order structure.

\subsection{Higher-order functions and recursion}
Another idea is to extend the language with first-class $\lambda$'s, supplying higher-order functions, or adding recursion. Quasi-Borel and diffeological spaces in themselves are a natural setting for higher-order semantics: indeed, as we have shown, $\CatDQBS$ is CCC. But to make a higher-order or recursive language, there are still many issues to resolve. Some of these are common to any PL problem, such as extending the domain with some sort of partial order structure to encode non-termination, or supplying a fixed point operator. Some are common to differentiable languages, such as the questions around differentiation through the limit. Others still are due to probabilistic structure (such as non-termination being connected to subprobability kernels). Some are unique to our approach: for example, that an increasing supremum of smooth quadratic forms may be only measurable or lower semicontinuous, and jump out of the cone of admissible metrics. In the case of the language $\PPL_\RR$, there are additionally issues with tails (we no longer have polynomial bounds even if every one-step primitive and every terminal metric is individually polynomially bounded), and for $\PPLscore$, score is no longer a martingale, as we noted above.

\subsection{Solomonoff-Jeffreys prior and Bayesian non-parametric regression}\label{ssec:jeffreys-solomonoff}
One of the motivations to study geometric probability is to build good (transformation-invariant) uninformative priors. In our case, this means defining a Jeffreys-like prior using score-velocity metric $\FisherDirac$ in place of pure Fisher information. On a single stratum $\sigma$ of a program parametrised by $\theta \in \Theta$, it could in principle be:
\[
\Jeffreys_\sigma(d\theta) = \sqrt{\det \FisherDirac(\theta)}\,d\theta / Z
\]
which would then build up to the prior on the full space, by summing over program skeletons, or factorising:
\[
\Pi(\dd \sigma, \dd \theta) = \Jeffreys_\sigma(d\theta)S(\sigma)
\]
Above, $S(\sigma)$ is a prior over program skeletons, such as a length prior $S(\sigma)\propto 2^{-|\sigma|}$, a prior from a probabilistic grammar over syntax trees, or similar. We note that the choice of this part is non-canonical, and computing this factor requires summing over the possible program derivations. It could also be difficult to reconcile with $\beta/\eta$ laws we have used in this paper.

Even disregarding these issues, there are several fundamental obstructions. First, the case of $Z = \infty$, i.e., an improper prior, which happens for many distributions - such as Gaussians. This is not necessarily an issue in classical Bayesian statistics, since there is simply one stratum within which sampling algorithms operate, but if we aim to jump between strata, each of them has to normalise separately. One fix could be to limit the programming language to compact types only: instead of real number types we would have the circle types, and Gaussian primitives would be replaced by von Mises distributions. A related problem is $Z = 0$, arising because $\FisherDirac$ is only degenerate-Riemannian. This in turn could potentially be solved by applying the Hausdorff-Jeffrey measure induced by the quotient Fisher metric, but the details are subtle and uncertain. Finally, for the prior to connect to minimal-description length or algorithmic information theory, and deserve the name Solomonoff-Jeffreys, the programming language should be Turing-complete, and thus solve the issues around higher-order functions and recursion or loops, already outlined in sections above.

We are nevertheless excited by the prospect of having such a prior. It could allow us to define genuine general non-parametric Bayesian regression in program spaces under an uninformative canonical repara\-metrisation-invariant (which in programs should be a stronger condition of equality-up-to-normalisation or optimisation-independent) prior.

\subsection{Trans-dimensional Riemannian Manifold Hamiltonian Monte Carlo}\label{ssec:rmhmc}
To sample from the posterior, or perform the SJ Bayesian non-parametric regression, we would like to use a Riemannian Diffeological HMC of roughly the shape of running the standard RM HMC inside one program skeleton, and then every so often attempting to jump between dimensions, randomly modifying the program skeleton. Our intuition is that an algorithm like this would retain the key property of HMC, namely, sampling from a typical set. This is because adding or removing dimensions would be done in a controlled fashion: in the $\ifte{b}{-}{-}$ case, we could initialise the branch splitting parameter $b$ at some small probability, not perturbing the overall denotation, but allowing the gradient to smoothly explore the new branch. We could make this reversible by also always making it possible to jump down in dimension.

Developing an algorithm like that requires solving a number of issues. For example, the semantics we've developed for $\ETrace{Y}$ are really stratified on its faces $\ETraceoIn{J}{Y}$, so the global space is not really a manifold, and the program-induced plots cannot smoothly reach the lower-strata (although they are topologically connected). However, setting apart the issues of formal justification of smoothness, we think that the algorithm could still be proved correct via the crude means of standard Metropolis-Hastings, and then simply tested in practice for probabilistic program synthesis.

\subsection{Natural gradient and variational inference}\label{ssec:variational}
The natural gradient can be written in a variational formulation. For a loss $\ell$ such that $D\ell$ annihilates $\ker G$:
\[
\NatGrad\ell = \arg\max_{\xi}\left\{D\ell(\xi)-\frac{1}{2} G(\xi,\xi)\right\} =  \arg\max_\xi\left\{D\ell(\xi)-\frac{1}{2} \E[S_\xi S_\xi + \kappa h(V_\xi, V_\xi)]\right\} 
\]
for $\xi \in T_mM / \ker G_m$. Because the quantity above is an expectation of the score and velocity energies, this objective can be estimated directly from program executions, without explicitly constructing or inverting the metric $G$. This suggests stochastic natural-gradient methods for variational inference over probabilistic programs. We leave the details of the implementation and properly developing this for future work.

\subsection{Extension to dualistic geometry}\label{ssec:dualistic-geometry}
A Riemannian metric such as that given by Fisher information is only the first ingredient of more interesting geometric structures. The next essential step is the study of dualistic geometry, which prescribes a pair of torsion-free dual connections on a statistical manifold, equivalently given by a symmetric cubic tensor $C$. In our setup, we can suppose that the output space $Y$ carries a pair $(h,C)$. Then, we can consider how a (normalized) finite jet trace $F=(\Omega,q,o;S,V):X\to Y$ pulls it back onto a pair $\mathcal B_F(h,C)=(\Affine_F(h),C^F)$ onto $X$, by the formulas:
\[
\Affine_F(h)(\xi,\xi)=\E\left[ S_\xi^2+h(V_\xi,V_\xi)\right]
\qquad
C^F(\xi,\xi,\xi)=\mathbb E_q\left[S_\xi^3 +3S_\xi h(V_\xi,V_\xi) +C(V_\xi,V_\xi,V_\xi)\right]
\]
This generalises the functor $\Affine$. Questions of how it relates to the weighted version $\AffineW$, how to extend it to $\PPL_\RR$, how this interacts with the Hessian structure described in~\Cref{ssec:hessian}, and whether dual coordinates have useful computational consequences are interesting future directions.

\subsection{Relationship to global divergence geometry}\label{ssec:global-divergence}
The compositional global divergence viewpoint has been developed in~\cite{perroneMarkovCategoriesEntropy2024}. In general, the move from the local Hessian approximation to the global divergence is non-injective: there are many global divergences giving the same Fisher information. One way to approach our subject from the divergence viewpoint is, for a jet trace $F:X \to Y$, to pull back a chosen smooth output divergence $D_Y$ to a divergence on $X$ by:
\[
(\mathfrak D_F D_Y)(x,x')=\KL(Q_x\|Q_{x'})+\E_{\omega\sim Q_x}\left[D_Y(o_x(\omega),o_{x'}(\omega))\right]
\]
This satisfies the composition rule:
\[
\mathfrak D_F \circ \mathfrak D_H = \mathfrak D_{H \circ F}
\]
and, if the local quadratic approximation of $D_Y$ is $h$, then for $\mathfrak D_F$ it is $\Affine_F(h)$. Expanding on this approach and connecting to prior work is another exciting direction.

\subsection{Chentsov's theorems for program-generated families}\label{ssec:chenstov}
We note that our characterisation theorem for the score-velocity metric is, at the moment, quite weak, compared to theorems characterising other standard geometries in probability.  Developing a stronger justification remains one of the most interesting future directions. Fisher information for finite-dimensional families has been famously characterised by~\cite{cencovStatisticalDecisionRules2000}, later extended by~\cite{ayInformationGeometry2017b} to arbitrary parametric families. However, we note that the theorem can still be extended in two directions - of more generality, or less generality. This is because the assumption about congruence invariance in the versions of Chentsov's theorem has to apply to all members of the family, so for smaller families the theorem leaves open the possibility of there being a metric that is invariant on a smaller (but still somehow interesting) subset. For example~\cite{dowtyChentsovsTheoremExponential2017a} proved it for exponential families very recently.
So, one approach would be to prove a similar theorem for certain program families, whatever they are (finite or infinite, smooth, compact, etc). Alternatively, one could attempt to prove a meta-theorem about all families of programs (subfamilies) that Chentsov-like theorems apply to, such as showing they are closed under certain operations but not others.

\subsection{Polynomially controlled maps}
In~\Cref{sssec:integrability-of-score}, in particular in~\Cref{example:moment-counterexample} we dealt with the problem of divergence of score and metrics in $\PPL_\RR$. In~\Cref{remark:primitives-sufficient-necessary}, we also pointed out that our restriction on primitives is sufficient but not necessary. The category we settled on was $\Jet\CatDQBS_\Trace$, generated by programs in $\PPL_\RR$, which is only an implicit description. This was a pragmatic choice, but not necessarily the most elegant one. What is needed is a principled theory of polynomially-bounded stochastic kernels and sized spaces. Such a theory might use notions such as Nash, quasi-Nash, or polynomially bounded definable geometry~\citep{debrouwereNoteCompositionOperators2022,shavivTemperedDistributionsSchwartz2020}, and it could potentially have a similar structure of concrete sheaves to $\CatDQBS$, but consider more complex compatibility conditions based on tail estimates. We leave it as another direction for future research.
\clearpage
\section*{Acknowledgements}
This work benefitted from a number of discussions the first author had with various researchers over the years. Tobias Fritz suggested framing the program geometry in the context of distribution theory, which partially led to the Wasserstein-Fisher-Rao geometry connection. Other people who provided helpful comments and suggestions included Pedro Amorim, Paolo Perrone, Laura Gonz\'{a}lez-Bravo, Ryszard Pawe\l{} Kostecki, and Mahito Sugiyama. We also benefitted from extensive use of AI, mostly ChatGPT o3, 5 Pro, and 6 Pro. It is difficult to exactly attribute the provenance of many results since they were iterated on over years, but the shape of the $\Affine$ and $\AffineW$ functors (as a generalisation of the purely Euclidean score-velocity metric developed earlier) was first suggested by an AI. The main paper text has been written by humans, but the supporting contents in the appendices and TikZ code for figures are partially LLM-generated. Research partly supported by ERC Consolidator Grant BLAST, AFOSR Project FA9550-21-1-0038, and the ARIA programme on Safeguarded AI and seed funding SynthStats.

\bibliographystyle{plainnat}
\bibliography{references}

\appendix

\clearpage
\section{Typing rules of $\PPL$ and $\PPL_\RR$}\label{appendix:typing}
\input{appendix/typing.tex}




\clearpage
\section{Diffeological and quasi-Borel spaces}\label{appendix:dqbs}
\input{appendix/dqbs}

\clearpage
\section{Example: Bernoulli regression}\label{appendix:bernoulli-example}
\input{appendix/bernoulli_example}

\clearpage
\section{Example: composition of Gaussian mixtures}\label{appendix:gaussian-mixture}

\input{appendix/gaussian_example}


\end{document}

%% file: figures/figure_sphere.tex
\begin{tikzpicture}[x=1cm,y=1cm]

  \node[program,anchor=north west] at (0,8.30) {%
    \textcolor{scoreCurve}{\textbf{Score family $p_1$}}\par\smallskip
    $\displaystyle
      \begin{aligned}
        (\alpha:\R)\vdash{}\;&\letbind{b}{\bern{\alpha}}{}\\[-1pt]
        &\ifte{b}{\return{1}}{}\\[-1pt]
        &\qquad \return{0}\vkto\R
      \end{aligned}$
  };

  \node[program,anchor=north west] at (0,4.18) {%
    \textcolor{transportCurve}{\textbf{Velocity family $p_2$}}\par\smallskip
    $\displaystyle
      \begin{aligned}
        (\beta:\R)\vdash{}\;&\letbind{b}{\bern{0}}{}\\[-1pt]
        &\ifte{b}{\return{1+\beta}}{}\\[-1pt]
        &\qquad \return{0}\vkto\R
      \end{aligned}$
  };

  \node[align=center,font=\small,anchor=north] at (11.55,8.30) {%
    \textbf{Two-parameter family $p_3$}\\[2pt]
    $\displaystyle
      \begin{aligned}
        (\alpha:\R,\beta:\R)\vdash{}\;&
          \letbind{b}{\bern{\alpha}}{}\\[-1pt]
        &\ifte{b}{\return{1 + \beta}}{}\\[-1pt]
        &\qquad \return{0}\vkto\R
      \end{aligned}$
  };

  %
  \begin{axis}[
    small curve,
    name=scoremini,
    at={(0.35cm,5.18cm)},
    anchor=south west,
    width=4.70cm,
    height=1.72cm,
    xmin=0.20,xmax=2.05,
    ymin=0.20,ymax=2.05
  ]
    \addplot[
      exact curve,
      scoreCurve,
      domain=-3.8:3.8
    ]
      ({2*sqrt(sigmoid(x))},
       {2*sqrt(1-sigmoid(x))});

    \addplot[
      only marks,
      mark=*,
      mark size=2.25pt,
      mark options={fill=white,draw=scoreCurve,line width=0.9pt}
    ] coordinates {(1.41421356,1.41421356)};

    \node[quietText,font=\scriptsize,anchor=west,
          xshift=5pt,yshift=2pt]
      at (axis cs:1.414,1.414) {$\alpha=0$};
  \end{axis}
  \node[quietText,font=\scriptsize,anchor=north]
    at ($(scoremini.south)+(0,-0.04)$)
    {$\alpha\longmapsto P(\alpha,1)$};

  \begin{axis}[
    small curve,
    name=transportmini,
    at={(0.35cm,0.67cm)},
    anchor=south west,
    width=4.70cm,
    height=1.72cm,
    xmin=-0.15,xmax=1.48,
    ymin=-0.92,ymax=1.48
  ]
    \addplot[
      exact curve,
      transportCurve,
      domain=-2.25:2.25
    ]
      ({sqrt(2)*cos(deg((1+x)/2))},
       {sqrt(2)*sin(deg((1+x)/2))});

    \addplot[
      only marks,
      mark=*,
      mark size=2.25pt,
      mark options={fill=white,draw=transportCurve,line width=0.9pt}
    ] coordinates {(1.24108916,0.67801010)};

    \node[quietText,font=\scriptsize,anchor=south west]
      at (axis cs:1.241,0.678) {$\beta=0$};
  \end{axis}
  \node[quietText,font=\scriptsize,anchor=north]
    at ($(transportmini.south)+(0,-0.04)$)
    {$\beta\longmapsto P(0,1+\beta)$};

  %
  %
  \begin{axis}[
    name=surfaceaxis,
    at={(11.75cm,3.65cm)},
    anchor=center,
    hide axis,
    x={(1.436cm,-0.670cm)},
    y={(-0.978cm,-0.674cm)},
    z={(0.207cm,1.470cm)},
    scale mode=none,
    clip=false,
    xmin=-0.65,xmax=2.08,
    ymin=-1.65,ymax=2.05,
    zmin=0,zmax=2.08
  ]
    \addplot3[
      surf,
      z buffer=sort,
      shader=interp,
      draw=none,
      opacity=0.62,
      point meta=z,
      colormap name=program surface,
      domain=-4:4,
      y domain=-1.4:3.4,
      samples=13,
      samples y=13
    ]
      ({2*sqrt(sigmoid(x))*cos(deg(y/2))},
       {2*sqrt(sigmoid(x))*sin(deg(y/2))},
       {2*sqrt(1-sigmoid(x))});

    \addplot3[
      mesh,
      z buffer=sort,
      draw=white,
      opacity=0.34,
      line width=0.32pt,
      domain=-4:4,
      y domain=-1.4:3.4,
      samples=7,
      samples y=7
    ]
      ({2*sqrt(sigmoid(x))*cos(deg(y/2))},
       {2*sqrt(sigmoid(x))*sin(deg(y/2))},
       {2*sqrt(1-sigmoid(x))});

    \draw[scoreCurve,line width=2.2pt,line cap=round]
      plot[domain=-3.8:3.8,samples=64,variable=\t]
      (axis cs:
        {2*sqrt(sigmoid(\t))*cos(deg(1/2))},
        {2*sqrt(sigmoid(\t))*sin(deg(1/2))},
        {2*sqrt(1-sigmoid(\t))});

    \draw[transportCurve,line width=2.2pt,line cap=round]
      plot[domain=-2.25:2.25,samples=64,variable=\t]
      (axis cs:
        {sqrt(2)*cos(deg((1+\t)/2))},
        {sqrt(2)*sin(deg((1+\t)/2))},
        {sqrt(2)});

    \addplot3[
      only marks,
      mark=*,
      mark size=3.0pt,
      mark options={fill=white,draw=black,line width=0.95pt}
    ] coordinates {(1.24108916,0.67801010,1.41421356)};

    \coordinate (intersection) at
      (axis cs:1.24108916,0.67801010,1.41421356);

    \node[font=\scriptsize,anchor=south west,inner sep=1.5pt]
      at (intersection) {$P(0,1)$};
    \node[scoreCurve,font=\scriptsize,anchor=north east]
      at (axis cs:1.714,0.936,0.690) {$P_{\mathrm s}(\alpha)$};
    \node[transportCurve,font=\scriptsize,anchor=south west]
      at (axis cs:0.446,1.342,1.414) {$P_{\mathrm t}(\beta)$};
  \end{axis}
  \node[quietText,font=\scriptsize,anchor=north]
    at ($(surfaceaxis.south)+(0,+1.0)$)
    {$(\alpha, \beta)\longmapsto P(\alpha,1+\beta)$};

  \draw[inclusion,scoreCurve]
    (3.70,5.72)
    arc[start angle=120,end angle=40,radius=3.55cm]
    node[pos=0.50,above,quietText,font=\scriptsize]
    {inclusion map};

  \draw[inclusion,transportCurve]
    (3.70,1.58)
    arc[start angle=240,end angle=320,radius=3.55cm]
    node[pos=0.50,below,quietText,font=\scriptsize]
    {inclusion map};

\end{tikzpicture}

%% file: figures/contraction-composition-pullback.tex

\begin{tikzpicture}[
  >=Latex,
  every node/.style={inner sep=0pt},
  panel/.style={align=center,text width=.255\linewidth,font=\small},
  arrowcol/.style={align=center,text width=.065\linewidth,font=\small}
]
  \node[align=center,text width=.97\linewidth] (program) {%
    {\footnotesize
    Program $p$ is given as:\quad
    $
    \begin{aligned}
      (a:\Rs,b:\Rs)\vdash{}\;&\letbind{z}{\gauss{a}{1}}{}\\[-1pt]
      &\letbind{y}{\gauss{\coerce{z}+b}{1}}{}\\[-1pt]
      &\qquad \return{\coerce{y}}\vkto\Rs
    \end{aligned}
    $
    }
  };

  \node[panel,below=3.5mm of program] (trace) {%
    {\bfseries Full trace}\\[-1pt]
    $\FisherDirac  =(\dd a)^2+(\dd b)^2$\\[-1pt]
    \includegraphics[width=\linewidth]{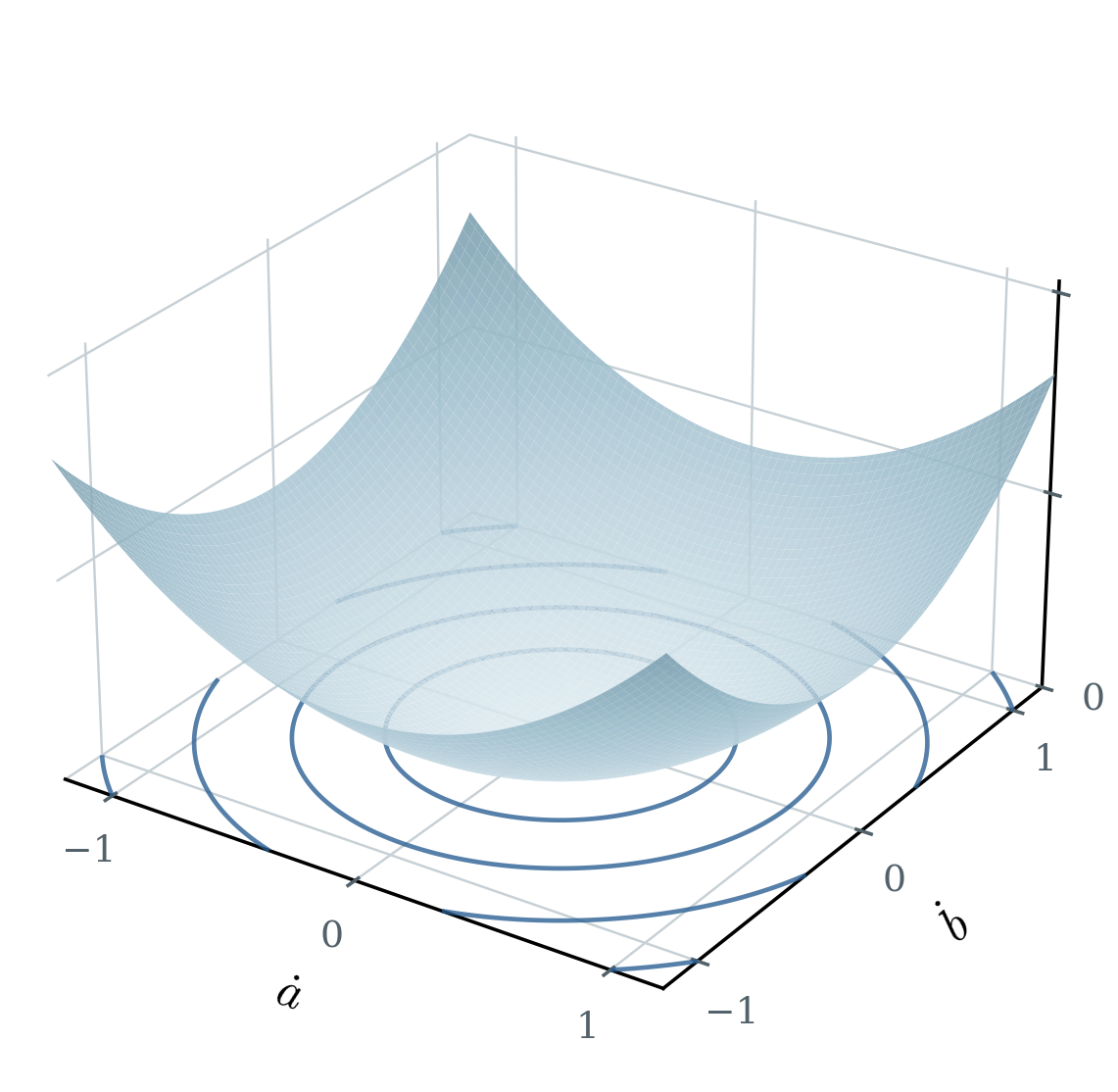}
  };

  \node[arrowcol,left=.010\linewidth of trace] (rarr) {%
    $\xrightarrow{\;r\;}$
  };

  \node[panel,left=.010\linewidth of rarr] (exp) {%
    {\bfseries Experiment}\\[-1pt]
    $r^*\FisherDirac = 2(\dd\mu)^2+\tfrac12(\dd\delta)^2$\\[-1pt]
    \includegraphics[width=\linewidth]{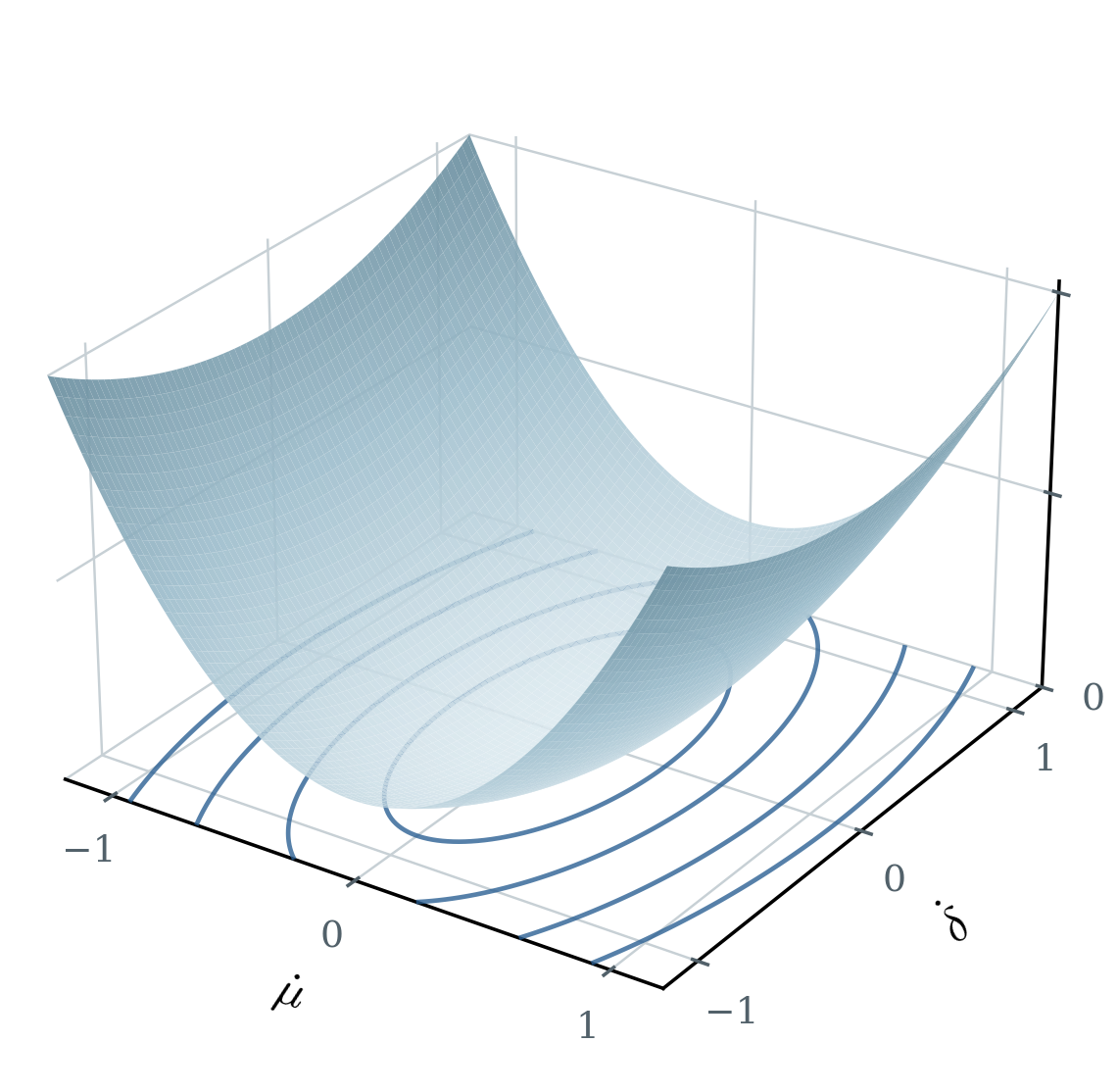}
  };

  \node[arrowcol,right=.010\linewidth of trace] (sarr) {%
    $\xrightarrow{\;s\;}$
  };

  \node[panel,right=.010\linewidth of sarr] (out) {%
    {\bfseries Return only}\\[-1pt]
    $\Gout =\tfrac12\bigl(\dd a+\dd b\bigr)^2$\\[-1pt]
    \includegraphics[width=\linewidth]{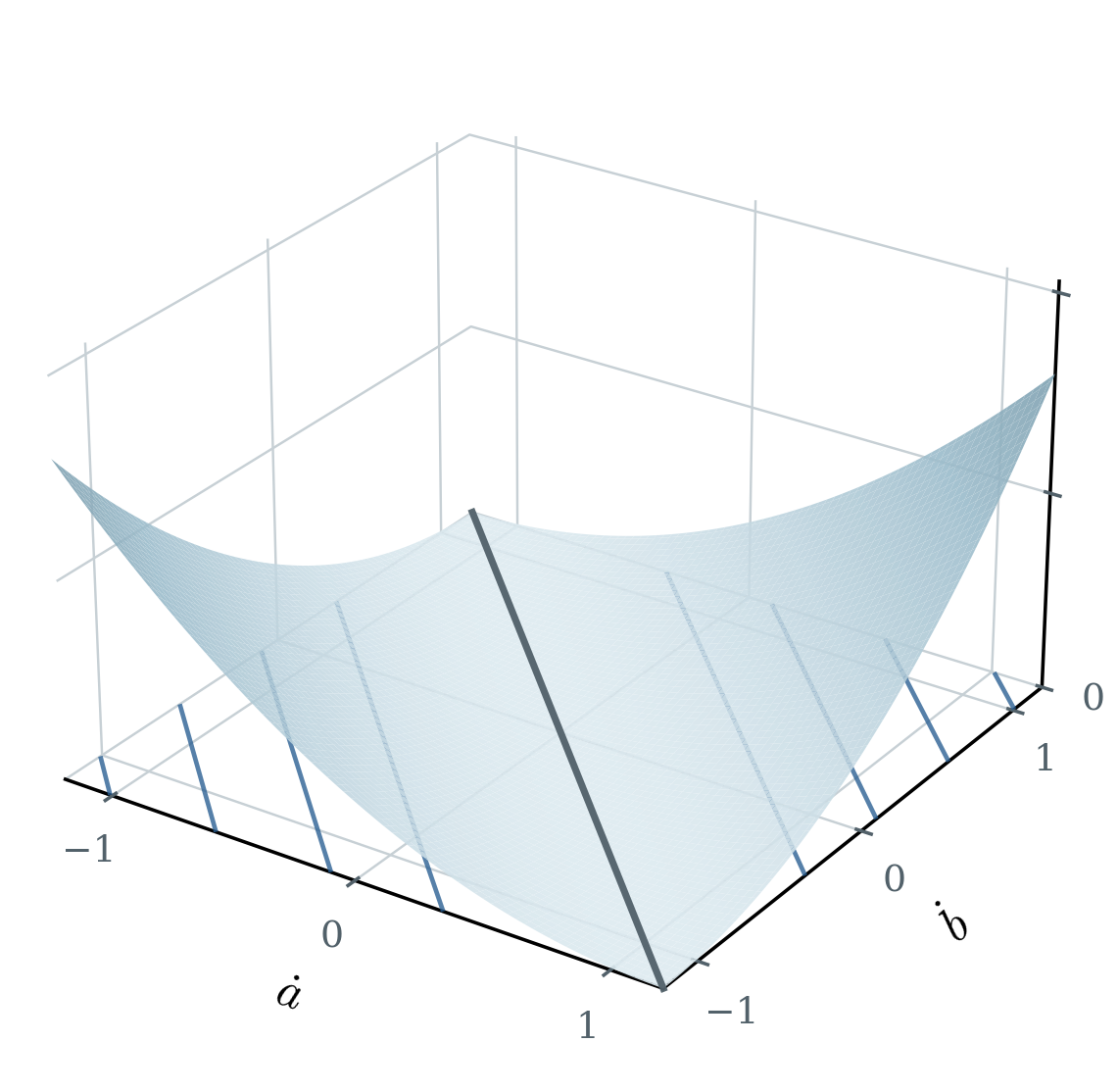}
  };
\end{tikzpicture}

%% file: figures/cor5-prop30-geometries.tex

\newcommand{\geometrycell}[2]{%
  \begin{minipage}[c][4.15cm][c]{3.42cm}
    \centering
    \includegraphics[trim=0 8pt 0 8pt,clip,width=3.28cm]{#1}\par\vspace{-5pt}
    {\scriptsize $#2$}
  \end{minipage}%
}
\newcommand{\relationcell}[1]{%
  \begin{minipage}[c][4.15cm][c]{0.42cm}
    \centering\Large $#1$
  \end{minipage}%
}
\newcommand{\programcell}[2]{%
  \begin{minipage}[c][4.15cm][c]{4.65cm}
    \centering
    \begin{minipage}{4.45cm}
      \raggedright\scriptsize #1
    \end{minipage}
    \par\vspace{5pt}
    {\scriptsize\itshape denoting}\par\vspace{2pt}
    {\scriptsize $\displaystyle #2$}
  \end{minipage}%
}
\newcommand{\progrule}{\noalign{\vskip2pt}\hline\noalign{\vskip2pt}}

\begin{minipage}{16.05cm}
\centering
\setlength{\tabcolsep}{1.2pt}
\renewcommand{\arraystretch}{1}
\begin{tabular}{@{}c@{\hspace{2pt}}|@{\hspace{2pt}}c@{}c@{}c@{}c@{}c@{}}
  \multicolumn{1}{c}{\small\bfseries Program} &
  \multicolumn{1}{c}{\small\bfseries $\GWFR$} & &
  \multicolumn{1}{c}{\small\bfseries $\Gout$} & &
  \multicolumn{1}{c}{\small\bfseries $\FisherDirac$}
  \\
  \progrule
  \programcell{%
    \(\begin{aligned}
      &(\alpha:\Rs,\,\beta:\Rs)\vdash\\[-1pt]
      &\quad \letbind{b}{\bern{\alpha}}{}\\[-1pt]
      &\qquad \ifte{b}{\return{1+\beta}}{}\\[-1pt]
      &\qquad\quad \return{0}\;\vkto\Rs
    \end{aligned}\)%
  }{%
    (1-p)\delta_0+p\delta_{1+\beta}\qquad p=\sigma(\alpha)
  }
  &
  \geometrycell{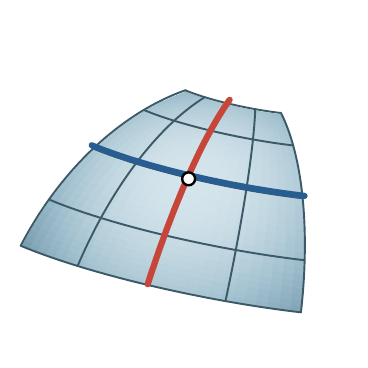}
    {p(1-p)\,d\alpha^2+p\,d\beta^2}
  & \relationcell{=}
  & \geometrycell{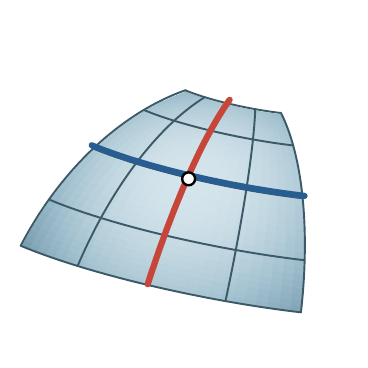}
    {p(1-p)\,d\alpha^2+p\,d\beta^2}
  & \relationcell{=}
  & \geometrycell{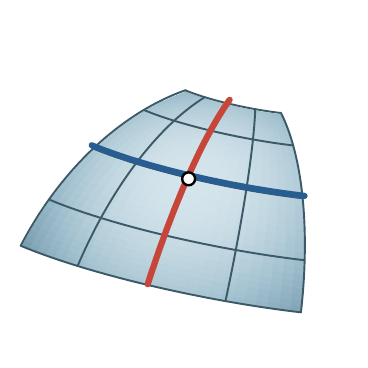}
    {p(1-p)\,d\alpha^2+p\,d\beta^2}
  \\
  \progrule
  \programcell{%
    \(\begin{aligned}
      &(\alpha:\Rs,\,\beta:\Rs)\vdash\\[-1pt]
      &\quad \letbind{b}{\bern{\alpha}}{}\\[-1pt]
      &\quad \letbind{c}{\bern{\alpha}}{}\\[-1pt]
      &\qquad \ifte{b}{\return{1+\beta}}{}\\[-1pt]
      &\qquad\quad \return{0}\;\vkto\Rs
    \end{aligned}\)%
  }{%
    (1-p)\delta_0+p\delta_{1+\beta}\qquad p=\sigma(\alpha)
  }
  &
  \geometrycell{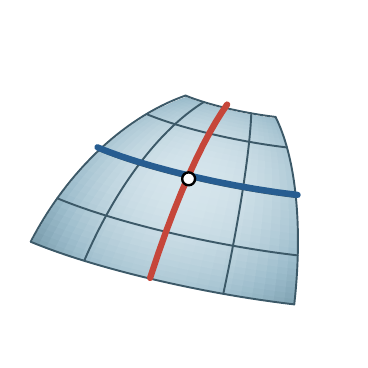}
    {p(1-p)\,d\alpha^2+p\,d\beta^2}
  & \relationcell{=}
  & \geometrycell{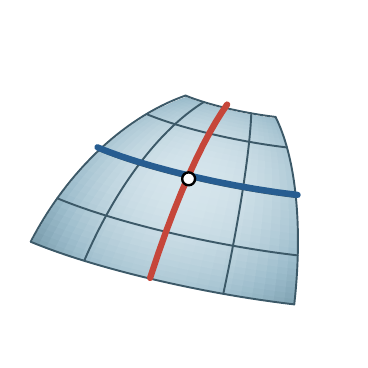}
    {p(1-p)\,d\alpha^2+p\,d\beta^2}
  & \relationcell{\prec}
  & \geometrycell{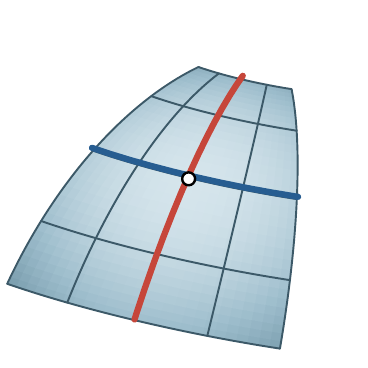}
    {2p(1-p)\,d\alpha^2+p\,d\beta^2}
  \\
  \progrule
  \programcell{%
    \(\begin{aligned}
      &(\alpha:\Rs,\,\beta:\Rs)\vdash\\[-1pt]
      &\quad \letbind{z}{\gauss{\alpha}{1}}{}\\[-1pt]
      &\quad \letbind{u}{\gauss{\alpha}{1}}{}\\[-1pt]
      &\quad \letbind{r}{\gauss{\beta}{1}}{}\\[-1pt]
      &\quad \return{(1+\beta)\cdot\coerce{z}}\;\vkto\Rs
    \end{aligned}\)%
  }{%
    \mathcal N\!\left((1+\beta)\alpha,(1+\beta)^2\right)
  }
  &
  \geometrycell{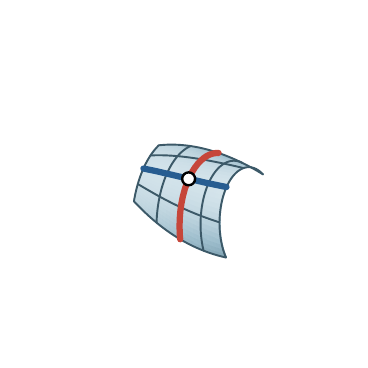}
    {\begin{gathered}
       \frac{(s\,d\alpha+\alpha\,d\beta)^2}{1+s^2}
       {}+\frac{2\,d\beta^2}{2+s^2}
     \end{gathered}}
  & \relationcell{\preceq}
  & \geometrycell{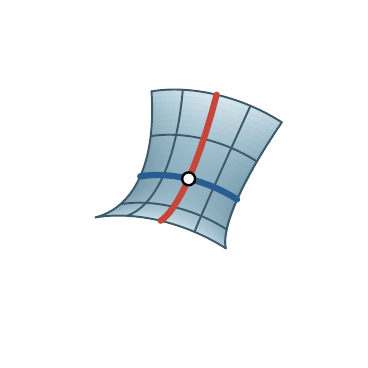}
    {d\alpha^2+(1+\alpha^2)\,d\beta^2}
  & \relationcell{\preceq}
  & \geometrycell{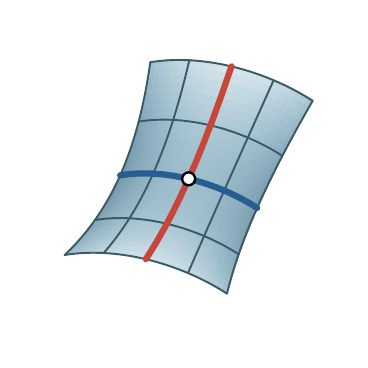}
    {2\,d\alpha^2+(2+\alpha^2)\,d\beta^2}
  \\
  \progrule
\end{tabular}
\end{minipage}

%% file: appendix/typing.tex
The structural rules apply to both languages. For $\PPL$, take $\Rs=\R$ and omit the rules involving $\Rc$.

\begin{figure}[hp]
\centering
\footnotesize

\begin{tabular}{@{}c c@{}}

\begin{minipage}[t]{0.47\linewidth}\centering
\begin{prooftree}
\AxiomC{}
\RightLabel{(Unit)}
\UnaryInfC{$\Gamma\vdash \unitelem:\unittype$}
\end{prooftree}
\end{minipage}
&
\begin{minipage}[t]{0.47\linewidth}\centering
\begin{prooftree}
\AxiomC{$(v:T)\in\Gamma$}
\RightLabel{(Var)}
\UnaryInfC{$\Gamma\vdash v:T$}
\end{prooftree}
\end{minipage}
\\[2mm]

\begin{minipage}[t]{0.47\linewidth}\centering
\begin{prooftree}
\AxiomC{$\Gamma\vdash t_1:T_1$}
\AxiomC{$\Gamma\vdash t_2:T_2$}
\RightLabel{(Pair)}
\BinaryInfC{$\Gamma\vdash (t_1,t_2):T_1\times T_2$}
\end{prooftree}
\end{minipage}
&
\begin{minipage}[t]{0.47\linewidth}\centering
\begin{prooftree}
\AxiomC{$\Gamma\vdash t:T_1\times T_2$}
\RightLabel{($\pi_1{}$)}
\UnaryInfC{$\Gamma\vdash \pi_1 t:T_1$}
\end{prooftree}
\end{minipage}
\\[2mm]

\begin{minipage}[t]{0.47\linewidth}\centering
\begin{prooftree}
\AxiomC{$\Gamma\vdash t:T_1\times T_2$}
\RightLabel{($\pi_2{}$)}
\UnaryInfC{$\Gamma\vdash \pi_2 t:T_2$}
\end{prooftree}
\end{minipage}
&
\begin{minipage}[t]{0.47\linewidth}\centering
\begin{prooftree}
\AxiomC{$\Gamma\vdash t:T_1$}
\RightLabel{(Inl)}
\UnaryInfC{$\Gamma\vdash \inleft\,t:T_1+T_2$}
\end{prooftree}
\end{minipage}
\\[2mm]

\begin{minipage}[t]{0.47\linewidth}\centering
\begin{prooftree}
\AxiomC{$\Gamma\vdash t:T_2$}
\RightLabel{(Inr)}
\UnaryInfC{$\Gamma\vdash \inright\,t:T_1+T_2$}
\end{prooftree}
\end{minipage}
&
\begin{minipage}[t]{0.47\linewidth}\centering
\begin{prooftree}
\AxiomC{$\Gamma\vdash b:\Bool$}
\AxiomC{$\Gamma\vdash t_1:T$}
\AxiomC{$\Gamma\vdash t_2:T$}
\RightLabel{(If)}
\TrinaryInfC{$\Gamma\vdash \ifte{b}{t_1}{t_2}:T$}
\end{prooftree}
\end{minipage}
\\[2mm]

\begin{minipage}[t]{0.47\linewidth}\centering
\begin{prooftree}
\AxiomC{$\Gamma\vdash t_1:U$}
\AxiomC{$\Gamma,(v:U)\vdash t_2:T$}
\RightLabel{(Let)}
\BinaryInfC{$\Gamma\vdash \letin{v}{t_1}{t_2}:T$}
\end{prooftree}
\end{minipage}

\end{tabular}

\caption{Typing rules for pure terms in $\PPL$.}
\label{fig:typing-safe-structural}
\end{figure}

\begin{figure}[hp]
\centering
\footnotesize
\setlength{\tabcolsep}{1.1em}
\renewcommand{\arraystretch}{1.25}

\begin{tabular}{@{}c c@{}}

\begin{minipage}[t]{0.47\linewidth}\centering
\begin{prooftree}
\AxiomC{$q\in\mathbb Q$}
\RightLabel{(Const$_{\square}$)}
\UnaryInfC{$\Gamma \vdash q_{\square}:\Rsquare$}
\end{prooftree}
\end{minipage}
&
\begin{minipage}[t]{0.47\linewidth}\centering
\begin{prooftree}
\AxiomC{$\Gamma \vdash t:\Rc$}
\RightLabel{(Coerce)}
\UnaryInfC{$\Gamma \vdash \coerce{t}:\Rs$}
\end{prooftree}
\end{minipage}
\\[2mm]

\begin{minipage}[t]{0.47\linewidth}\centering
\begin{prooftree}
\AxiomC{$\Gamma \vdash t:\Rc$}
\RightLabel{(IsNonNegative)}
\UnaryInfC{$\Gamma \vdash \isnneg{t}:\Bool$}
\end{prooftree}
\end{minipage}
\\[2mm]

\begin{minipage}[t]{0.47\linewidth}\centering
\begin{prooftree}
\AxiomC{$\Gamma \vdash c_1:\Rsquare$}
\AxiomC{$\Gamma \vdash c_2:\Rsquare$}
\RightLabel{($+_{\square}$)}
\BinaryInfC{$\Gamma \vdash c_1+c_2:\Rsquare$}
\end{prooftree}
\end{minipage}
&
\begin{minipage}[t]{0.47\linewidth}\centering
\begin{prooftree}
\AxiomC{$\Gamma \vdash c_1:\Rsquare$}
\AxiomC{$\Gamma \vdash c_2:\Rsquare$}
\RightLabel{($\cdot_{\square}$)}
\BinaryInfC{$\Gamma \vdash c_1\cdot c_2:\Rsquare$}
\end{prooftree}
\end{minipage}
\\[2mm]

\begin{minipage}[t]{0.47\linewidth}\centering
\begin{prooftree}
\AxiomC{$\Gamma \vdash s:\Rsquare$}
\RightLabel{($-_{\square}$)}
\UnaryInfC{$\Gamma \vdash -s:\Rsquare$}
\end{prooftree}
\end{minipage}
&
\begin{minipage}[t]{0.47\linewidth}\centering
\begin{prooftree}
\AxiomC{$
\begin{array}{c}
f\in\Prim(k,m)\\
\Gamma \vdash s_1:\Rsquare
\quad\cdots\quad
\Gamma \vdash s_k:\Rsquare
\end{array}
$}
\RightLabel{(Prim$_{\square}$)}
\UnaryInfC{$\Gamma \vdash f(s_1,\ldots,s_k):\Rsquare^m$}
\end{prooftree}
\end{minipage}
\\[2mm]
\end{tabular}

\caption{Typing rules for smooth and control real terms in $\PPL_\RR$. Rules with $\square$ are repeated twice, once for $\Rsquare = \Rs$ and once for $\Rsquare = \Rc$. For each language, $\Prim(k,m)$ denotes its chosen primitive symbols interpreted as smooth maps $\RR^k\to\RR^m$, with a separate typed copy for each $\square$. In both languages, these primitives must also be polynomially bounded in the sense of~\Cref{def:polynomially-bounded}. Note that only control $\Rc$ terms can be cast to $\Bool$. They can also be converted to $\Rs$, but not the other way around.}
\label{fig:typing-safe-reals}
\end{figure}







\begin{figure}[p]
\centering
\footnotesize
\setlength{\tabcolsep}{1.1em}
\renewcommand{\arraystretch}{1.25}

\begin{tabular}{@{}c c@{}}

\begin{minipage}[t]{0.47\linewidth}\centering
\begin{prooftree}
\AxiomC{$\Gamma \vdash M\vkto U$}
\AxiomC{$\Gamma,(v:U) \vdash N\vkto T$}
\RightLabel{(Bind)}
\BinaryInfC{$\Gamma \vdash \letbind{v}{M}{N}\vkto T$}
\end{prooftree}
\end{minipage}
&
\begin{minipage}[t]{0.47\linewidth}\centering
\begin{prooftree}
\AxiomC{$\Gamma \vdash t:T$}
\RightLabel{(Return)}
\UnaryInfC{$\Gamma \vdash \return{t}\vkto T$}
\end{prooftree}
\end{minipage}
\\[2mm]

\begin{minipage}[t]{0.47\linewidth}\centering
\begin{prooftree}
\AxiomC{$\Gamma \vdash x:\Rs$}
\RightLabel{(Bern)}
\UnaryInfC{$\Gamma \vdash \bern{x} \vkto \Bool$}
\end{prooftree}
\end{minipage}
&
\begin{minipage}[t]{0.47\linewidth}\centering
\begin{prooftree}
\AxiomC{$\Gamma \vdash x:\Rs$}
\RightLabel{(Gauss)}
\UnaryInfC{$\Gamma \vdash \gauss{x}{1_{}}\vkto \Rc$}
\end{prooftree}
\end{minipage}
\end{tabular}

\begin{prooftree}
\AxiomC{$\Gamma \vdash b:\Bool$}
\AxiomC{$\Gamma \vdash p_1\vkto T$}
\AxiomC{$\Gamma \vdash p_2\vkto T$}
\RightLabel{(If)}
\TrinaryInfC{$\Gamma \vdash \ifte{b}{p_1}{p_2}\vkto T$}
\end{prooftree}

\caption{Typing rules for probabilistic computations in $\PPL_\RR$.}
\label{fig:typing-safe-programs}
\end{figure}

%% file: appendix/dqbs.tex
\label{sec:diffeology-preliminaries}

A smooth manifold is locally modelled on a Euclidean space of one fixed
dimension. The spaces of programs used below are built instead from countable
coproducts of parameter manifolds and from quotients which identify different
programs with the same denotation. Their dimensions can jump, and the quotient
maps can have critical points. Diffeological spaces retain a useful notion of
smooth parameterised family without requiring manifold charts of fixed rank.

Diffeology controls smooth parameter dependence. Quasi-Borel structure
controls random elements and probability kernels. The two structures are kept
separate and combined only by requiring smooth plots to be measurable.

\subsection{Plots and smooth maps}

\begin{definition}[Diffeology]
Let $X$ be a set. A diffeology on $X$ is a collection $\mathcal D_X$ of maps
\[
    p:U\longrightarrow X,
\]
called plots, where $U$ ranges over open subsets of $\RR^n$ for all
$n\geq 0$, satisfying the following axioms.
\begin{enumerate}
    \item Every constant map is a plot.
    \item If $p:U\to X$ is a plot and $f:V\to U$ is an ordinary smooth map,
    then $p\circ f:V\to X$ is a plot.
    \item Plots are local. If $U=\bigcup_i U_i$ is an open cover and
    $p|_{U_i}:U_i\to X$ is a plot for every $i$, then $p$ is a plot.
\end{enumerate}
A diffeological space is a pair $(X,\mathcal D_X)$.
\end{definition}

The second axiom says that a smooth family remains smooth after a smooth
change of parameters. The third says that smoothness can be checked locally
on the parameter domain.

\begin{definition}[Smooth map]
A map $F:X\to Y$ between diffeological spaces is smooth if
\[
    F\circ p:U\longrightarrow Y
\]
is a plot of $Y$ for every plot $p:U\to X$ of $X$.
\end{definition}

Identities and composites are smooth, so diffeological spaces and smooth maps
form a category, denoted $\mathbf{Diff}$.

\begin{example}[Standard and discrete diffeologies]
\leavevmode
\begin{enumerate}
    \item A smooth manifold has its standard diffeology: the plots are the
    ordinary smooth maps from Euclidean domains. A map between manifolds is
    diffeologically smooth exactly when it is smooth in the usual sense.
    \item The discrete diffeology on a set $X$ is the smallest diffeology on
    $X$. Its plots are the locally constant maps $U\to X$.
    \item The coarse diffeology on $X$ is the largest diffeology: every map
    from a Euclidean domain is a plot.
\end{enumerate}
\end{example}

\begin{proposition}[Maps involving a discrete diffeology]
Let $X_{\mathrm{disc}}$ denote a set with the discrete diffeology.
\begin{enumerate}
    \item Every set map $X_{\mathrm{disc}}\to Y$ is smooth.
    \item A smooth map from a connected manifold $M$ to $X_{\mathrm{disc}}$
    is constant. More generally, a smooth map from a manifold to
    $X_{\mathrm{disc}}$ is locally constant.
\end{enumerate}
\end{proposition}

\begin{proof}
The composite of a locally constant plot with any set map is locally constant,
and every locally constant map is a plot. Conversely, a smooth map
$F:M\to X_{\mathrm{disc}}$ must send every ordinary smooth plot of $M$ to a
locally constant plot. Applying this to coordinate charts shows that $F$ is
locally constant.
\end{proof}

\subsection{Subspaces, products, coproducts, and quotients}

The usual set-theoretic constructions carry canonical diffeologies.

\begin{definition}[Generated diffeologies]
Let $X$ and $Y$ be diffeological spaces.
\begin{enumerate}
    \item A subset $A\subseteq X$ has the subset diffeology: a map
    $p:U\to A$ is a plot exactly when its composite with $A\hookrightarrow X$
    is a plot of $X$.
    \item The product $X\times Y$ has the product diffeology: a map
    $p:U\to X\times Y$ is a plot exactly when both coordinate maps are plots.
    \item A coproduct $\coprod_{i\in I}X_i$ has the coproduct, or sum,
    diffeology: a map $p:U\to\coprod_iX_i$ is a plot if every point of $U$ has
    a neighbourhood $V$ on which $p$ lands in one summand $X_i$ and, viewed
    as a map $V\to X_i$, is a plot.
\end{enumerate}
\end{definition}

The coproduct diffeology is particularly important for presentation spaces.
A smooth family in a coproduct locally has a fixed tag. Thus a program
skeleton is locally constant along an ordinary smooth parameter family, even
though different neighbourhoods may use different skeletons.

\begin{proposition}[Universal property of the final diffeology]
Let $X$ carry the final diffeology generated by $(e_i)_{i\in I}$. A set map
$F:X\to Z$ is smooth if and only if every composite
\[
    F\circ e_i:X_i\longrightarrow Z
\]
is smooth.
\end{proposition}

\begin{proof}
One direction follows by composition. For the other, let $p:U\to X$ be a
plot. Locally, $p=e_i\circ\widetilde p$, so
$F\circ p=(F\circ e_i)\circ\widetilde p$ is locally a plot. The locality axiom
then implies that $F\circ p$ is a plot on all of $U$.
\end{proof}

If $q:X\to X/{\sim}$ is a quotient map, its quotient diffeology is the final
diffeology generated by $q$. A plot in the quotient is therefore a map which
locally admits a smooth lift to $X$. No constant-rank or regular-value
hypothesis is needed.

\subsection{Topology, internal tangents, and plotwise forms}

Every diffeological space has an underlying topology, but the diffeology
contains more information than this topology.

\begin{definition}[$D$-topology]
A subset $O\subseteq X$ is $D$-open if
\[
    p^{-1}(O)\subseteq U
\]
is open for every plot $p:U\to X$. The resulting topology is called the
$D$-topology of $X$.
\end{definition}

Every smooth map is continuous for the $D$-topologies. Nevertheless, two
different diffeologies can induce the same topology, so smooth structure
should be defined through plots rather than reconstructed from open sets.

\subsection{Quasi-Borel spaces and probability}

Diffeology does not by itself specify which random elements or probability
kernels are admissible. For this we use a quasi-Borel structure.

\begin{definition}[Quasi-Borel space]
A quasi-Borel space is a set $X$ equipped with a collection
\[
    \Meas_X\subseteq X^\RR
\]
of maps $\alpha:\RR\to X$, called random elements, satisfying:
\begin{enumerate}
    \item every constant map belongs to $\Meas_X$;
    \item if $\alpha\in\Meas_X$ and $f:\RR\to\RR$ is Borel measurable, then
    $\alpha\circ f\in\Meas_X$;
    \item if $(A_i)_{i\in\N}$ is a Borel measurable partition of $\RR$ and
    $\alpha_i\in\Meas_X$, then the map
    \[
        \alpha(r)=\alpha_i(r)\quad\text{for }r\in A_i
    \]
    belongs to $\Meas_X$.
\end{enumerate}
A map $F:X\to Y$ is a quasi-Borel morphism if
$F\circ\alpha\in\Meas_Y$ for every $\alpha\in\Meas_X$.
\end{definition}

Every standard Borel space has a canonical quasi-Borel structure in which the
random elements are exactly the Borel measurable maps $\RR\to X$. On standard
Borel spaces, quasi-Borel morphisms are the usual measurable maps. The benefit
of the quasi-Borel presentation is that it also gives well-behaved spaces of
functions and random functions.

A probability law on a quasi-Borel space $X$ is represented by a random
element $\alpha\in\Meas_X$ together with an ordinary probability measure
$\nu$ on $\RR$. Two representations are identified when they give the same
integral against every bounded quasi-Borel morphism $X\to\RR$. We denote the resulting
space of laws by $\Prob(X)$.

\begin{proposition}[Quasi-Borel function spaces and probability]
The category $\mathbf{QBS}$ of quasi-Borel spaces is cartesian closed. Its
exponential $Y^X$ has underlying set $\Hom_{\mathbf{QBS}}(X,Y)$, and a map
$\alpha:\RR\to Y^X$ is a random element exactly when its uncurrying
\[
    \widetilde\alpha:\RR\times X\longrightarrow Y,
    \qquad
    \widetilde\alpha(r,x)=\alpha(r)(x),
\]
is a quasi-Borel morphism.

The construction $\Prob$ extends to a commutative affine probability monad on
$\mathbf{QBS}$. Its Kleisli morphisms $X\to\Prob(Y)$ are quasi-Borel
probability kernels, and the Kleisli category $\operatorname{Kl}(\Prob)$ is a
Markov category. On standard Borel spaces this recovers ordinary probability
measures and Markov kernels.
\end{proposition}

%% file: appendix/bernoulli_example.tex
\subsubsection*{Bernoulli regression as the score-only case}
\label{subsec:bernoulli-regression}

Consider the program modelling $d$-dimensional Bernoulli regression:
\[
\theta:\Rs^d, x:\Rs^d
\vdash
p\df\bern{\theta\transp x}
\vkto\Bool
\]
The output $b\in\{0,1\}$ has probability $P(b=1\mid\theta,x)=\logistic(\theta\transp x)$.
Let us denote:
\[
s=\theta\transp x
\qquad
p_{\theta,x}=\logistic(s)
\]
The trace and output then coincide. For a tangent $(\xi_\theta,\xi_x)\in T_\theta\RR^d\oplus T_x\RR^d$:
\[
Ds_{(\theta,x)}[(\xi_\theta,\xi_x)]
=\xi_\theta\transp x+\theta\transp \xi_x
\]
and hence
\[
S_{(\theta,x),(\xi_\theta,\xi_x)}(b)
=(b-p_{\theta,x})(\xi_\theta\transp x+\theta\transp \xi_x),
\qquad
V=0.
\]
Therefore the score-transport form on the whole context is
\[
(G_p^{\kappa,h})_{(\theta,x)}((\xi_\theta,\xi_x),(\eta_\theta,\eta_x))
=p_{\theta,x}(1-p_{\theta,x})
(\xi_\theta\transp x+\theta\transp \xi_x)
(\eta_\theta\transp x+\theta\transp \eta_x).
\]
Thus real covariates are differentiated exactly like real parameters, and holding $x$ fixed later is a pullback of this tensor.

\paragraph{Fixed-covariate experiment.}
Suppose observed data are $(x_i,b_i)_{i=1}^n$, and the unknown state is $m=\theta$. The experiment maps are then:
\[
r_i(\theta)=(\theta,x_i),
\qquad
Dr_i[\xi]=(\xi,0).
\]
The experiment metric is:
\[
G_D(\theta)
=\sum_{i=1}^n
p_i(1-p_i)x_ix_i\transp,
\qquad
p_i=\logistic(\theta\transp x_i)
\]
The log-likelihood covector is:
\[
\alpha_D(\xi)
=\sum_{i=1}^n(b_i-p_i)x_i\transp \xi
\]
so:
\[
\nabla\ell_D(\theta)=\sum_{i=1}^n(b_i-p_i)x_i
\qquad
\NatGrad\ell_D
=\left(\sum_i p_i(1-p_i)x_ix_i\transp\right)^\dagger
\left(\sum_i(b_i-p_i)x_i\right)
\]

\paragraph{Errors-in-variables variant.}
If the covariates are also unknown, we can take:
\[
m=(\theta,x_1,\ldots,x_n)
\qquad
r_i(m)=(\theta,x_i)
\]
Then:
\[
Dr_i[(\xi_\theta,\xi_{x_1},\ldots,\xi_{x_n})]=(\xi_\theta,\xi_{x_i})
\]
and the same full-context score-transport form differentiates both $\theta$ and the latent covariates $x_i$. Whether a real variable is called a parameter, input, or latent datum is determined by the experiment map, not by the program semantics.

%% file: appendix/gaussian_example.tex
As an example of the whole construction, we present a calculation of the semantics on a more complex program of a Gaussian mixture. Let us compose two programs, $\Gamma_1 \vdash p_1 \vkto \Rc$ given by:
\[
\theta : \Rs \vdash \gauss{\theta}{1} \vkto \Rc
\]
and $\Gamma_2 \vdash p_2 \vkto \Rs $ given by:
\begin{align*}
a: \Rs, b : \Rs, z: \Rc \vdash\ & \ifte{\isnneg{z}\\}
    {&
     \letbind{v}{\gauss{b}{1}}{\return{\chi(\coerce{v})}\\}
    }
    {& \letbind{u}{\gauss{a}{1}}{\return{-_s\chi(\coerce{u})}}}\vkto \Rs
\end{align*}
where we write $\chi(x) = x + \sqrt{1 + x^2}$ to be a smooth diffeomorphism $\chi: \RR \to (0, \infty)$. (It can be assumed that $\chi \in \Prim(\PPL_\RR)$ as it has polynomially bounded tails, see~\Cref{remark:bounded-tails-primitives}).

Let $F_1:\RRs^3\to\RRc$ be the (weakened) denotation of $p_1$, and let $F_2:\RRs^3\times\RRc\to\RRs$ be the denotation of $p_2$. The bind rule gives the composite:
\[
F=\Trace_\RR(\letbind{z}{p_1}{p_2})
=F_2\circ(\id_{\RRs^3}\otimes F_1)\circ\Delta_{{\RRs}^3}:\RRs^3\to\RRs
\]

The trace space is isomorphic to $\Omega=\RR^2\sqcup\RR^2$. Writing the two components as $(z,u)_-$ and $(z,v)_+$, the density and output maps are:
\begin{align*}
q_-(\theta,a,b;z,u)
&=\Ind_{\{z<0\}}\varphi(z-\theta)\varphi(u-a),
& o_-(z,u)&=-\chi(u),\\
q_+(\theta,a,b;z,v)
&=\Ind_{\{z\geq0\}}\varphi(z-\theta)\varphi(v-b),
& o_+(z,v)&=\chi(v),
\end{align*}
where $\varphi$ is the standard Gaussian density. For a tangent $\xi=(\xi_\theta,\xi_a,\xi_b)$, the score-velocity pairs on the two support components are
\begin{align*}
S^-_\xi(z,u)
&=(z-\theta)\xi_\theta+(u-a)\xi_a,
&V^-_\xi(z,u)&=0,\\
S^+_\xi(z,v)
&=(z-\theta)\xi_\theta+(v-b)\xi_b,
&V^+_\xi(z,v)&=0.
\end{align*}
Put
\[
w(\theta)=\Phi(-\theta),
\qquad
1-w(\theta)=\Phi(\theta).
\]
With the Euclidean output tensor $h=\dd y^2$, the trace-level score-transport metric $G^{\mathrm{tr}}\df\Affine_F(\kappa h)$ has no transport summand and matrix
\[
G^{\mathrm{tr}}(\theta,a,b)
=\begin{pmatrix}
1&0&0\\
0&w(\theta)&0\\
0&0&1-w(\theta)
\end{pmatrix}.
\]

Let $\mu^-_a$ be the law of $-\chi(U)$ for $U\sim N(a,1)$, and let $\mu^+_b$ be the law of $\chi(V)$ for $V\sim N(b,1)$. Their supports are respectively $(-\infty,0)$ and $(0,\infty)$, and the realised output law is
\[
\mu_{\theta,a,b}
=w(\theta)\mu^-_a+(1-w(\theta))\mu^+_b.
\]
The output reveals the branch and the second Gaussian trace, but only the sign of $z$. Since $w'(\theta)=-\varphi(\theta)$, the conditional output scores are
\[
s^{\mathrm{out}}_-(y)
=\begin{pmatrix}
\dfrac{w'}{w}\\[3pt]
\chi^{-1}(-y)-a\\
0
\end{pmatrix}
\quad(y<0),
\qquad
s^{\mathrm{out}}_+(y)
=\begin{pmatrix}
-\dfrac{w'}{1-w}\\[3pt]
0\\
\chi^{-1}(y)-b
\end{pmatrix}
\quad(y>0).
\]
The projected score-transport lift is $(s^{\mathrm{out}},0)$, and hence its matrix is
\[
G^{\mathrm{out}}(\theta,a,b)
=\begin{pmatrix}
\dfrac{\varphi(\theta)^2}{\Phi(-\theta)\Phi(\theta)}&0&0\\[7pt]
0&\Phi(-\theta)&0\\
0&0&\Phi(\theta)
\end{pmatrix}.
\]
Thus
\[
G^{\WFR,\kappa}\preceq G^{\mathrm{out}}\preceq G^{\mathrm{tr}},
\qquad
G^{\mathrm{out}}_{\theta\theta}
=\frac{\varphi(\theta)^2}{\Phi(-\theta)\Phi(\theta)}
<1
=G^{\mathrm{tr}}_{\theta\theta}.
\]
The second inequality is the loss of the magnitude of $z$ under observation. The first need not be an equality, since this is a continuous output family and $(s^{\mathrm{out}},0)$ need not be the minimum-energy reaction-transport representation.

\subsubsection*{Parameter sharing}
Suppose $a=b=m$. The reparametrisation is
\[
r:\RR^2\to\RR^3,
\qquad
r(\theta,m)=(\theta,m,m),
\qquad
Dr=\begin{pmatrix}
1&0\\
0&1\\
0&1
\end{pmatrix}.
\]
The two experiment metrics pull back to
\[
r^*G^{\mathrm{tr}}
=Dr\transp G^{\mathrm{tr}}Dr
=\begin{pmatrix}
1&0\\
0&1
\end{pmatrix},
\qquad
r^*G^{\mathrm{out}}
=Dr\transp G^{\mathrm{out}}Dr
=\begin{pmatrix}
\dfrac{\varphi(\theta)^2}{\Phi(-\theta)\Phi(\theta)}&0\\[7pt]
0&1
\end{pmatrix}.
\]
Thus parameter sharing is ordinary deterministic pullback of the chosen score-transport geometry.

\subsubsection*{Different implementation of the same kernel}
For the same Euclidean output tensor, the same realised output family is obtained by sampling $u,v\sim N(0,1)$ and returning $-\chi(a+\coerce{u})$ or $\chi(b+\coerce{v})$. In that presentation, $a,b$ disappear from the score and enter the output velocity. Its trace-level metric is:
\[
\widetilde G^{\mathrm{tr}}
=\diag\left(
1,
\kappa w(\theta)c(a),
\kappa(1-w(\theta))c(b)
\right),
\qquad
c(t)=\E_{U\sim N(0,1)}\left[\chi'(t+U)^2\right].
\]
The realised kernels agree, but the presentation-level score-transport metrics need not, because the score-transport is intensional. Both dominate the same extensional $\WFR$ form.